\documentclass[10pt,aps,prl,twocolumn,superscriptaddress,floatfix,longbibliography]{revtex4-1}

\pdfoutput=1
\usepackage{amssymb}
\usepackage{amsmath}
\usepackage{graphicx}
\usepackage{dcolumn}
\usepackage{xcolor}
\usepackage{bm}
\usepackage{enumitem}
\usepackage{amsfonts}%
\usepackage{bbold}
\usepackage[colorlinks=true, allcolors=blue]{hyperref}
\usepackage{amsmath}

\usepackage{ulem}

\providecommand{\U}[1]{\protect\rule{.1in}{.1in}}

\begin{document}

\title{Coexistence of $s$-wave superconductivity and phase separation in the \\half-filled extended Hubbard model with attractive interactions}

\author{E. Linn\'er}
\affiliation{CPHT, CNRS, {\'E}cole polytechnique, Institut Polytechnique de Paris, 91120 Palaiseau, France}

\author{C. Dutreix}
\affiliation{
Univ. Bordeaux, CNRS, LOMA, UMR 5798, F-33400 Talence, France}

\author{S. Biermann}
\affiliation{CPHT, CNRS, {\'E}cole polytechnique, Institut Polytechnique de Paris, 91120 Palaiseau, France}
\affiliation{Coll\`ege de France, 11 place Marcelin Berthelot, 75005 Paris, France}
\affiliation{Department of Physics, Division of Mathematical Physics, Lund University, Professorsgatan 1, 22363 Lund, Sweden}
\affiliation{European Theoretical Spectroscopy Facility, 91128 Palaiseau, France}

\author{E. A. Stepanov}
\affiliation{CPHT, CNRS, {\'E}cole polytechnique, Institut Polytechnique de Paris, 91120 Palaiseau, France}

\begin{abstract}
Understanding competing instabilities in systems with correlated fermions remains one of the holy grails of modern condensed matter physics.
Among the fermionic lattice models used to this effect, the extended Hubbard model occupies a prime place due to the potential relevance of its repulsive and attractive versions for both electronic materials and artificial systems.
Using the recently introduced multi-channel fluctuating field approach, we address the interplay of charge density wave, $s$-wave superconductivity, and phase separation fluctuations in the attractive extended Hubbard model.
Despite of the fact that this model has been intensively studied for decades, our novel approach has allowed us to identify a novel phase that is characterised by the coexistence of $s$-wave superconductivity and phase separation.
Our findings resonate with previous observations of interplaying phase separation and superconducting phases in electronic systems, most importantly in high-temperature superconductors.
\end{abstract}

\maketitle

Materials with strong electronic correlations exhibit sophisticated phase diagrams incorporating a complex selection of collective ordering phenomena.
The latter are associated with a variety of interplaying instabilities, e.g. charge, spin, or pairing fluctuations.
They may appear either in a mutually exclusive form~\cite{PhysRevB.29.5089,PhysRevLett.53.2327}, or with a stabilisation of additional intermediate phases~\cite{PhysRevLett.53.2327,doi:10.1143/JPSJ.68.3123,PhysRevB.61.16377,PhysRevLett.92.236401}, e.g., when different fluctuations coexist~\cite{PhysRevB.14.2989}.
An interplay between collective charge, spin and pairing fluctuations occur already in the single-band extended Hubbard model~\cite{rspa.1963.0204, PhysRevLett.10.159,10.1143/PTP.30.275, rspa.1964.0190, PhysRevB.3.2662, PhysRevB.14.2989}.
Their competition is determined by two parameters: the local $U$ and non-local $V$ interactions between electrons.
Thus, the model is a suitable framework for a well-controlled investigation of competing instabilities in correlated electronic systems.
The local interaction stabilizes collective spin and pseudo-spin fluctuations in the repulsive~\cite{PhysRev.157.295, PhysRevB.80.245102, PhysRevLett.114.236402, PhysRevB.96.041105} and attractive~\cite{10.1143/PTP.48.2171,PhysRevB.14.2989,PhysRevLett.62.1407} regimes, respectively.
Here pseudo-spin fluctuations are associated with $\eta$-pairing, combining the charge density and $s$-wave pairing degrees of freedom~\cite{PhysRevLett.63.2144, PhysRevLett.65.120}.
The spin and pseudo-spin fluctuations may compete with charge fluctuations that are driven by the non-local interaction~\cite{PhysRevB.3.2662, Vonsovsky_1979,PhysRevB.76.085115}. 
Strong charge fluctuations may result in the development of the charge density wave (CDW) and phase separation (PS) phases in the repulsive and attractive $V$ cases, respectively.

Significant insights into the collective electronic behavior in the repulsive $U,V$ regime of the extended Hubbard model exist due to extensive research conducted since the 1970's~\cite{PhysRevB.3.2662, PhysRevB.14.2989, doi:10.1080/00018737900101375, emery1979, PhysRevB.29.5089, PhysRevB.29.5096, PhysRevLett.53.2327,doi:10.1143/JPSJ.68.3123,PhysRevB.61.16377,PhysRevB.39.9397, PhysRevB.42.465, PhysRevB.48.7140, PhysRevB.70.235107, PhysRevLett.92.236401, PhysRevB.74.035113, PhysRevB.76.085115, PhysRevB.87.125149, PhysRevB.90.235105, Werner_2016, PhysRevB.94.165141, PhysRevB.94.205110, PhysRevB.95.115149, van_Loon_2018, RevModPhys.90.025003, PhysRevB.99.115112, PhysRevB.99.115124, PhysRevB.99.245146, PhysRevB.100.075108, stepanov2021coexisting, PhysRevB.102.195109, PhysRevB.104.085129, PhysRevB.106.195121, 2022arXiv221005540L, https://doi.org/10.48550/arxiv.2301.07162}.
In contrast, much less attention has been paid to the regime of attractive $U$, dominated by charge fluctuations and $s$-wave superconductivity (\mbox{s-SC})~\cite{PhysRevB.14.2989, PhysRevB.23.1447, PhysRevLett.62.1407, PhysRevLett.66.946, PhysRevLett.69.2001, RevModPhys.62.113, PhysRevLett.86.4612, PhysRevLett.88.126403, PhysRevLett.114.236402, PhysRevB.61.7028, PhysRevB.49.3548, PhysRevB.72.235118, PhysRevB.72.024517, PhysRevB.76.085115, Toschi_2005, PhysRevB.72.024517, PhysRevB.81.014523, PhysRevA.84.023638, PhysRevA.89.053604, van_Loon_2018, PhysRevB.94.155114, PhysRevB.99.045137, Xiang_2019, PhysRevB.106.195121, https://doi.org/10.48550/arxiv.2206.01119}.
Although the Coulomb interaction is repulsive, it is known that coupling electrons to external degrees of freedom, e.g. phonons, may lead to an effective electronic system with attractive interactions.
Specific examples are doped fullerenes~\cite{PhysRevLett.67.3452} and one-dimensional copper oxide chains~\cite{PhysRevLett.127.197003}, Ba$_{1-x}$K$_{x}$BiO$_3$ materials~\cite{PhysRevB.57.14453, PhysRevX.3.021011}, LaAlO$_3$/SrTiO$_3$ interfaces~\cite{nature14398,PhysRevX.6.041042,PhysRevLett.117.096801,s41467-017-00495-7}, and select $d$- and $f$- transition metals~\cite{PhysRevB.37.10674, PhysRevB.90.155108}.
In addition, fermionic systems with attractive local interactions are realizable in cold atom experiments~\cite{doi:10.1146/annurev-conmatphys-070909-104059}.

In this Letter, we focus on the leading collective electronic fluctuations in the half-filled extended Hubbard model with an attractive local interaction $U$. 
We consider both, the repulsive and attractive cases for the non-local interaction $V$ between neighboring sites on a square lattice.   
This allows us to investigate the interplay between the CDW, PS, and \mbox{s-SC} instabilities that appear in the system.
To this aim, we employ the multi-channel fluctuating field (MCFF) approach~\cite{2022arXiv221005540L}, based on the earlier introduced fluctuating local field method~\cite{PhysRevE.97.052120,PhysRevB.102.224423, PhysRevB.105.035118, s10948-022-06303-8}. 
Within this approach, a trial system incorporating the main leading collective fluctuations is constructed based on a variational optimisation with respect to a reference system.
The construction allows us to treat competing fluctuations without any explicit symmetry breaking, thus respecting the Mermin-Wagner theorem~\cite{PhysRev.158.383, PhysRevLett.17.1133, PhysRev.171.513}.
Note, a ``phase'' will in the current work refer to a broader definition including short-range ordering, i.e. transforming into a true phase within a quasi-two-dimensional system.
We find that the emergence of a phase combining CDW and s-SC fluctuations is correctly captured at vanishing non-local $V$, signaling the emergent pseudo-spin SU(2) symmetry of the model.
In addition, we discover a novel coexistence phase composed of PS and s-SC fluctuations spanning a relatively broad region of the attractive $U$-$V$ phase diagram.
Our results obtained within a simple quantum lattice system call for further investigations of novel collective phenomena due to interplaying fluctuations in realistic materials.

We consider the single-band extended Hubbard model at half-filling on a square lattice, defined by the Hamiltonian:
\begin{align}
\hat{H} = 
- t \sum_{\langle i,j \rangle,\sigma} \hat{c}^\dagger_{i \sigma} \hat{c}^{\phantom{\dagger}}_{j \sigma} 
+ U \sum_{i} \hat{n}_{i \uparrow} \hat{n}_{i \downarrow} 
+ \frac{V}{2} \sum_{\langle i,j\rangle,\sigma\sigma'} \hat{n}_{i \sigma} \hat{n}_{j \sigma'},
\label{Eq:EHubbard_Ham}
\end{align}
where the $\hat{c}_{i\sigma}^{(\dagger)}$ operators correspond to annihilation (creation) of electrons and ${\hat{n}^{\phantom{\dagger}}_{i \sigma} = \hat{c}^\dagger_{i \sigma} \hat{c}^{\phantom{\dagger}}_{i \sigma}}$ are the electronic densities, with the subscripts denoting the position $i$ and spin projection ${\sigma \in \{ \uparrow, \downarrow \}}$.
The kinetics is modeled by a nearest-neighbor hopping amplitude $t$ and the interaction is modeled by the on-site $U$ and the nearest-neighbor $V$ interactions.
Our considerations are limited to attractive $U$, while $V$ may be both repulsive and attractive.

The attractive $U$ regime is dominated by charge and $s$-wave pairing fluctuations.
A natural description combining the two channels is the pseudo-spin, conveniently written using the Nambu basis: {${\hat{\psi}_{\bf{k},\omega,\uparrow} =  \hat{c}_{\bf{k}\omega\uparrow}}$}, {${\hat{\psi}_{\bf{k},\omega,\downarrow} =  \hat{c}^\dagger_{-\bf{k}+\bf{Q},\omega\downarrow}}$}, {${\hat{\psi}^\dagger_{\bf{k},\omega,\uparrow} =  \hat{c}^\dagger_{\bf{k}\omega\uparrow}}$}, and {${\hat{\psi}^\dagger_{\bf{k},\omega,\downarrow} =  \hat{c}_{-\bf{k}+\bf{Q},\omega\downarrow}}$}.
Within this basis, the pseudo-spin density operator is defined as:
\begin{align}
\hat{n}^{\varsigma}_{{\bf{Q}}} & \equiv \frac{1}{\beta{}N}\sum_{{\bf k}, \nu, \sigma\sigma'} \hat{\psi}^{\dagger}_{\bf{k+Q},\nu\sigma}\sigma^{\varsigma}_{\sigma\sigma'}\hat{\psi}^{\phantom{\dagger}}_{{\bf k}\nu\sigma'},
\end{align}
with the inverse temperature $\beta$ and number of sites $N$, and where the subscripts denote the momentum $\bf{k}$ and the fermionic Matsubara frequency $\nu$.
Here the mode is specified by the channel ${\varsigma \in \{x,y,z\}}$, where ${\bf Q}$ the ordering vector, and $\sigma^{\varsigma}$ are the Pauli spin matrices.
Hence, $\hat{n}^{\varsigma}_{{\bf{Q}}}$ refers to the $s$-wave pairing ($\varsigma \in \{x,y\}$) and the charge fluctuations ($\varsigma \in \{z\}$).
The Nambu basis allows for a clear exhibition of the emergence of the SU(2) pseudo-spin symmetry at half-filling in the absence of the non-local interaction $V$~\cite{PhysRevLett.63.2144, PhysRevLett.65.120}.
In fact, the staggered particle-hole symmetry of the Hubbard model ($V=0$) relates the spin and pseudo-spin degrees of freedom~\cite{PhysRevLett.63.2144, PhysRevLett.65.120}.
Within the charge and $s$-wave pairing channels, our work focuses on the leading instabilities: the CDW, PS and s-SC orderings.
All three orderings are determined by their respective order parameters, given by the expectation value of the (static) operator $\hat{n}^{\varsigma}_{{\bf{Q}}}$.
Here, s-SC and CDW are associated with momenta $\bf{Q} = (\pi,\pi)$, and PS are associated with momenta ${\bf{Q}} = (0,0)^+$.

To study the competing instabilities, we employ the multi-channel fluctuating field (MCFF) method~\cite{2022arXiv221005540L}.
The decisive advantage of this numerical method is the ability to account for the leading fluctuations and their interplay exactly. 
This approach is based on the construction of an effective action ${\mathcal{S}^{*}}$,
where the fluctuations in the charge (CDW, PS) and superconducting channels (s-SC) are incorporated via the associate classical fields $\phi^\varsigma_{\bf{Q}}$ coupled to the respective components of $\hat{n}^{\varsigma}_{{\bf{Q}}}$ (see Supplemental Material (SM)~\cite{SM} for details).
This construction is determined by the Peierls-Feynman-Bogoliubov variational principle~\cite{PhysRev.54.918, Bogolyubov:1958zv, feynman1972}, with the extended Hubbard model as a reference system.
Within the MCFF approach, the interplay between different fluctuations may be determined by a single-channel free energy ${{\cal F}(\phi^\varsigma_{\bf{Q}})}$.
The functional ${{\cal F}(\phi^\varsigma_{\bf{Q}})}$ is constructed with respect to a classical field $\phi^\varsigma_{\bf{Q}}$, after integrating out analytically the fermionic degrees of freedom and numerically the remaining classical fields. 
Phase transitions are then identified by the development of global minima of the single-channel free energy at ${\phi^\varsigma_{\bf{Q}} \neq 0}$, akin to a Mexican hat potential.
In contrast, a local minimum at ${\phi^\varsigma_{\bf{Q}} \neq 0}$ signals metastable collective fluctuations.
To obtain further insight into the interplay between collective fluctuations, it is also useful to calculate the corresponding order parameters $\langle \hat{n}^\varsigma_{\bf{Q}}\rangle$.
This can be done by substituting the ${{\cal F}(\phi^\varsigma_{\bf{Q}})}$ saddle-point value of the classical field ${\phi^\varsigma_{\bf{Q}}}$ in the effective action ${\mathcal{S}^{*}}$ (see SM~\cite{SM} for details).

\begin{figure}[t!]
\includegraphics[width=1.\linewidth]{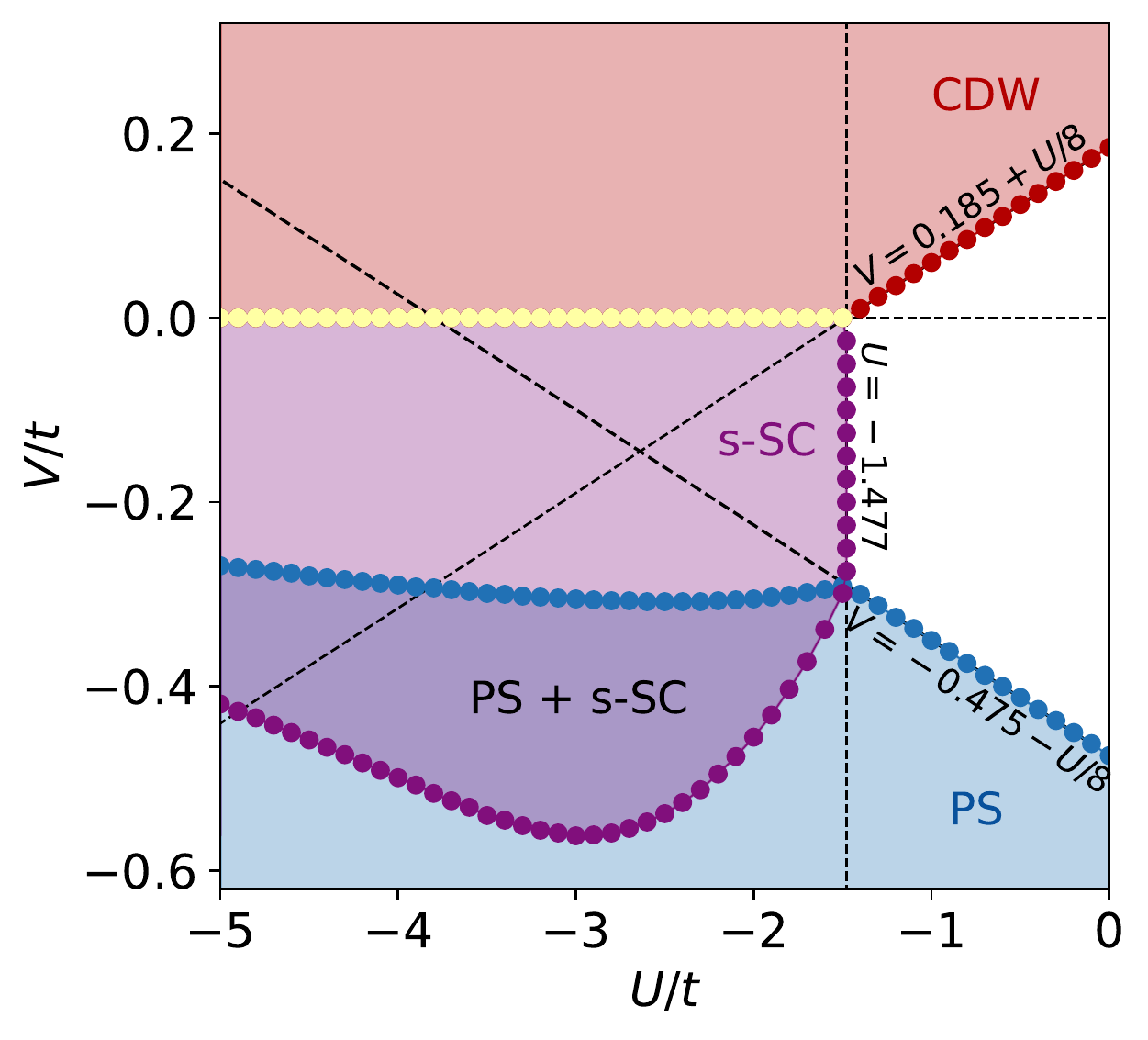}
\caption{Phase diagram of the half-filled extended Hubbard model for attractive $U$. It is obtained from the MCFF method for a ${128\times 128}$ square lattice with periodic boundary conditions at inverse temperature ${\beta=10/t}$. This shows the existence of a novel phase ``\mbox{PS + s-SC}'' where PS and \mbox{s-SC} coexist, in addition to the conventional CDW (red), s-SC (purple), and PS (blue) phases.
The yellow line specifies the CDW and \mbox{s-SC} coexistence in the attractive Hubbard model ($V=0$).
\label{figure1}}
\end{figure}

We perform calculations at ${\beta = 10/t}$ for the half-filled extended Hubbard model close to the thermodynamic limit for a square lattice of ${N=128\times 128}$ sites with periodic boundary conditions.
Fig.~\ref{figure1} shows the resulting $U$-$V$ phase diagram. It consists of six phases: normal metal (white), CDW (red), \mbox{s-SC} (purple), PS (blue), and two phases, where \mbox{s-SC} coexists with either CDW (yellow line), or with PS (labelled ``\mbox{PS + s-SC}'') orderings.
At weak coupling (${|U| \lesssim 1.5}$), the CDW and PS phase boundaries follow the ${V = 0.185 + U/8}$ and ${V = -0.475 - U/8}$ lines, respectively.
These asymptotics are identical to the perturbative estimates for the phase boundaries obtained using the random phase approximation (RPA) in the ${U\to0}$ limit.
We note, that the MCFF approach correctly captures the exact ${U\to0}$ limit for the CDW phase boundary, as observed previously in Ref.~\cite{2022arXiv221005540L}.
In contrast, the PS boundary is slightly overestimated, as an extrapolated ${U\to0}$ dual boson result for the PS transition point gives ${V^{\rm PS}_{U=0} \simeq -0.54}$~\cite{van_Loon_2018}.
In agreement with the fluctuating exchange (FLEX) result obtained for ${V=0}$, the MCFF \mbox{s-SC} phase boundary in the weak coupling regime follows the ${U^{\rm s-SC}_{V=0} = -1.478}$ line.
FLEX is known to overestimate the strength of antiferromagnetic (AFM) fluctuations at ${V=0}$.
Therefore, by the staggered particle-hole symmetry of the Hubbard model relating the spin and pseudo-spin degrees of freedom~\cite{PhysRevLett.63.2144, PhysRevLett.65.120}, FLEX is also expected to overestimate the strength of the coexisting CDW and \mbox{s-SC} fluctuations at ${V=0}$.
Exploiting this symmetry, in the thermodynamic limit the exact diagrammatic Monte Carlo solution gives ${U^{\rm DiagMC}_{V=0}\simeq -2.5}$ value at ${\beta=10/t}$ for this transition point~\cite{PhysRevLett.124.017003}.

\begin{figure}[t!]
\includegraphics[width=1.\linewidth]{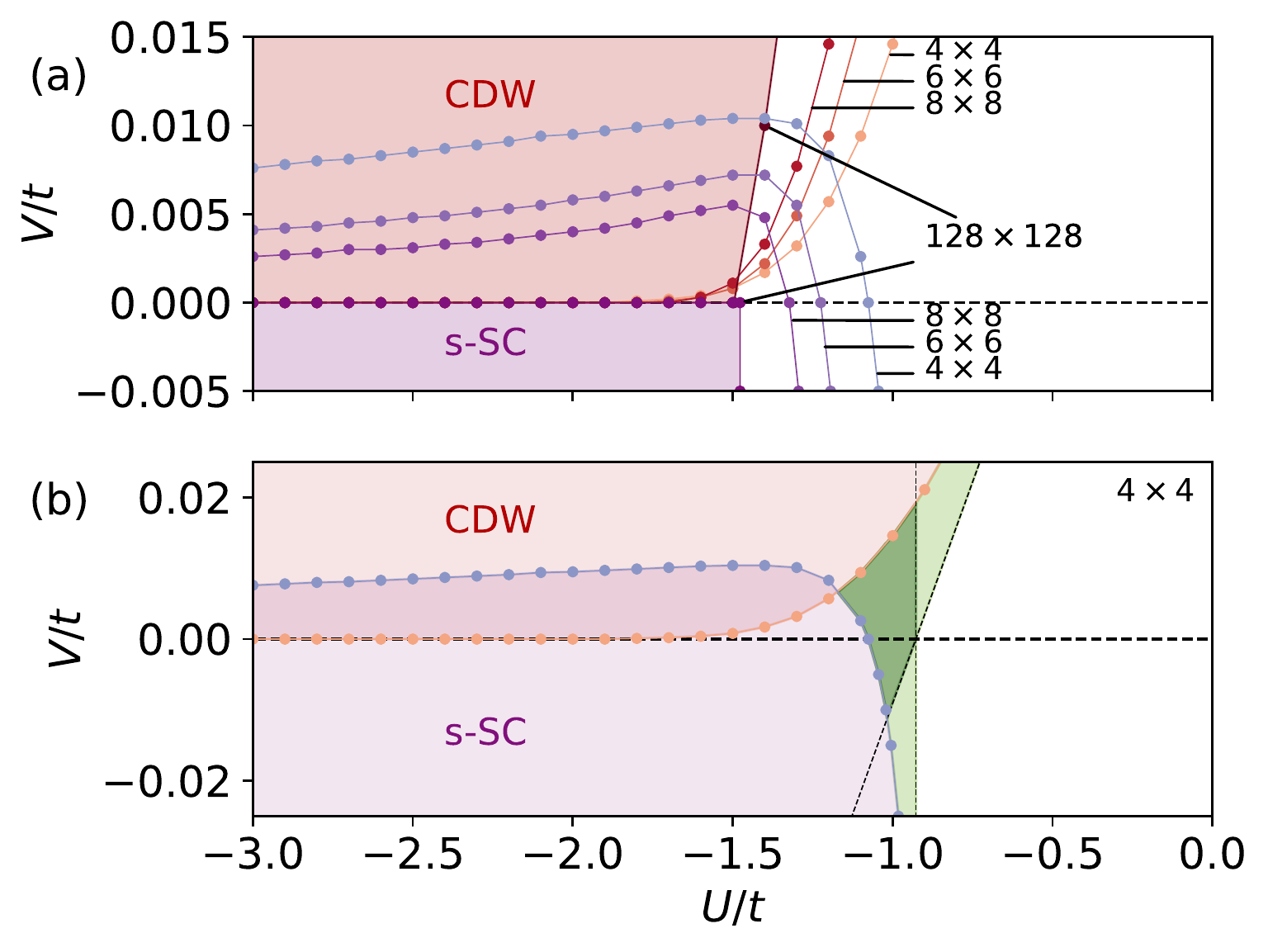}
\caption{CDW (red) and s-SC (purple) ordering boundaries predicted by the MCFF approach for the half-filled extended Hubbard model obtained for ${\beta=10/t}$: (a) for ${4\times 4}$, ${6\times 6}$, ${8\times 8}$, and ${128\times 128}$ plaquettes. (b) for a ${4\times 4}$ plaquette with the green region enclosed by the thin black dashed lines that depict asymptotics for the non-interplaying CDW and \mbox{s-SC} instabilities, displaying the region, where the CDW and \mbox{s-SC} orderings stabilize without interplay.
Dark green denotes the region where stabilization of either the CDW or \mbox{s-SC} phase destroys the other ordering.
\label{figure2}}
\end{figure}

Turning to the intermediate coupling regime, the CDW and \mbox{s-SC} fluctuations develop a coexisting phase along the ${V=0}$ line displayed in yellow color in Fig.~\ref{figure1}.
This coexistence is associated with the emergent pseudo-spin symmetry between CDW and \mbox{s-SC} order parameters.
Beyond this line the finite non-local interaction $V$ favors the formation of either the CDW (${V>0}$) or \mbox{s-SC} (${V<0}$) phase.
Remarkably, we find that at ${V\neq0}$ the CDW and \mbox{s-SC} phases are mutually exclusive only in the thermodynamic limit. 
For small-size plaquettes of ${4\times4}$, ${6\times6}$, and ${8\times8}$ lattice sites we find that the CDW and \mbox{s-SC} orderings can coexist also in the vicinity of ${V=0}$, and the coexistence region decreases with increasing the size of the system (Fig.~\ref{figure2}\,a).
This convergence check allows us to identify that the coexistence region in the vicinity of ${V=0}$ converges towards a single transition line occurring along ${V = 0}$ for ${U\leq -1.447}$ in the thermodynamic limit.
Thus, the transition between the CDW and s-SC phases appearing as a direct first-order phase transition is composed of two first-order phase transitions passing through the intermediate coexistence phase constrained by the pseudo-spin SU(2) symmetry~\cite{PhysRevB.14.2989}.

Another interesting effect can be found in the region of the phase diagram depicted in Fig.~\ref{figure2}\,b by green color.
It displays a region where CDW or \mbox{s-SC} orderings are separately stable without interplay between the modes. 
The dark green area denotes the overlap region of the non-competing CDW and \mbox{s-SC} orderings.
In the MCFF method, the CDW phase transition in the presence of the \mbox{s-SC} fluctuations is studied by integrating out the \mbox{s-SC} modes and investigating the behavior of the free energy ${{\cal F}(\phi^\varsigma_{\bf{Q}})}$ for the remaining CDW mode, and \textit{vice versa}.
In the region where the integrated \mbox{s-SC} mode is ordered, the MCFF analysis of the CDW transition corresponds to the investigation of the stability of the CDW ordering in the presence of the \mbox{s-SC} phase.
In this regard, the integration of an ordered mode can be seen as an observation/measurement of this ordering in the system.
We note that green regions in Fig.~\ref{figure2}\,b lie outside the CDW and \mbox{s-SC} phases that are obtained considering the interplay between the two fluctuations.
Therefore, our results suggest that stabilising one of the two orderings in the dark green region immediately destroys the other one, which can be seen as a destruction of a quantum superposition of the two orderings by an observer.
Remarkably, we find that no such non-trivial ``green'' phases exist in the thermodynamic limit, where quantum effects are suppressed.

\begin{figure}[t!]
\includegraphics[width=1.\linewidth]{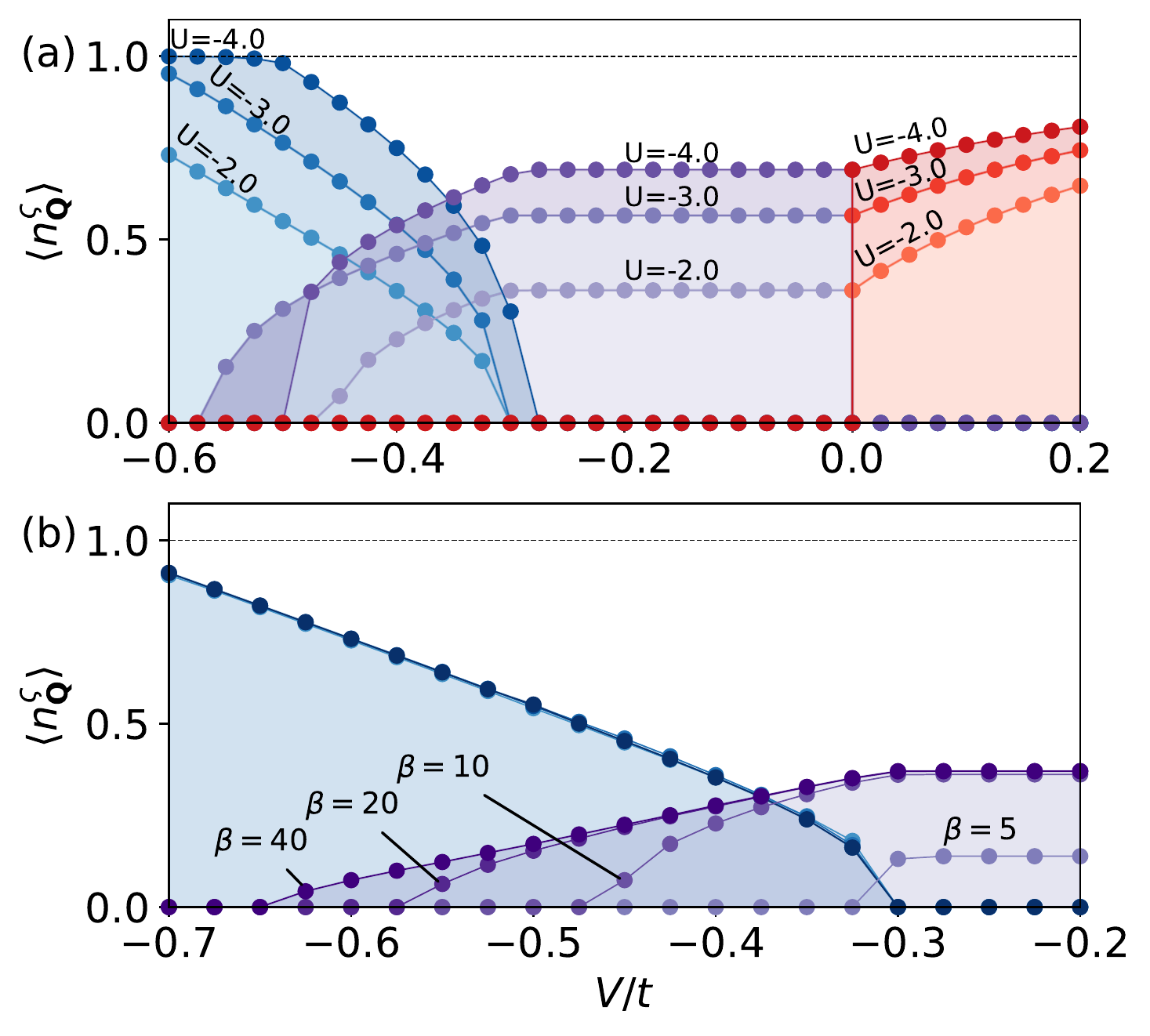}
\caption{Normalized order parameters ${\langle n^{\varsigma}_{\bf{Q}}\rangle}$ computed for the half-filled ${128\times 128}$ system using the MCFF approach. (a) CDW (red), \mbox{s-SC} (purple), and PS (blue) order parameters are calculated at ${\beta=10/t}$ for ${U = \{-2,-3,-4\}}$ for a range of $V$. (b) \mbox{s-SC} (purple) and PS (blue) order parameters calculated at ${\beta{}t =\{5, 10, 20, 40\}}$ at ${U= -2}$ for a range of $V$.
\label{figure3}}
\end{figure}

Further, we observe the emergence of a novel phase that comprises coexisting PS and \mbox{s-SC} orderings.
This \mbox{PS+s-SC} phase can be found in the regime of intermediate couplings of the attractive $U,V$ extended Hubbard model (Fig.~\ref{figure1}).
In contrast to the previously considered coexisting CDW and \mbox{s-SC} orderings, the novel coexistence phase does not collapse to a single transition line in the thermodynamic limit, thus acquiring a finite width in $V$ for a given $U$.
We observe the width to be a non-monotonic function of $U$, with a maximal width occurring near ${U = -3}$.
To obtain insight into the interplay between PS and \mbox{s-SC} ordering, in Fig.~\ref{figure3}\,a we show the normalized CDW, \mbox{s-SC}, and PS order parameters $\langle n^\varsigma_{\bf{Q}}\rangle$ that are computed for $U=\{-2,-3,-4\}$ over a range of $V$.
We observe a suppression of PS fluctuations in the weak coupling regime ${V\gtrsim-0.3}$ due to \mbox{s-SC} fluctuations, and {\it vice versa} at strong $V$.
The competition between these two modes originates from the fact that the PS ordering on a lattice corresponds to the formation of broad puddles with uniform filling larger or smaller than the average filling of the system.
Instead, the pairing process of \mbox{s-SC} fluctuations is energetically most favorable at half-filling.
Due to the stability of the \mbox{s-SC} fluctuations for a relatively large range of fillings~\cite{PhysRevLett.66.946, PhysRevB.99.045137}, the \mbox{s-SC} ordering can be formed inside the PS puddles, which results in a novel coexistence phase.
As $U$ increases, the region of \mbox{s-SC} fluctuations becomes more stable with respect to stronger PS fluctuations, leading to an increasing width of the coexistence region. 
However, the opposite trend occurs above a critical $U$ as strong PS fluctuations leaves the system effectively in an empty or fully-filled sites configuration with ${\langle n_{\rm PS} \rangle = 1}$, completely suppressing any \mbox{s-SC} fluctuations.
Note, that the CDW ordering on a square lattice corresponds to a checkerboard pattern of alternating lattice sites with higher and lower electronic densities.
This does not allow for the formation of the \mbox{s-SC} ordering inside the CDW phase due to the strong inhomogeneity of the filling, except along the degenerate ${V=0}$ line due to symmetry constraints.

In preparation of the current work, we noticed a recent determinant quantum Monte Carlo (DQMC) study of the zero-temperature $U$-$V$ phase diagram of the half-filled extended Hubbard model~\cite{PhysRevB.106.195121}.
In this work, few points of coexisting PS and \mbox{s-SC} orderings were identified evidencing our observations.
However, due to the sparsity of the grid in the $U$-$V$ space, the DQMC results do not allow one to make a definite statement on the presence of the coexistence phase in the system.
In fact, the authors of this work interpret this coexistence as a signature of a first-order transition between the \mbox{s-SC} and PS phases.
Indeed, first-order transitions are usually accompanied by regions of metastable collective fluctuations appearing as coexistence regions~\cite{2022arXiv221005540L}.
However, in the current work we do not observe metastable collective fluctuations associated with any first-order transition, although the MCFF method allows for their detection in other contexts~\cite{2022arXiv221005540L}.
This fact allows us to argue for a true coexistence phase stable in the thermodynamic limit enclosed by two apparent second-order transition lines.
An order parameter for this novel phase may be defined as the product of the \mbox{s-SC} and PS order parameters. 
To further connect our finite-temperature calculations to the zero-temperature DQMC results, we compute the \mbox{s-SC} and PS order parameters $\langle n^\varsigma_{\bf{Q}}\rangle$ for ${U=-2}$ over a range of $V$ at different inverse temperatures ${\beta{}t = \{5, 10, 20, 40\}}$.
Fig.~\ref{figure3}\,b shows that the stability of the \mbox{s-SC} fluctuations increases with decreasing temperature, as PS fluctuations remain nearly temperature-independent.
Thus, we expect the novel phase of coexisting PS and \mbox{s-SC} ordering to remain stable at zero temperature and to connect to the results observed in Ref.~\onlinecite{PhysRevB.106.195121}.

Exploring the predicted phase diagram experimentally and switching between the different phases in realistic materials can be performed, e.g., by applying an external laser field.
In the high-frequency regime of the driving, the applied laser field effectively decreases the hopping amplitude $t$ of electrons~\cite{ITIN2014822, PhysRevLett.115.075301, PhysRevLett.116.125301, PhysRevB.95.024306, PhysRevB.93.241404, PhysRevLett.118.157201, PhysRevB.102.220301}, which effectively enhances the interactions ${U/t}$ and ${V/t}$.
In the low-frequency regime of the field, driving phonon degrees of freedom may lead to enhancement of the electron-phonon coupling~\cite{PhysRevB.95.205111}, which would increase the strength of effective attractive electronic interactions~\cite{PhysRevB.52.4806, PhysRevLett.94.026401, PhysRevLett.99.146404}.
This can potentially allow one to propagate within the $U$-$V$ phase diagram and access the novel coexistence phase.

Interplay between SC and PS fluctuations has been observed in high-temperature superconducting materials, such as copper oxides~\cite{Bednorz1986,PhysRevB.38.11337,PhysRevB.40.7391,EMERY1993597,PhysRevB.52.15575,nature09260,pnas.1208492109,Campi2015,10.1038/nmat3719,sciadv.1600664} and iron-based superconductors~\cite{ja063355c,nature06972,PhysRevB.84.060511,Bendele2014,PhysRevB.90.214516,PhysRevB.91.020503}, but the microscopic mechanisms of the observed phenomena remain elusive. 
In doped copper oxides, it has been argued early on~\cite{PhysRevB.40.7391, EMERY1993597} that dilute holes in an antiferromagnet have a strong tendency to phase-separate.
Experimentally, interfaces of La$_{2-x}$Sr$_x$CuO$_4$-La$_2$CuO$_4$~\cite{10.1038/nmat3719} display an intriguing insensitivity of the critical temperature of the SC phase over an extended range of doping. 
These findings have been rationalised by invoking interlayer phase separation~\cite{sciadv.1600664}.
Our findings of coexisting SC and PS at half-filling give yet another indication hinting at the possibly very fundamental role of phase separation in the physics of superconducting correlated fermionic systems.\\

\begin{acknowledgments}
The authors thank Maria Chatzieleftheriou for useful discussions and Alexey Rubtsov for inspiration.
E.L., S.B., and E.A.S. acknowledge the help of the CPHT computer support team and the support from IDRIS/GENCI Orsay under project number A0130901393. C.D. also acknowledges the supports from Quantum Matter Bordeaux and the SMR department under the projects TED and CDS-QM.
\end{acknowledgments}

\bibliography{Bib_MCFF}

\begin{thebibliography}{115}%
\makeatletter
\providecommand \@ifxundefined [1]{%
 \@ifx{#1\undefined}
}%
\providecommand \@ifnum [1]{%
 \ifnum #1\expandafter \@firstoftwo
 \else \expandafter \@secondoftwo
 \fi
}%
\providecommand \@ifx [1]{%
 \ifx #1\expandafter \@firstoftwo
 \else \expandafter \@secondoftwo
 \fi
}%
\providecommand \natexlab [1]{#1}%
\providecommand \enquote  [1]{``#1''}%
\providecommand \bibnamefont  [1]{#1}%
\providecommand \bibfnamefont [1]{#1}%
\providecommand \citenamefont [1]{#1}%
\providecommand \href@noop [0]{\@secondoftwo}%
\providecommand \href [0]{\begingroup \@sanitize@url \@href}%
\providecommand \@href[1]{\@@startlink{#1}\@@href}%
\providecommand \@@href[1]{\endgroup#1\@@endlink}%
\providecommand \@sanitize@url [0]{\catcode `\\12\catcode `\$12\catcode
  `\&12\catcode `\#12\catcode `\^12\catcode `\_12\catcode `\%12\relax}%
\providecommand \@@startlink[1]{}%
\providecommand \@@endlink[0]{}%
\providecommand \url  [0]{\begingroup\@sanitize@url \@url }%
\providecommand \@url [1]{\endgroup\@href {#1}{\urlprefix }}%
\providecommand \urlprefix  [0]{URL }%
\providecommand \Eprint [0]{\href }%
\providecommand \doibase [0]{http://dx.doi.org/}%
\providecommand \selectlanguage [0]{\@gobble}%
\providecommand \bibinfo  [0]{\@secondoftwo}%
\providecommand \bibfield  [0]{\@secondoftwo}%
\providecommand \translation [1]{[#1]}%
\providecommand \BibitemOpen [0]{}%
\providecommand \bibitemStop [0]{}%
\providecommand \bibitemNoStop [0]{.\EOS\space}%
\providecommand \EOS [0]{\spacefactor3000\relax}%
\providecommand \BibitemShut  [1]{\csname bibitem#1\endcsname}%
\let\auto@bib@innerbib\@empty
%</preamble>
\bibitem [{\citenamefont {Fourcade}\ and\ \citenamefont
  {Spronken}(1984{\natexlab{a}})}]{PhysRevB.29.5089}%
  \BibitemOpen
  \bibfield  {author} {\bibinfo {author} {\bibfnamefont {B.}~\bibnamefont
  {Fourcade}}\ and\ \bibinfo {author} {\bibfnamefont {G.}~\bibnamefont
  {Spronken}},\ }\bibfield  {title} {\enquote {\bibinfo {title} {{Real-space
  scaling methods applied to the one-dimensional extended Hubbard model. I. The
  real-space renormalization-group method}},}\ }\href {\doibase
  10.1103/PhysRevB.29.5089} {\bibfield  {journal} {\bibinfo  {journal} {Phys.
  Rev. B}\ }\textbf {\bibinfo {volume} {29}},\ \bibinfo {pages} {5089--5095}
  (\bibinfo {year} {1984}{\natexlab{a}})}\BibitemShut {NoStop}%
\bibitem [{\citenamefont {Hirsch}(1984)}]{PhysRevLett.53.2327}%
  \BibitemOpen
  \bibfield  {author} {\bibinfo {author} {\bibfnamefont {J.~E.}\ \bibnamefont
  {Hirsch}},\ }\bibfield  {title} {\enquote {\bibinfo {title}
  {{Charge-Density-Wave to Spin-Density-Wave Transition in the Extended Hubbard
  Model}},}\ }\href {\doibase 10.1103/PhysRevLett.53.2327} {\bibfield
  {journal} {\bibinfo  {journal} {Phys. Rev. Lett.}\ }\textbf {\bibinfo
  {volume} {53}},\ \bibinfo {pages} {2327--2330} (\bibinfo {year}
  {1984})}\BibitemShut {NoStop}%
\bibitem [{\citenamefont {Nakamura}(1999)}]{doi:10.1143/JPSJ.68.3123}%
  \BibitemOpen
  \bibfield  {author} {\bibinfo {author} {\bibfnamefont {M.}~\bibnamefont
  {Nakamura}},\ }\bibfield  {title} {\enquote {\bibinfo {title} {{Mechanism of
  CDW-SDW Transition in One Dimension}},}\ }\href {\doibase
  10.1143/JPSJ.68.3123} {\bibfield  {journal} {\bibinfo  {journal} {J. Phys.
  Soc. Jpn.}\ }\textbf {\bibinfo {volume} {68}},\ \bibinfo {pages} {3123--3126}
  (\bibinfo {year} {1999})}\BibitemShut {NoStop}%
\bibitem [{\citenamefont {Nakamura}(2000)}]{PhysRevB.61.16377}%
  \BibitemOpen
  \bibfield  {author} {\bibinfo {author} {\bibfnamefont {M.}~\bibnamefont
  {Nakamura}},\ }\bibfield  {title} {\enquote {\bibinfo {title} {{Tricritical
  behavior in the extended Hubbard chains}},}\ }\href {\doibase
  10.1103/PhysRevB.61.16377} {\bibfield  {journal} {\bibinfo  {journal} {Phys.
  Rev. B}\ }\textbf {\bibinfo {volume} {61}},\ \bibinfo {pages} {16377--16392}
  (\bibinfo {year} {2000})}\BibitemShut {NoStop}%
\bibitem [{\citenamefont {Sandvik}\ \emph {et~al.}(2004)\citenamefont
  {Sandvik}, \citenamefont {Balents},\ and\ \citenamefont
  {Campbell}}]{PhysRevLett.92.236401}%
  \BibitemOpen
  \bibfield  {author} {\bibinfo {author} {\bibfnamefont {A.~W.}\ \bibnamefont
  {Sandvik}}, \bibinfo {author} {\bibfnamefont {L.}~\bibnamefont {Balents}}, \
  and\ \bibinfo {author} {\bibfnamefont {D.~K.}\ \bibnamefont {Campbell}},\
  }\bibfield  {title} {\enquote {\bibinfo {title} {{Ground State Phases of the
  Half-Filled One-Dimensional Extended Hubbard Model}},}\ }\href {\doibase
  10.1103/PhysRevLett.92.236401} {\bibfield  {journal} {\bibinfo  {journal}
  {Phys. Rev. Lett.}\ }\textbf {\bibinfo {volume} {92}},\ \bibinfo {pages}
  {236401} (\bibinfo {year} {2004})}\BibitemShut {NoStop}%
\bibitem [{\citenamefont {Emery}(1976)}]{PhysRevB.14.2989}%
  \BibitemOpen
  \bibfield  {author} {\bibinfo {author} {\bibfnamefont {V.~J.}\ \bibnamefont
  {Emery}},\ }\bibfield  {title} {\enquote {\bibinfo {title} {{Theory of the
  quasi-one-dimensional electron gas with strong ``on-site'' interactions}},}\
  }\href {\doibase 10.1103/PhysRevB.14.2989} {\bibfield  {journal} {\bibinfo
  {journal} {Phys. Rev. B}\ }\textbf {\bibinfo {volume} {14}},\ \bibinfo
  {pages} {2989--2994} (\bibinfo {year} {1976})}\BibitemShut {NoStop}%
\bibitem [{\citenamefont {Hubbard}(1963)}]{rspa.1963.0204}%
  \BibitemOpen
  \bibfield  {author} {\bibinfo {author} {\bibfnamefont {J.}~\bibnamefont
  {Hubbard}},\ }\bibfield  {title} {\enquote {\bibinfo {title} {{Electron
  correlations in narrow energy bands}},}\ }\href {\doibase
  10.1098/rspa.1963.0204} {\bibfield  {journal} {\bibinfo  {journal} {Proc. R.
  Soc. Lond. A}\ }\textbf {\bibinfo {volume} {276}},\ \bibinfo {pages}
  {238–257} (\bibinfo {year} {1963})}\BibitemShut {NoStop}%
\bibitem [{\citenamefont {Gutzwiller}(1963)}]{PhysRevLett.10.159}%
  \BibitemOpen
  \bibfield  {author} {\bibinfo {author} {\bibfnamefont {M.~C.}\ \bibnamefont
  {Gutzwiller}},\ }\bibfield  {title} {\enquote {\bibinfo {title} {{Effect of
  Correlation on the Ferromagnetism of Transition Metals}},}\ }\href {\doibase
  10.1103/PhysRevLett.10.159} {\bibfield  {journal} {\bibinfo  {journal} {Phys.
  Rev. Lett.}\ }\textbf {\bibinfo {volume} {10}},\ \bibinfo {pages} {159--162}
  (\bibinfo {year} {1963})}\BibitemShut {NoStop}%
\bibitem [{\citenamefont {Kanamori}(1963)}]{10.1143/PTP.30.275}%
  \BibitemOpen
  \bibfield  {author} {\bibinfo {author} {\bibfnamefont {J.}~\bibnamefont
  {Kanamori}},\ }\bibfield  {title} {\enquote {\bibinfo {title} {{Electron
  Correlation and Ferromagnetism of Transition Metals}},}\ }\href {\doibase
  10.1143/PTP.30.275} {\bibfield  {journal} {\bibinfo  {journal} {Prog. Theor.
  Phys.}\ }\textbf {\bibinfo {volume} {30}},\ \bibinfo {pages} {275--289}
  (\bibinfo {year} {1963})}\BibitemShut {NoStop}%
\bibitem [{\citenamefont {Hubbard}(1964)}]{rspa.1964.0190}%
  \BibitemOpen
  \bibfield  {author} {\bibinfo {author} {\bibfnamefont {J.}~\bibnamefont
  {Hubbard}},\ }\bibfield  {title} {\enquote {\bibinfo {title} {{Electron
  correlations in narrow energy bands III. An improved solution}},}\ }\href
  {\doibase 10.1098/rspa.1964.0190} {\bibfield  {journal} {\bibinfo  {journal}
  {Proc. R. Soc. Lond. A}\ }\textbf {\bibinfo {volume} {281}},\ \bibinfo
  {pages} {401–419} (\bibinfo {year} {1964})}\BibitemShut {NoStop}%
\bibitem [{\citenamefont {Bari}(1971)}]{PhysRevB.3.2662}%
  \BibitemOpen
  \bibfield  {author} {\bibinfo {author} {\bibfnamefont {R.~A.}\ \bibnamefont
  {Bari}},\ }\bibfield  {title} {\enquote {\bibinfo {title} {{Effects of
  Short-Range Interactions on Electron-Charge Ordering and Lattice Distortions
  in the Localized State}},}\ }\href {\doibase 10.1103/PhysRevB.3.2662}
  {\bibfield  {journal} {\bibinfo  {journal} {Phys. Rev. B}\ }\textbf {\bibinfo
  {volume} {3}},\ \bibinfo {pages} {2662--2670} (\bibinfo {year}
  {1971})}\BibitemShut {NoStop}%
\bibitem [{\citenamefont {Harris}\ and\ \citenamefont
  {Lange}(1967)}]{PhysRev.157.295}%
  \BibitemOpen
  \bibfield  {author} {\bibinfo {author} {\bibfnamefont {A.~B.}\ \bibnamefont
  {Harris}}\ and\ \bibinfo {author} {\bibfnamefont {R.~V.}\ \bibnamefont
  {Lange}},\ }\bibfield  {title} {\enquote {\bibinfo {title} {{Single-Particle
  Excitations in Narrow Energy Bands}},}\ }\href {\doibase
  10.1103/PhysRev.157.295} {\bibfield  {journal} {\bibinfo  {journal} {Phys.
  Rev.}\ }\textbf {\bibinfo {volume} {157}},\ \bibinfo {pages} {295--314}
  (\bibinfo {year} {1967})}\BibitemShut {NoStop}%
\bibitem [{\citenamefont {Gull}\ \emph {et~al.}(2009)\citenamefont {Gull},
  \citenamefont {Parcollet}, \citenamefont {Werner},\ and\ \citenamefont
  {Millis}}]{PhysRevB.80.245102}%
  \BibitemOpen
  \bibfield  {author} {\bibinfo {author} {\bibfnamefont {E.}~\bibnamefont
  {Gull}}, \bibinfo {author} {\bibfnamefont {O.}~\bibnamefont {Parcollet}},
  \bibinfo {author} {\bibfnamefont {P.}~\bibnamefont {Werner}}, \ and\ \bibinfo
  {author} {\bibfnamefont {A.~J.}\ \bibnamefont {Millis}},\ }\bibfield  {title}
  {\enquote {\bibinfo {title} {{Momentum-sector-selective metal-insulator
  transition in the eight-site dynamical mean-field approximation to the
  Hubbard model in two dimensions}},}\ }\href {\doibase
  10.1103/PhysRevB.80.245102} {\bibfield  {journal} {\bibinfo  {journal} {Phys.
  Rev. B}\ }\textbf {\bibinfo {volume} {80}},\ \bibinfo {pages} {245102}
  (\bibinfo {year} {2009})}\BibitemShut {NoStop}%
\bibitem [{\citenamefont {Gunnarsson}\ \emph {et~al.}(2015)\citenamefont
  {Gunnarsson}, \citenamefont {Sch\"afer}, \citenamefont {LeBlanc},
  \citenamefont {Gull}, \citenamefont {Merino}, \citenamefont {Sangiovanni},
  \citenamefont {Rohringer},\ and\ \citenamefont
  {Toschi}}]{PhysRevLett.114.236402}%
  \BibitemOpen
  \bibfield  {author} {\bibinfo {author} {\bibfnamefont {O.}~\bibnamefont
  {Gunnarsson}}, \bibinfo {author} {\bibfnamefont {T.}~\bibnamefont
  {Sch\"afer}}, \bibinfo {author} {\bibfnamefont {J.~P.~F.}\ \bibnamefont
  {LeBlanc}}, \bibinfo {author} {\bibfnamefont {E.}~\bibnamefont {Gull}},
  \bibinfo {author} {\bibfnamefont {J.}~\bibnamefont {Merino}}, \bibinfo
  {author} {\bibfnamefont {G.}~\bibnamefont {Sangiovanni}}, \bibinfo {author}
  {\bibfnamefont {G.}~\bibnamefont {Rohringer}}, \ and\ \bibinfo {author}
  {\bibfnamefont {A.}~\bibnamefont {Toschi}},\ }\bibfield  {title} {\enquote
  {\bibinfo {title} {{Fluctuation Diagnostics of the Electron Self-Energy:
  Origin of the Pseudogap Physics}},}\ }\href {\doibase
  10.1103/PhysRevLett.114.236402} {\bibfield  {journal} {\bibinfo  {journal}
  {Phys. Rev. Lett.}\ }\textbf {\bibinfo {volume} {114}},\ \bibinfo {pages}
  {236402} (\bibinfo {year} {2015})}\BibitemShut {NoStop}%
\bibitem [{\citenamefont {Wu}\ \emph {et~al.}(2017)\citenamefont {Wu},
  \citenamefont {Ferrero}, \citenamefont {Georges},\ and\ \citenamefont
  {Kozik}}]{PhysRevB.96.041105}%
  \BibitemOpen
  \bibfield  {author} {\bibinfo {author} {\bibfnamefont {W.}~\bibnamefont
  {Wu}}, \bibinfo {author} {\bibfnamefont {M.}~\bibnamefont {Ferrero}},
  \bibinfo {author} {\bibfnamefont {A.}~\bibnamefont {Georges}}, \ and\
  \bibinfo {author} {\bibfnamefont {E.}~\bibnamefont {Kozik}},\ }\bibfield
  {title} {\enquote {\bibinfo {title} {{Controlling Feynman diagrammatic
  expansions: Physical nature of the pseudogap in the two-dimensional Hubbard
  model}},}\ }\href {\doibase 10.1103/PhysRevB.96.041105} {\bibfield  {journal}
  {\bibinfo  {journal} {Phys. Rev. B}\ }\textbf {\bibinfo {volume} {96}},\
  \bibinfo {pages} {041105} (\bibinfo {year} {2017})}\BibitemShut {NoStop}%
\bibitem [{\citenamefont {Shiba}(1972)}]{10.1143/PTP.48.2171}%
  \BibitemOpen
  \bibfield  {author} {\bibinfo {author} {\bibfnamefont {H.}~\bibnamefont
  {Shiba}},\ }\bibfield  {title} {\enquote {\bibinfo {title} {{Thermodynamic
  Properties of the One-Dimensional Half-Filled-Band Hubbard Model. II:
  Application of the Grand Canonical Method}},}\ }\href {\doibase
  10.1143/PTP.48.2171} {\bibfield  {journal} {\bibinfo  {journal} {Prog. Theor.
  Phys.}\ }\textbf {\bibinfo {volume} {48}},\ \bibinfo {pages} {2171--2186}
  (\bibinfo {year} {1972})}\BibitemShut {NoStop}%
\bibitem [{\citenamefont {Scalettar}\ \emph {et~al.}(1989)\citenamefont
  {Scalettar}, \citenamefont {Loh}, \citenamefont {Gubernatis}, \citenamefont
  {Moreo}, \citenamefont {White}, \citenamefont {Scalapino}, \citenamefont
  {Sugar},\ and\ \citenamefont {Dagotto}}]{PhysRevLett.62.1407}%
  \BibitemOpen
  \bibfield  {author} {\bibinfo {author} {\bibfnamefont {R.~T.}\ \bibnamefont
  {Scalettar}}, \bibinfo {author} {\bibfnamefont {E.~Y.}\ \bibnamefont {Loh}},
  \bibinfo {author} {\bibfnamefont {J.~E.}\ \bibnamefont {Gubernatis}},
  \bibinfo {author} {\bibfnamefont {A.}~\bibnamefont {Moreo}}, \bibinfo
  {author} {\bibfnamefont {S.~R.}\ \bibnamefont {White}}, \bibinfo {author}
  {\bibfnamefont {D.~J.}\ \bibnamefont {Scalapino}}, \bibinfo {author}
  {\bibfnamefont {R.~L.}\ \bibnamefont {Sugar}}, \ and\ \bibinfo {author}
  {\bibfnamefont {E.}~\bibnamefont {Dagotto}},\ }\bibfield  {title} {\enquote
  {\bibinfo {title} {{Phase diagram of the two-dimensional negative-U Hubbard
  model}},}\ }\href {\doibase 10.1103/PhysRevLett.62.1407} {\bibfield
  {journal} {\bibinfo  {journal} {Phys. Rev. Lett.}\ }\textbf {\bibinfo
  {volume} {62}},\ \bibinfo {pages} {1407--1410} (\bibinfo {year}
  {1989})}\BibitemShut {NoStop}%
\bibitem [{\citenamefont {Yang}(1989)}]{PhysRevLett.63.2144}%
  \BibitemOpen
  \bibfield  {author} {\bibinfo {author} {\bibfnamefont {C.~N.}\ \bibnamefont
  {Yang}},\ }\bibfield  {title} {\enquote {\bibinfo {title} {{\ensuremath{\eta}
  pairing and off-diagonal long-range order in a Hubbard model}},}\ }\href
  {\doibase 10.1103/PhysRevLett.63.2144} {\bibfield  {journal} {\bibinfo
  {journal} {Phys. Rev. Lett.}\ }\textbf {\bibinfo {volume} {63}},\ \bibinfo
  {pages} {2144--2147} (\bibinfo {year} {1989})}\BibitemShut {NoStop}%
\bibitem [{\citenamefont {Zhang}(1990)}]{PhysRevLett.65.120}%
  \BibitemOpen
  \bibfield  {author} {\bibinfo {author} {\bibfnamefont {S.}~\bibnamefont
  {Zhang}},\ }\bibfield  {title} {\enquote {\bibinfo {title} {{Pseudospin
  symmetry and new collective modes of the Hubbard model}},}\ }\href {\doibase
  10.1103/PhysRevLett.65.120} {\bibfield  {journal} {\bibinfo  {journal} {Phys.
  Rev. Lett.}\ }\textbf {\bibinfo {volume} {65}},\ \bibinfo {pages} {120--122}
  (\bibinfo {year} {1990})}\BibitemShut {NoStop}%
\bibitem [{\citenamefont {Vonsovsky}\ and\ \citenamefont
  {Katsnelson}(1979)}]{Vonsovsky_1979}%
  \BibitemOpen
  \bibfield  {author} {\bibinfo {author} {\bibfnamefont {S.~V.}\ \bibnamefont
  {Vonsovsky}}\ and\ \bibinfo {author} {\bibfnamefont {M.~I.}\ \bibnamefont
  {Katsnelson}},\ }\bibfield  {title} {\enquote {\bibinfo {title} {{Some types
  of instabilities in the electron energy spectrum of the polar model of the
  crystal. I. The maximum-polarity state}},}\ }\href {\doibase
  10.1088/0022-3719/12/11/015} {\bibfield  {journal} {\bibinfo  {journal} {J.
  Phys. C: Solid State Phys.}\ }\textbf {\bibinfo {volume} {12}},\ \bibinfo
  {pages} {2043--2053} (\bibinfo {year} {1979})}\BibitemShut {NoStop}%
\bibitem [{\citenamefont {Davoudi}\ and\ \citenamefont
  {Tremblay}(2007)}]{PhysRevB.76.085115}%
  \BibitemOpen
  \bibfield  {author} {\bibinfo {author} {\bibfnamefont {B.}~\bibnamefont
  {Davoudi}}\ and\ \bibinfo {author} {\bibfnamefont {A.-M.~S.}\ \bibnamefont
  {Tremblay}},\ }\bibfield  {title} {\enquote {\bibinfo {title}
  {{Non-perturbative treatment of charge and spin fluctuations in the
  two-dimensional extended Hubbard model: Extended two-particle self-consistent
  approach}},}\ }\href {\doibase 10.1103/PhysRevB.76.085115} {\bibfield
  {journal} {\bibinfo  {journal} {Phys. Rev. B}\ }\textbf {\bibinfo {volume}
  {76}},\ \bibinfo {pages} {085115} (\bibinfo {year} {2007})}\BibitemShut
  {NoStop}%
\bibitem [{\citenamefont {S\'olyom}(1979)}]{doi:10.1080/00018737900101375}%
  \BibitemOpen
  \bibfield  {author} {\bibinfo {author} {\bibfnamefont {J.}~\bibnamefont
  {S\'olyom}},\ }\bibfield  {title} {\enquote {\bibinfo {title} {{The Fermi gas
  model of one-dimensional conductors}},}\ }\href {\doibase
  10.1080/00018737900101375} {\bibfield  {journal} {\bibinfo  {journal} {Adv.
  Phys.}\ }\textbf {\bibinfo {volume} {28}},\ \bibinfo {pages} {201--303}
  (\bibinfo {year} {1979})}\BibitemShut {NoStop}%
\bibitem [{\citenamefont {Emery}(1979)}]{emery1979}%
  \BibitemOpen
  \bibfield  {author} {\bibinfo {author} {\bibfnamefont {V.~J.}\ \bibnamefont
  {Emery}},\ }\enquote {\bibinfo {title} {{Theory of the One-Dimensional
  Electron Gas}},}\ in\ \href {\doibase 10.1007/978-1-4613-2895-7_6} {\emph
  {\bibinfo {booktitle} {Highly Conducting One-Dimensional Solids}}},\ \bibinfo
  {editor} {edited by\ \bibinfo {editor} {\bibfnamefont {Jozef~T.}\
  \bibnamefont {Devreese}}, \bibinfo {editor} {\bibfnamefont {Roger~P.}\
  \bibnamefont {Evrard}}, \ and\ \bibinfo {editor} {\bibfnamefont {Victor~E.}\
  \bibnamefont {van Doren}}}\ (\bibinfo  {publisher} {Springer US},\ \bibinfo
  {address} {Boston, MA},\ \bibinfo {year} {1979})\ pp.\ \bibinfo {pages}
  {247--303}\BibitemShut {NoStop}%
\bibitem [{\citenamefont {Fourcade}\ and\ \citenamefont
  {Spronken}(1984{\natexlab{b}})}]{PhysRevB.29.5096}%
  \BibitemOpen
  \bibfield  {author} {\bibinfo {author} {\bibfnamefont {B.}~\bibnamefont
  {Fourcade}}\ and\ \bibinfo {author} {\bibfnamefont {G.}~\bibnamefont
  {Spronken}},\ }\bibfield  {title} {\enquote {\bibinfo {title} {{Real-space
  scaling methods applied to the one-dimensional extended Hubbard model. II.
  The finite-cell scaling method}},}\ }\href {\doibase
  10.1103/PhysRevB.29.5096} {\bibfield  {journal} {\bibinfo  {journal} {Phys.
  Rev. B}\ }\textbf {\bibinfo {volume} {29}},\ \bibinfo {pages} {5096--5102}
  (\bibinfo {year} {1984}{\natexlab{b}})}\BibitemShut {NoStop}%
\bibitem [{\citenamefont {Zhang}\ and\ \citenamefont
  {Callaway}(1989)}]{PhysRevB.39.9397}%
  \BibitemOpen
  \bibfield  {author} {\bibinfo {author} {\bibfnamefont {Y.}~\bibnamefont
  {Zhang}}\ and\ \bibinfo {author} {\bibfnamefont {J.}~\bibnamefont
  {Callaway}},\ }\bibfield  {title} {\enquote {\bibinfo {title} {{Extended
  Hubbard model in two dimensions}},}\ }\href {\doibase
  10.1103/PhysRevB.39.9397} {\bibfield  {journal} {\bibinfo  {journal} {Phys.
  Rev. B}\ }\textbf {\bibinfo {volume} {39}},\ \bibinfo {pages} {9397--9404}
  (\bibinfo {year} {1989})}\BibitemShut {NoStop}%
\bibitem [{\citenamefont {Callaway}\ \emph {et~al.}(1990)\citenamefont
  {Callaway}, \citenamefont {Chen}, \citenamefont {Kanhere},\ and\
  \citenamefont {Li}}]{PhysRevB.42.465}%
  \BibitemOpen
  \bibfield  {author} {\bibinfo {author} {\bibfnamefont {J.}~\bibnamefont
  {Callaway}}, \bibinfo {author} {\bibfnamefont {D.~P.}\ \bibnamefont {Chen}},
  \bibinfo {author} {\bibfnamefont {D.~G.}\ \bibnamefont {Kanhere}}, \ and\
  \bibinfo {author} {\bibfnamefont {Q.}~\bibnamefont {Li}},\ }\bibfield
  {title} {\enquote {\bibinfo {title} {{Small-cluster calculations for the
  simple and extended Hubbard models}},}\ }\href {\doibase
  10.1103/PhysRevB.42.465} {\bibfield  {journal} {\bibinfo  {journal} {Phys.
  Rev. B}\ }\textbf {\bibinfo {volume} {42}},\ \bibinfo {pages} {465--474}
  (\bibinfo {year} {1990})}\BibitemShut {NoStop}%
\bibitem [{\citenamefont {Yan}(1993)}]{PhysRevB.48.7140}%
  \BibitemOpen
  \bibfield  {author} {\bibinfo {author} {\bibfnamefont {Xin-Zhong}\
  \bibnamefont {Yan}},\ }\bibfield  {title} {\enquote {\bibinfo {title}
  {{Theory of the extended Hubbard model at half filling}},}\ }\href {\doibase
  10.1103/PhysRevB.48.7140} {\bibfield  {journal} {\bibinfo  {journal} {Phys.
  Rev. B}\ }\textbf {\bibinfo {volume} {48}},\ \bibinfo {pages} {7140--7147}
  (\bibinfo {year} {1993})}\BibitemShut {NoStop}%
\bibitem [{\citenamefont {Aichhorn}\ \emph {et~al.}(2004)\citenamefont
  {Aichhorn}, \citenamefont {Evertz}, \citenamefont {von~der Linden},\ and\
  \citenamefont {Potthoff}}]{PhysRevB.70.235107}%
  \BibitemOpen
  \bibfield  {author} {\bibinfo {author} {\bibfnamefont {M.}~\bibnamefont
  {Aichhorn}}, \bibinfo {author} {\bibfnamefont {H.~G.}\ \bibnamefont
  {Evertz}}, \bibinfo {author} {\bibfnamefont {W.}~\bibnamefont {von~der
  Linden}}, \ and\ \bibinfo {author} {\bibfnamefont {M.}~\bibnamefont
  {Potthoff}},\ }\bibfield  {title} {\enquote {\bibinfo {title} {{Charge
  ordering in extended Hubbard models: Variational cluster approach}},}\ }\href
  {\doibase 10.1103/PhysRevB.70.235107} {\bibfield  {journal} {\bibinfo
  {journal} {Phys. Rev. B}\ }\textbf {\bibinfo {volume} {70}},\ \bibinfo
  {pages} {235107} (\bibinfo {year} {2004})}\BibitemShut {NoStop}%
\bibitem [{\citenamefont {Davoudi}\ and\ \citenamefont
  {Tremblay}(2006)}]{PhysRevB.74.035113}%
  \BibitemOpen
  \bibfield  {author} {\bibinfo {author} {\bibfnamefont {B.}~\bibnamefont
  {Davoudi}}\ and\ \bibinfo {author} {\bibfnamefont {A.-M.~S.}\ \bibnamefont
  {Tremblay}},\ }\bibfield  {title} {\enquote {\bibinfo {title}
  {{Nearest-neighbor repulsion and competing charge and spin order in the
  extended Hubbard model}},}\ }\href {\doibase 10.1103/PhysRevB.74.035113}
  {\bibfield  {journal} {\bibinfo  {journal} {Phys. Rev. B}\ }\textbf {\bibinfo
  {volume} {74}},\ \bibinfo {pages} {035113} (\bibinfo {year}
  {2006})}\BibitemShut {NoStop}%
\bibitem [{\citenamefont {Ayral}\ \emph {et~al.}(2013)\citenamefont {Ayral},
  \citenamefont {Biermann},\ and\ \citenamefont {Werner}}]{PhysRevB.87.125149}%
  \BibitemOpen
  \bibfield  {author} {\bibinfo {author} {\bibfnamefont {T.}~\bibnamefont
  {Ayral}}, \bibinfo {author} {\bibfnamefont {S.}~\bibnamefont {Biermann}}, \
  and\ \bibinfo {author} {\bibfnamefont {P.}~\bibnamefont {Werner}},\
  }\bibfield  {title} {\enquote {\bibinfo {title} {{Screening and nonlocal
  correlations in the extended Hubbard model from self-consistent combined $GW$
  and dynamical mean field theory}},}\ }\href {\doibase
  10.1103/PhysRevB.87.125149} {\bibfield  {journal} {\bibinfo  {journal} {Phys.
  Rev. B}\ }\textbf {\bibinfo {volume} {87}},\ \bibinfo {pages} {125149}
  (\bibinfo {year} {2013})}\BibitemShut {NoStop}%
\bibitem [{\citenamefont {Hafermann}\ \emph {et~al.}(2014)\citenamefont
  {Hafermann}, \citenamefont {van Loon}, \citenamefont {Katsnelson},
  \citenamefont {Lichtenstein},\ and\ \citenamefont
  {Parcollet}}]{PhysRevB.90.235105}%
  \BibitemOpen
  \bibfield  {author} {\bibinfo {author} {\bibfnamefont {H.}~\bibnamefont
  {Hafermann}}, \bibinfo {author} {\bibfnamefont {E.~G. C.~P.}\ \bibnamefont
  {van Loon}}, \bibinfo {author} {\bibfnamefont {M.~I.}\ \bibnamefont
  {Katsnelson}}, \bibinfo {author} {\bibfnamefont {A.~I.}\ \bibnamefont
  {Lichtenstein}}, \ and\ \bibinfo {author} {\bibfnamefont {O.}~\bibnamefont
  {Parcollet}},\ }\bibfield  {title} {\enquote {\bibinfo {title} {{Collective
  charge excitations of strongly correlated electrons, vertex corrections, and
  gauge invariance}},}\ }\href {\doibase 10.1103/PhysRevB.90.235105} {\bibfield
   {journal} {\bibinfo  {journal} {Phys. Rev. B}\ }\textbf {\bibinfo {volume}
  {90}},\ \bibinfo {pages} {235105} (\bibinfo {year} {2014})}\BibitemShut
  {NoStop}%
\bibitem [{\citenamefont {Werner}\ and\ \citenamefont
  {Casula}(2016)}]{Werner_2016}%
  \BibitemOpen
  \bibfield  {author} {\bibinfo {author} {\bibfnamefont {P.}~\bibnamefont
  {Werner}}\ and\ \bibinfo {author} {\bibfnamefont {M.}~\bibnamefont
  {Casula}},\ }\bibfield  {title} {\enquote {\bibinfo {title} {{Dynamical
  screening in correlated electron systems—from lattice models to realistic
  materials}},}\ }\href {\doibase 10.1088/0953-8984/28/38/383001} {\bibfield
  {journal} {\bibinfo  {journal} {J. Phys. Condens. Matter}\ }\textbf {\bibinfo
  {volume} {28}},\ \bibinfo {pages} {383001} (\bibinfo {year}
  {2016})}\BibitemShut {NoStop}%
\bibitem [{\citenamefont {van Loon}\ \emph {et~al.}(2016)\citenamefont {van
  Loon}, \citenamefont {Sch\"uler}, \citenamefont {Katsnelson},\ and\
  \citenamefont {Wehling}}]{PhysRevB.94.165141}%
  \BibitemOpen
  \bibfield  {author} {\bibinfo {author} {\bibfnamefont {E.~G. C.~P.}\
  \bibnamefont {van Loon}}, \bibinfo {author} {\bibfnamefont {M.}~\bibnamefont
  {Sch\"uler}}, \bibinfo {author} {\bibfnamefont {M.~I.}\ \bibnamefont
  {Katsnelson}}, \ and\ \bibinfo {author} {\bibfnamefont {T.~O.}\ \bibnamefont
  {Wehling}},\ }\bibfield  {title} {\enquote {\bibinfo {title} {{Capturing
  nonlocal interaction effects in the Hubbard model: Optimal mappings and
  limits of applicability}},}\ }\href {\doibase 10.1103/PhysRevB.94.165141}
  {\bibfield  {journal} {\bibinfo  {journal} {Phys. Rev. B}\ }\textbf {\bibinfo
  {volume} {94}},\ \bibinfo {pages} {165141} (\bibinfo {year}
  {2016})}\BibitemShut {NoStop}%
\bibitem [{\citenamefont {Stepanov}\ \emph {et~al.}(2016)\citenamefont
  {Stepanov}, \citenamefont {Huber}, \citenamefont {van Loon}, \citenamefont
  {Lichtenstein},\ and\ \citenamefont {Katsnelson}}]{PhysRevB.94.205110}%
  \BibitemOpen
  \bibfield  {author} {\bibinfo {author} {\bibfnamefont {E.~A.}\ \bibnamefont
  {Stepanov}}, \bibinfo {author} {\bibfnamefont {A.}~\bibnamefont {Huber}},
  \bibinfo {author} {\bibfnamefont {E.~G. C.~P.}\ \bibnamefont {van Loon}},
  \bibinfo {author} {\bibfnamefont {A.~I.}\ \bibnamefont {Lichtenstein}}, \
  and\ \bibinfo {author} {\bibfnamefont {M.~I.}\ \bibnamefont {Katsnelson}},\
  }\bibfield  {title} {\enquote {\bibinfo {title} {{From local to nonlocal
  correlations: The Dual Boson perspective}},}\ }\href {\doibase
  10.1103/PhysRevB.94.205110} {\bibfield  {journal} {\bibinfo  {journal} {Phys.
  Rev. B}\ }\textbf {\bibinfo {volume} {94}},\ \bibinfo {pages} {205110}
  (\bibinfo {year} {2016})}\BibitemShut {NoStop}%
\bibitem [{\citenamefont {Terletska}\ \emph {et~al.}(2017)\citenamefont
  {Terletska}, \citenamefont {Chen},\ and\ \citenamefont
  {Gull}}]{PhysRevB.95.115149}%
  \BibitemOpen
  \bibfield  {author} {\bibinfo {author} {\bibfnamefont {H.}~\bibnamefont
  {Terletska}}, \bibinfo {author} {\bibfnamefont {T.}~\bibnamefont {Chen}}, \
  and\ \bibinfo {author} {\bibfnamefont {E.}~\bibnamefont {Gull}},\ }\bibfield
  {title} {\enquote {\bibinfo {title} {{Charge ordering and correlation effects
  in the extended Hubbard model}},}\ }\href {\doibase
  10.1103/PhysRevB.95.115149} {\bibfield  {journal} {\bibinfo  {journal} {Phys.
  Rev. B}\ }\textbf {\bibinfo {volume} {95}},\ \bibinfo {pages} {115149}
  (\bibinfo {year} {2017})}\BibitemShut {NoStop}%
\bibitem [{\citenamefont {van Loon}\ and\ \citenamefont
  {Katsnelson}(2018)}]{van_Loon_2018}%
  \BibitemOpen
  \bibfield  {author} {\bibinfo {author} {\bibfnamefont {E.~G. C.~P.}\
  \bibnamefont {van Loon}}\ and\ \bibinfo {author} {\bibfnamefont {M.~I.}\
  \bibnamefont {Katsnelson}},\ }\bibfield  {title} {\enquote {\bibinfo {title}
  {{The extended Hubbard model with attractive interactions}},}\ }\href
  {\doibase 10.1088/1742-6596/1136/1/012006} {\bibfield  {journal} {\bibinfo
  {journal} {J. Phys.: Conf. Ser.}\ }\textbf {\bibinfo {volume} {1136}},\
  \bibinfo {pages} {012006} (\bibinfo {year} {2018})}\BibitemShut {NoStop}%
\bibitem [{\citenamefont {Rohringer}\ \emph {et~al.}(2018)\citenamefont
  {Rohringer}, \citenamefont {Hafermann}, \citenamefont {Toschi}, \citenamefont
  {Katanin}, \citenamefont {Antipov}, \citenamefont {Katsnelson}, \citenamefont
  {Lichtenstein}, \citenamefont {Rubtsov},\ and\ \citenamefont
  {Held}}]{RevModPhys.90.025003}%
  \BibitemOpen
  \bibfield  {author} {\bibinfo {author} {\bibfnamefont {G.}~\bibnamefont
  {Rohringer}}, \bibinfo {author} {\bibfnamefont {H.}~\bibnamefont
  {Hafermann}}, \bibinfo {author} {\bibfnamefont {A.}~\bibnamefont {Toschi}},
  \bibinfo {author} {\bibfnamefont {A.~A.}\ \bibnamefont {Katanin}}, \bibinfo
  {author} {\bibfnamefont {A.~E.}\ \bibnamefont {Antipov}}, \bibinfo {author}
  {\bibfnamefont {M.~I.}\ \bibnamefont {Katsnelson}}, \bibinfo {author}
  {\bibfnamefont {A.~I.}\ \bibnamefont {Lichtenstein}}, \bibinfo {author}
  {\bibfnamefont {A.~N.}\ \bibnamefont {Rubtsov}}, \ and\ \bibinfo {author}
  {\bibfnamefont {K.}~\bibnamefont {Held}},\ }\bibfield  {title} {\enquote
  {\bibinfo {title} {{Diagrammatic routes to nonlocal correlations beyond
  dynamical mean field theory}},}\ }\href {\doibase
  10.1103/RevModPhys.90.025003} {\bibfield  {journal} {\bibinfo  {journal}
  {Rev. Mod. Phys.}\ }\textbf {\bibinfo {volume} {90}},\ \bibinfo {pages}
  {025003} (\bibinfo {year} {2018})}\BibitemShut {NoStop}%
\bibitem [{\citenamefont {Katanin}(2019)}]{PhysRevB.99.115112}%
  \BibitemOpen
  \bibfield  {author} {\bibinfo {author} {\bibfnamefont {A.~A.}\ \bibnamefont
  {Katanin}},\ }\bibfield  {title} {\enquote {\bibinfo {title} {{Extended
  dynamical mean field theory combined with the two-particle irreducible
  functional renormalization-group approach as a tool to study strongly
  correlated systems}},}\ }\href {\doibase 10.1103/PhysRevB.99.115112}
  {\bibfield  {journal} {\bibinfo  {journal} {Phys. Rev. B}\ }\textbf {\bibinfo
  {volume} {99}},\ \bibinfo {pages} {115112} (\bibinfo {year}
  {2019})}\BibitemShut {NoStop}%
\bibitem [{\citenamefont {Stepanov}\ \emph {et~al.}(2019)\citenamefont
  {Stepanov}, \citenamefont {Huber}, \citenamefont {Lichtenstein},\ and\
  \citenamefont {Katsnelson}}]{PhysRevB.99.115124}%
  \BibitemOpen
  \bibfield  {author} {\bibinfo {author} {\bibfnamefont {E.~A.}\ \bibnamefont
  {Stepanov}}, \bibinfo {author} {\bibfnamefont {A.}~\bibnamefont {Huber}},
  \bibinfo {author} {\bibfnamefont {A.~I.}\ \bibnamefont {Lichtenstein}}, \
  and\ \bibinfo {author} {\bibfnamefont {M.~I.}\ \bibnamefont {Katsnelson}},\
  }\bibfield  {title} {\enquote {\bibinfo {title} {{Effective Ising model for
  correlated systems with charge ordering}},}\ }\href {\doibase
  10.1103/PhysRevB.99.115124} {\bibfield  {journal} {\bibinfo  {journal} {Phys.
  Rev. B}\ }\textbf {\bibinfo {volume} {99}},\ \bibinfo {pages} {115124}
  (\bibinfo {year} {2019})}\BibitemShut {NoStop}%
\bibitem [{\citenamefont {Paki}\ \emph {et~al.}(2019)\citenamefont {Paki},
  \citenamefont {Terletska}, \citenamefont {Iskakov},\ and\ \citenamefont
  {Gull}}]{PhysRevB.99.245146}%
  \BibitemOpen
  \bibfield  {author} {\bibinfo {author} {\bibfnamefont {J.}~\bibnamefont
  {Paki}}, \bibinfo {author} {\bibfnamefont {H.}~\bibnamefont {Terletska}},
  \bibinfo {author} {\bibfnamefont {S.}~\bibnamefont {Iskakov}}, \ and\
  \bibinfo {author} {\bibfnamefont {E.}~\bibnamefont {Gull}},\ }\bibfield
  {title} {\enquote {\bibinfo {title} {{Charge order and antiferromagnetism in
  the extended Hubbard model}},}\ }\href {\doibase 10.1103/PhysRevB.99.245146}
  {\bibfield  {journal} {\bibinfo  {journal} {Phys. Rev. B}\ }\textbf {\bibinfo
  {volume} {99}},\ \bibinfo {pages} {245146} (\bibinfo {year}
  {2019})}\BibitemShut {NoStop}%
\bibitem [{\citenamefont {Pudleiner}\ \emph {et~al.}(2019)\citenamefont
  {Pudleiner}, \citenamefont {Kauch}, \citenamefont {Held},\ and\ \citenamefont
  {Li}}]{PhysRevB.100.075108}%
  \BibitemOpen
  \bibfield  {author} {\bibinfo {author} {\bibfnamefont {P.}~\bibnamefont
  {Pudleiner}}, \bibinfo {author} {\bibfnamefont {A.}~\bibnamefont {Kauch}},
  \bibinfo {author} {\bibfnamefont {K.}~\bibnamefont {Held}}, \ and\ \bibinfo
  {author} {\bibfnamefont {G.}~\bibnamefont {Li}},\ }\bibfield  {title}
  {\enquote {\bibinfo {title} {{Competition between antiferromagnetic and
  charge density wave fluctuations in the extended Hubbard model}},}\ }\href
  {\doibase 10.1103/PhysRevB.100.075108} {\bibfield  {journal} {\bibinfo
  {journal} {Phys. Rev. B}\ }\textbf {\bibinfo {volume} {100}},\ \bibinfo
  {pages} {075108} (\bibinfo {year} {2019})}\BibitemShut {NoStop}%
\bibitem [{\citenamefont {Stepanov}\ \emph {et~al.}(2022)\citenamefont
  {Stepanov}, \citenamefont {Harkov}, \citenamefont {R\"osner}, \citenamefont
  {Lichtenstein}, \citenamefont {Katsnelson},\ and\ \citenamefont
  {Rudenko}}]{stepanov2021coexisting}%
  \BibitemOpen
  \bibfield  {author} {\bibinfo {author} {\bibfnamefont {E.~A.}\ \bibnamefont
  {Stepanov}}, \bibinfo {author} {\bibfnamefont {V.}~\bibnamefont {Harkov}},
  \bibinfo {author} {\bibfnamefont {M.}~\bibnamefont {R\"osner}}, \bibinfo
  {author} {\bibfnamefont {A.~I.}\ \bibnamefont {Lichtenstein}}, \bibinfo
  {author} {\bibfnamefont {M.~I.}\ \bibnamefont {Katsnelson}}, \ and\ \bibinfo
  {author} {\bibfnamefont {A.~N.}\ \bibnamefont {Rudenko}},\ }\bibfield
  {title} {\enquote {\bibinfo {title} {{Coexisting charge density wave and
  ferromagnetic instabilities in monolayer InSe}},}\ }\href {\doibase
  10.1038/s41524-022-00798-4} {\bibfield  {journal} {\bibinfo  {journal} {npj
  Comput. Mater.}\ }\textbf {\bibinfo {volume} {8}},\ \bibinfo {pages} {118}
  (\bibinfo {year} {2022})}\BibitemShut {NoStop}%
\bibitem [{\citenamefont {Vandelli}\ \emph {et~al.}(2020)\citenamefont
  {Vandelli}, \citenamefont {Harkov}, \citenamefont {Stepanov}, \citenamefont
  {Gukelberger}, \citenamefont {Kozik}, \citenamefont {Rubio},\ and\
  \citenamefont {Lichtenstein}}]{PhysRevB.102.195109}%
  \BibitemOpen
  \bibfield  {author} {\bibinfo {author} {\bibfnamefont {M.}~\bibnamefont
  {Vandelli}}, \bibinfo {author} {\bibfnamefont {V.}~\bibnamefont {Harkov}},
  \bibinfo {author} {\bibfnamefont {E.~A.}\ \bibnamefont {Stepanov}}, \bibinfo
  {author} {\bibfnamefont {J.}~\bibnamefont {Gukelberger}}, \bibinfo {author}
  {\bibfnamefont {E.}~\bibnamefont {Kozik}}, \bibinfo {author} {\bibfnamefont
  {A.}~\bibnamefont {Rubio}}, \ and\ \bibinfo {author} {\bibfnamefont {A.~I.}\
  \bibnamefont {Lichtenstein}},\ }\bibfield  {title} {\enquote {\bibinfo
  {title} {{Dual boson diagrammatic Monte Carlo approach applied to the
  extended Hubbard model}},}\ }\href {\doibase 10.1103/PhysRevB.102.195109}
  {\bibfield  {journal} {\bibinfo  {journal} {Phys. Rev. B}\ }\textbf {\bibinfo
  {volume} {102}},\ \bibinfo {pages} {195109} (\bibinfo {year}
  {2020})}\BibitemShut {NoStop}%
\bibitem [{\citenamefont {Terletska}\ \emph {et~al.}(2021)\citenamefont
  {Terletska}, \citenamefont {Iskakov}, \citenamefont {Maier},\ and\
  \citenamefont {Gull}}]{PhysRevB.104.085129}%
  \BibitemOpen
  \bibfield  {author} {\bibinfo {author} {\bibfnamefont {H.}~\bibnamefont
  {Terletska}}, \bibinfo {author} {\bibfnamefont {S.}~\bibnamefont {Iskakov}},
  \bibinfo {author} {\bibfnamefont {T.}~\bibnamefont {Maier}}, \ and\ \bibinfo
  {author} {\bibfnamefont {E.}~\bibnamefont {Gull}},\ }\bibfield  {title}
  {\enquote {\bibinfo {title} {{Dynamical cluster approximation study of
  electron localization in the extended Hubbard model}},}\ }\href {\doibase
  10.1103/PhysRevB.104.085129} {\bibfield  {journal} {\bibinfo  {journal}
  {Phys. Rev. B}\ }\textbf {\bibinfo {volume} {104}},\ \bibinfo {pages}
  {085129} (\bibinfo {year} {2021})}\BibitemShut {NoStop}%
\bibitem [{\citenamefont {Yao}\ \emph {et~al.}(2022)\citenamefont {Yao},
  \citenamefont {Wang},\ and\ \citenamefont {Wang}}]{PhysRevB.106.195121}%
  \BibitemOpen
  \bibfield  {author} {\bibinfo {author} {\bibfnamefont {M.}~\bibnamefont
  {Yao}}, \bibinfo {author} {\bibfnamefont {D.}~\bibnamefont {Wang}}, \ and\
  \bibinfo {author} {\bibfnamefont {Q.-H.}\ \bibnamefont {Wang}},\ }\bibfield
  {title} {\enquote {\bibinfo {title} {{Determinant quantum Monte Carlo for the
  half-filled Hubbard model with nonlocal density-density interactions}},}\
  }\href {\doibase 10.1103/PhysRevB.106.195121} {\bibfield  {journal} {\bibinfo
   {journal} {Phys. Rev. B}\ }\textbf {\bibinfo {volume} {106}},\ \bibinfo
  {pages} {195121} (\bibinfo {year} {2022})}\BibitemShut {NoStop}%
\bibitem [{\citenamefont {{Linn{\'e}r}}\ \emph {et~al.}(2022)\citenamefont
  {{Linn{\'e}r}}, \citenamefont {{Lichtenstein}}, \citenamefont {{Biermann}},\
  and\ \citenamefont {{Stepanov}}}]{2022arXiv221005540L}%
  \BibitemOpen
  \bibfield  {author} {\bibinfo {author} {\bibfnamefont {E.}~\bibnamefont
  {{Linn{\'e}r}}}, \bibinfo {author} {\bibfnamefont {A.~I.}\ \bibnamefont
  {{Lichtenstein}}}, \bibinfo {author} {\bibfnamefont {S.}~\bibnamefont
  {{Biermann}}}, \ and\ \bibinfo {author} {\bibfnamefont {E.~A.}\ \bibnamefont
  {{Stepanov}}},\ }\bibfield  {title} {\enquote {\bibinfo {title}
  {{Multi-channel fluctuating field approach to competing instabilities in
  interacting electronic systems}},}\ }\href@noop {} {\bibfield  {journal}
  {\bibinfo  {journal} {arXiv e-prints}\ } (\bibinfo {year} {2022})},\ \Eprint
  {http://arxiv.org/abs/2210.05540} {arXiv:2210.05540} \BibitemShut {NoStop}%
\bibitem [{\citenamefont {Vandelli}\ \emph {et~al.}(2023)\citenamefont
  {Vandelli}, \citenamefont {Galler}, \citenamefont {Rubio}, \citenamefont
  {Lichtenstein}, \citenamefont {Biermann},\ and\ \citenamefont
  {Stepanov}}]{https://doi.org/10.48550/arxiv.2301.07162}%
  \BibitemOpen
  \bibfield  {author} {\bibinfo {author} {\bibfnamefont {M.}~\bibnamefont
  {Vandelli}}, \bibinfo {author} {\bibfnamefont {A.}~\bibnamefont {Galler}},
  \bibinfo {author} {\bibfnamefont {A.}~\bibnamefont {Rubio}}, \bibinfo
  {author} {\bibfnamefont {A.~I.}\ \bibnamefont {Lichtenstein}}, \bibinfo
  {author} {\bibfnamefont {S.}~\bibnamefont {Biermann}}, \ and\ \bibinfo
  {author} {\bibfnamefont {E.~A.}\ \bibnamefont {Stepanov}},\ }\href {\doibase
  10.48550/ARXIV.2301.07162} {\enquote {\bibinfo {title} {{Doping-dependent
  charge- and spin-density wave orderings in a monolayer of Pb adatoms on
  Si(111)}},}\ } (\bibinfo {year} {2023})\BibitemShut {NoStop}%
\bibitem [{\citenamefont {Robaszkiewicz}\ \emph {et~al.}(1981)\citenamefont
  {Robaszkiewicz}, \citenamefont {Micnas},\ and\ \citenamefont
  {Chao}}]{PhysRevB.23.1447}%
  \BibitemOpen
  \bibfield  {author} {\bibinfo {author} {\bibfnamefont {S.}~\bibnamefont
  {Robaszkiewicz}}, \bibinfo {author} {\bibfnamefont {R.}~\bibnamefont
  {Micnas}}, \ and\ \bibinfo {author} {\bibfnamefont {K.~A.}\ \bibnamefont
  {Chao}},\ }\bibfield  {title} {\enquote {\bibinfo {title} {{Thermodynamic
  properties of the extended Hubbard model with strong intra-atomic attraction
  and an arbitrary electron density}},}\ }\href {\doibase
  10.1103/PhysRevB.23.1447} {\bibfield  {journal} {\bibinfo  {journal} {Phys.
  Rev. B}\ }\textbf {\bibinfo {volume} {23}},\ \bibinfo {pages} {1447--1458}
  (\bibinfo {year} {1981})}\BibitemShut {NoStop}%
\bibitem [{\citenamefont {Moreo}\ and\ \citenamefont
  {Scalapino}(1991)}]{PhysRevLett.66.946}%
  \BibitemOpen
  \bibfield  {author} {\bibinfo {author} {\bibfnamefont {A.}~\bibnamefont
  {Moreo}}\ and\ \bibinfo {author} {\bibfnamefont {D.~J.}\ \bibnamefont
  {Scalapino}},\ }\bibfield  {title} {\enquote {\bibinfo {title}
  {{Two-dimensional negative-U Hubbard model}},}\ }\href {\doibase
  10.1103/PhysRevLett.66.946} {\bibfield  {journal} {\bibinfo  {journal} {Phys.
  Rev. Lett.}\ }\textbf {\bibinfo {volume} {66}},\ \bibinfo {pages} {946--948}
  (\bibinfo {year} {1991})}\BibitemShut {NoStop}%
\bibitem [{\citenamefont {Randeria}\ \emph {et~al.}(1992)\citenamefont
  {Randeria}, \citenamefont {Trivedi}, \citenamefont {Moreo},\ and\
  \citenamefont {Scalettar}}]{PhysRevLett.69.2001}%
  \BibitemOpen
  \bibfield  {author} {\bibinfo {author} {\bibfnamefont {M.}~\bibnamefont
  {Randeria}}, \bibinfo {author} {\bibfnamefont {N.}~\bibnamefont {Trivedi}},
  \bibinfo {author} {\bibfnamefont {A.}~\bibnamefont {Moreo}}, \ and\ \bibinfo
  {author} {\bibfnamefont {R.~T.}\ \bibnamefont {Scalettar}},\ }\bibfield
  {title} {\enquote {\bibinfo {title} {{Pairing and spin gap in the normal
  state of short coherence length superconductors}},}\ }\href {\doibase
  10.1103/PhysRevLett.69.2001} {\bibfield  {journal} {\bibinfo  {journal}
  {Phys. Rev. Lett.}\ }\textbf {\bibinfo {volume} {69}},\ \bibinfo {pages}
  {2001--2004} (\bibinfo {year} {1992})}\BibitemShut {NoStop}%
\bibitem [{\citenamefont {Micnas}\ \emph {et~al.}(1990)\citenamefont {Micnas},
  \citenamefont {Ranninger},\ and\ \citenamefont
  {Robaszkiewicz}}]{RevModPhys.62.113}%
  \BibitemOpen
  \bibfield  {author} {\bibinfo {author} {\bibfnamefont {R.}~\bibnamefont
  {Micnas}}, \bibinfo {author} {\bibfnamefont {J.}~\bibnamefont {Ranninger}}, \
  and\ \bibinfo {author} {\bibfnamefont {S.}~\bibnamefont {Robaszkiewicz}},\
  }\bibfield  {title} {\enquote {\bibinfo {title} {Superconductivity in
  narrow-band systems with local nonretarded attractive interactions},}\ }\href
  {\doibase 10.1103/RevModPhys.62.113} {\bibfield  {journal} {\bibinfo
  {journal} {Rev. Mod. Phys.}\ }\textbf {\bibinfo {volume} {62}},\ \bibinfo
  {pages} {113--171} (\bibinfo {year} {1990})}\BibitemShut {NoStop}%
\bibitem [{\citenamefont {Keller}\ \emph {et~al.}(2001)\citenamefont {Keller},
  \citenamefont {Metzner},\ and\ \citenamefont
  {Schollw\"ock}}]{PhysRevLett.86.4612}%
  \BibitemOpen
  \bibfield  {author} {\bibinfo {author} {\bibfnamefont {M.}~\bibnamefont
  {Keller}}, \bibinfo {author} {\bibfnamefont {W.}~\bibnamefont {Metzner}}, \
  and\ \bibinfo {author} {\bibfnamefont {U.}~\bibnamefont {Schollw\"ock}},\
  }\bibfield  {title} {\enquote {\bibinfo {title} {{Dynamical Mean-Field Theory
  for Pairing and Spin Gap in the Attractive Hubbard Model}},}\ }\href
  {\doibase 10.1103/PhysRevLett.86.4612} {\bibfield  {journal} {\bibinfo
  {journal} {Phys. Rev. Lett.}\ }\textbf {\bibinfo {volume} {86}},\ \bibinfo
  {pages} {4612--4615} (\bibinfo {year} {2001})}\BibitemShut {NoStop}%
\bibitem [{\citenamefont {Capone}\ \emph {et~al.}(2002)\citenamefont {Capone},
  \citenamefont {Castellani},\ and\ \citenamefont
  {Grilli}}]{PhysRevLett.88.126403}%
  \BibitemOpen
  \bibfield  {author} {\bibinfo {author} {\bibfnamefont {M.}~\bibnamefont
  {Capone}}, \bibinfo {author} {\bibfnamefont {C.}~\bibnamefont {Castellani}},
  \ and\ \bibinfo {author} {\bibfnamefont {M.}~\bibnamefont {Grilli}},\
  }\bibfield  {title} {\enquote {\bibinfo {title} {{First-Order Pairing
  Transition and Single-Particle Spectral Function in the Attractive Hubbard
  Model}},}\ }\href {\doibase 10.1103/PhysRevLett.88.126403} {\bibfield
  {journal} {\bibinfo  {journal} {Phys. Rev. Lett.}\ }\textbf {\bibinfo
  {volume} {88}},\ \bibinfo {pages} {126403} (\bibinfo {year}
  {2002})}\BibitemShut {NoStop}%
\bibitem [{\citenamefont {Aligia}(2000)}]{PhysRevB.61.7028}%
  \BibitemOpen
  \bibfield  {author} {\bibinfo {author} {\bibfnamefont {A.~A.}\ \bibnamefont
  {Aligia}},\ }\bibfield  {title} {\enquote {\bibinfo {title} {{Phase diagram
  of the one-dimensional extended attractive Hubbard model for large
  nearest-neighbor repulsion}},}\ }\href {\doibase 10.1103/PhysRevB.61.7028}
  {\bibfield  {journal} {\bibinfo  {journal} {Phys. Rev. B}\ }\textbf {\bibinfo
  {volume} {61}},\ \bibinfo {pages} {7028--7032} (\bibinfo {year}
  {2000})}\BibitemShut {NoStop}%
\bibitem [{\citenamefont {Dagotto}\ \emph {et~al.}(1994)\citenamefont
  {Dagotto}, \citenamefont {Riera}, \citenamefont {Chen}, \citenamefont
  {Moreo}, \citenamefont {Nazarenko}, \citenamefont {Alcaraz},\ and\
  \citenamefont {Ortolani}}]{PhysRevB.49.3548}%
  \BibitemOpen
  \bibfield  {author} {\bibinfo {author} {\bibfnamefont {E.}~\bibnamefont
  {Dagotto}}, \bibinfo {author} {\bibfnamefont {J.}~\bibnamefont {Riera}},
  \bibinfo {author} {\bibfnamefont {Y.~C.}\ \bibnamefont {Chen}}, \bibinfo
  {author} {\bibfnamefont {A.}~\bibnamefont {Moreo}}, \bibinfo {author}
  {\bibfnamefont {A.}~\bibnamefont {Nazarenko}}, \bibinfo {author}
  {\bibfnamefont {F.}~\bibnamefont {Alcaraz}}, \ and\ \bibinfo {author}
  {\bibfnamefont {F.}~\bibnamefont {Ortolani}},\ }\bibfield  {title} {\enquote
  {\bibinfo {title} {{Superconductivity near phase separation in models of
  correlated electrons}},}\ }\href {\doibase 10.1103/PhysRevB.49.3548}
  {\bibfield  {journal} {\bibinfo  {journal} {Phys. Rev. B}\ }\textbf {\bibinfo
  {volume} {49}},\ \bibinfo {pages} {3548--3565} (\bibinfo {year}
  {1994})}\BibitemShut {NoStop}%
\bibitem [{\citenamefont {Toschi}\ \emph
  {et~al.}(2005{\natexlab{a}})\citenamefont {Toschi}, \citenamefont {Capone},\
  and\ \citenamefont {Castellani}}]{PhysRevB.72.235118}%
  \BibitemOpen
  \bibfield  {author} {\bibinfo {author} {\bibfnamefont {A.}~\bibnamefont
  {Toschi}}, \bibinfo {author} {\bibfnamefont {M.}~\bibnamefont {Capone}}, \
  and\ \bibinfo {author} {\bibfnamefont {C.}~\bibnamefont {Castellani}},\
  }\bibfield  {title} {\enquote {\bibinfo {title} {{Energetic balance of the
  superconducting transition across the BCS---Bose Einstein crossover in the
  attractive Hubbard model}},}\ }\href {\doibase 10.1103/PhysRevB.72.235118}
  {\bibfield  {journal} {\bibinfo  {journal} {Phys. Rev. B}\ }\textbf {\bibinfo
  {volume} {72}},\ \bibinfo {pages} {235118} (\bibinfo {year}
  {2005}{\natexlab{a}})}\BibitemShut {NoStop}%
\bibitem [{\citenamefont {Garg}\ \emph {et~al.}(2005)\citenamefont {Garg},
  \citenamefont {Krishnamurthy},\ and\ \citenamefont
  {Randeria}}]{PhysRevB.72.024517}%
  \BibitemOpen
  \bibfield  {author} {\bibinfo {author} {\bibfnamefont {A.}~\bibnamefont
  {Garg}}, \bibinfo {author} {\bibfnamefont {H.~R.}\ \bibnamefont
  {Krishnamurthy}}, \ and\ \bibinfo {author} {\bibfnamefont {M.}~\bibnamefont
  {Randeria}},\ }\bibfield  {title} {\enquote {\bibinfo {title} {{BCS-BEC
  crossover at $T=0$: A dynamical mean-field theory approach}},}\ }\href
  {\doibase 10.1103/PhysRevB.72.024517} {\bibfield  {journal} {\bibinfo
  {journal} {Phys. Rev. B}\ }\textbf {\bibinfo {volume} {72}},\ \bibinfo
  {pages} {024517} (\bibinfo {year} {2005})}\BibitemShut {NoStop}%
\bibitem [{\citenamefont {Toschi}\ \emph
  {et~al.}(2005{\natexlab{b}})\citenamefont {Toschi}, \citenamefont {Barone},
  \citenamefont {Capone},\ and\ \citenamefont {Castellani}}]{Toschi_2005}%
  \BibitemOpen
  \bibfield  {author} {\bibinfo {author} {\bibfnamefont {A}~\bibnamefont
  {Toschi}}, \bibinfo {author} {\bibfnamefont {P}~\bibnamefont {Barone}},
  \bibinfo {author} {\bibfnamefont {M}~\bibnamefont {Capone}}, \ and\ \bibinfo
  {author} {\bibfnamefont {C}~\bibnamefont {Castellani}},\ }\bibfield  {title}
  {\enquote {\bibinfo {title} {{Pairing and superconductivity from weak to
  strong coupling in the attractive Hubbard model}},}\ }\href {\doibase
  10.1088/1367-2630/7/1/007} {\bibfield  {journal} {\bibinfo  {journal} {New J.
  Phys.}\ }\textbf {\bibinfo {volume} {7}},\ \bibinfo {pages} {7} (\bibinfo
  {year} {2005}{\natexlab{b}})}\BibitemShut {NoStop}%
\bibitem [{\citenamefont {Privitera}\ \emph {et~al.}(2010)\citenamefont
  {Privitera}, \citenamefont {Capone},\ and\ \citenamefont
  {Castellani}}]{PhysRevB.81.014523}%
  \BibitemOpen
  \bibfield  {author} {\bibinfo {author} {\bibfnamefont {A.}~\bibnamefont
  {Privitera}}, \bibinfo {author} {\bibfnamefont {M.}~\bibnamefont {Capone}}, \
  and\ \bibinfo {author} {\bibfnamefont {C.}~\bibnamefont {Castellani}},\
  }\bibfield  {title} {\enquote {\bibinfo {title} {{Finite-density corrections
  to the unitary Fermi gas: A lattice perspective from dynamical mean-field
  theory}},}\ }\href {\doibase 10.1103/PhysRevB.81.014523} {\bibfield
  {journal} {\bibinfo  {journal} {Phys. Rev. B}\ }\textbf {\bibinfo {volume}
  {81}},\ \bibinfo {pages} {014523} (\bibinfo {year} {2010})}\BibitemShut
  {NoStop}%
\bibitem [{\citenamefont {Koga}\ and\ \citenamefont
  {Werner}(2011)}]{PhysRevA.84.023638}%
  \BibitemOpen
  \bibfield  {author} {\bibinfo {author} {\bibfnamefont {A.}~\bibnamefont
  {Koga}}\ and\ \bibinfo {author} {\bibfnamefont {P.}~\bibnamefont {Werner}},\
  }\bibfield  {title} {\enquote {\bibinfo {title} {{Low-temperature properties
  of the infinite-dimensional attractive Hubbard model}},}\ }\href {\doibase
  10.1103/PhysRevA.84.023638} {\bibfield  {journal} {\bibinfo  {journal} {Phys.
  Rev. A}\ }\textbf {\bibinfo {volume} {84}},\ \bibinfo {pages} {023638}
  (\bibinfo {year} {2011})}\BibitemShut {NoStop}%
\bibitem [{\citenamefont {Amaricci}\ \emph {et~al.}(2014)\citenamefont
  {Amaricci}, \citenamefont {Privitera},\ and\ \citenamefont
  {Capone}}]{PhysRevA.89.053604}%
  \BibitemOpen
  \bibfield  {author} {\bibinfo {author} {\bibfnamefont {A.}~\bibnamefont
  {Amaricci}}, \bibinfo {author} {\bibfnamefont {A.}~\bibnamefont {Privitera}},
  \ and\ \bibinfo {author} {\bibfnamefont {M.}~\bibnamefont {Capone}},\
  }\bibfield  {title} {\enquote {\bibinfo {title} {{Inhomogeneous BCS-BEC
  crossover for trapped cold atoms in optical lattices}},}\ }\href {\doibase
  10.1103/PhysRevA.89.053604} {\bibfield  {journal} {\bibinfo  {journal} {Phys.
  Rev. A}\ }\textbf {\bibinfo {volume} {89}},\ \bibinfo {pages} {053604}
  (\bibinfo {year} {2014})}\BibitemShut {NoStop}%
\bibitem [{\citenamefont {Tagliavini}\ \emph {et~al.}(2016)\citenamefont
  {Tagliavini}, \citenamefont {Capone},\ and\ \citenamefont
  {Toschi}}]{PhysRevB.94.155114}%
  \BibitemOpen
  \bibfield  {author} {\bibinfo {author} {\bibfnamefont {A.}~\bibnamefont
  {Tagliavini}}, \bibinfo {author} {\bibfnamefont {M.}~\bibnamefont {Capone}},
  \ and\ \bibinfo {author} {\bibfnamefont {A.}~\bibnamefont {Toschi}},\
  }\bibfield  {title} {\enquote {\bibinfo {title} {{Detecting a preformed pair
  phase: Response to a pairing forcing field}},}\ }\href {\doibase
  10.1103/PhysRevB.94.155114} {\bibfield  {journal} {\bibinfo  {journal} {Phys.
  Rev. B}\ }\textbf {\bibinfo {volume} {94}},\ \bibinfo {pages} {155114}
  (\bibinfo {year} {2016})}\BibitemShut {NoStop}%
\bibitem [{\citenamefont {Del~Re}\ \emph {et~al.}(2019)\citenamefont {Del~Re},
  \citenamefont {Capone},\ and\ \citenamefont {Toschi}}]{PhysRevB.99.045137}%
  \BibitemOpen
  \bibfield  {author} {\bibinfo {author} {\bibfnamefont {L.}~\bibnamefont
  {Del~Re}}, \bibinfo {author} {\bibfnamefont {M.}~\bibnamefont {Capone}}, \
  and\ \bibinfo {author} {\bibfnamefont {A.}~\bibnamefont {Toschi}},\
  }\bibfield  {title} {\enquote {\bibinfo {title} {{Dynamical vertex
  approximation for the attractive Hubbard model}},}\ }\href {\doibase
  10.1103/PhysRevB.99.045137} {\bibfield  {journal} {\bibinfo  {journal} {Phys.
  Rev. B}\ }\textbf {\bibinfo {volume} {99}},\ \bibinfo {pages} {045137}
  (\bibinfo {year} {2019})}\BibitemShut {NoStop}%
\bibitem [{\citenamefont {Xiang}\ \emph {et~al.}(2019)\citenamefont {Xiang},
  \citenamefont {Liu}, \citenamefont {Yuan}, \citenamefont {Cao},\ and\
  \citenamefont {Tang}}]{Xiang_2019}%
  \BibitemOpen
  \bibfield  {author} {\bibinfo {author} {\bibfnamefont {Y.-Y.}\ \bibnamefont
  {Xiang}}, \bibinfo {author} {\bibfnamefont {X.-J.}\ \bibnamefont {Liu}},
  \bibinfo {author} {\bibfnamefont {Y.-H.}\ \bibnamefont {Yuan}}, \bibinfo
  {author} {\bibfnamefont {J.}~\bibnamefont {Cao}}, \ and\ \bibinfo {author}
  {\bibfnamefont {C.-M.}\ \bibnamefont {Tang}},\ }\bibfield  {title} {\enquote
  {\bibinfo {title} {{Doping dependence of the phase diagram in one-dimensional
  extended Hubbard model: a functional renormalization group study}},}\ }\href
  {\doibase 10.1088/1361-648X/aafd4d} {\bibfield  {journal} {\bibinfo
  {journal} {J. Phys. Condens. Matter}\ }\textbf {\bibinfo {volume} {31}},\
  \bibinfo {pages} {125601} (\bibinfo {year} {2019})}\BibitemShut {NoStop}%
\bibitem [{\citenamefont {Chen}\ \emph {et~al.}(2022)\citenamefont {Chen},
  \citenamefont {Wang},\ and\ \citenamefont
  {Chen}}]{https://doi.org/10.48550/arxiv.2206.01119}%
  \BibitemOpen
  \bibfield  {author} {\bibinfo {author} {\bibfnamefont {W.-C.}\ \bibnamefont
  {Chen}}, \bibinfo {author} {\bibfnamefont {Y.}~\bibnamefont {Wang}}, \ and\
  \bibinfo {author} {\bibfnamefont {C.-C.}\ \bibnamefont {Chen}},\ }\href
  {\doibase 10.48550/ARXIV.2206.01119} {\enquote {\bibinfo {title}
  {Superconducting phases of the square-lattice extended hubbard model},}\ }
  (\bibinfo {year} {2022})\BibitemShut {NoStop}%
\bibitem [{\citenamefont {Zhang}\ \emph {et~al.}(1991)\citenamefont {Zhang},
  \citenamefont {Ogata},\ and\ \citenamefont {Rice}}]{PhysRevLett.67.3452}%
  \BibitemOpen
  \bibfield  {author} {\bibinfo {author} {\bibfnamefont {F.~C.}\ \bibnamefont
  {Zhang}}, \bibinfo {author} {\bibfnamefont {Masao}\ \bibnamefont {Ogata}}, \
  and\ \bibinfo {author} {\bibfnamefont {T.~M.}\ \bibnamefont {Rice}},\
  }\bibfield  {title} {\enquote {\bibinfo {title} {{Attractive interaction and
  superconductivity for ${\mathrm{K}}_{3}$${\mathrm{C}}_{60}$}},}\ }\href
  {\doibase 10.1103/PhysRevLett.67.3452} {\bibfield  {journal} {\bibinfo
  {journal} {Phys. Rev. Lett.}\ }\textbf {\bibinfo {volume} {67}},\ \bibinfo
  {pages} {3452--3455} (\bibinfo {year} {1991})}\BibitemShut {NoStop}%
\bibitem [{\citenamefont {Wang}\ \emph {et~al.}(2021)\citenamefont {Wang},
  \citenamefont {Chen}, \citenamefont {Shi}, \citenamefont {Moritz},
  \citenamefont {Shen},\ and\ \citenamefont
  {Devereaux}}]{PhysRevLett.127.197003}%
  \BibitemOpen
  \bibfield  {author} {\bibinfo {author} {\bibfnamefont {Y.}~\bibnamefont
  {Wang}}, \bibinfo {author} {\bibfnamefont {Z.}~\bibnamefont {Chen}}, \bibinfo
  {author} {\bibfnamefont {T.}~\bibnamefont {Shi}}, \bibinfo {author}
  {\bibfnamefont {B.}~\bibnamefont {Moritz}}, \bibinfo {author} {\bibfnamefont
  {Z.-X.}\ \bibnamefont {Shen}}, \ and\ \bibinfo {author} {\bibfnamefont
  {T.~P.}\ \bibnamefont {Devereaux}},\ }\bibfield  {title} {\enquote {\bibinfo
  {title} {{Phonon-Mediated Long-Range Attractive Interaction in
  One-Dimensional Cuprates}},}\ }\href {\doibase
  10.1103/PhysRevLett.127.197003} {\bibfield  {journal} {\bibinfo  {journal}
  {Phys. Rev. Lett.}\ }\textbf {\bibinfo {volume} {127}},\ \bibinfo {pages}
  {197003} (\bibinfo {year} {2021})}\BibitemShut {NoStop}%
\bibitem [{\citenamefont {Meregalli}\ and\ \citenamefont
  {Savrasov}(1998)}]{PhysRevB.57.14453}%
  \BibitemOpen
  \bibfield  {author} {\bibinfo {author} {\bibfnamefont {V.}~\bibnamefont
  {Meregalli}}\ and\ \bibinfo {author} {\bibfnamefont {S.~Y.}\ \bibnamefont
  {Savrasov}},\ }\bibfield  {title} {\enquote {\bibinfo {title}
  {{Electron-phonon coupling and properties of doped
  ${\mathrm{BaBiO}}_{3}$}},}\ }\href {\doibase 10.1103/PhysRevB.57.14453}
  {\bibfield  {journal} {\bibinfo  {journal} {Phys. Rev. B}\ }\textbf {\bibinfo
  {volume} {57}},\ \bibinfo {pages} {14453--14469} (\bibinfo {year}
  {1998})}\BibitemShut {NoStop}%
\bibitem [{\citenamefont {Yin}\ \emph {et~al.}(2013)\citenamefont {Yin},
  \citenamefont {Kutepov},\ and\ \citenamefont {Kotliar}}]{PhysRevX.3.021011}%
  \BibitemOpen
  \bibfield  {author} {\bibinfo {author} {\bibfnamefont {Z.~P.}\ \bibnamefont
  {Yin}}, \bibinfo {author} {\bibfnamefont {A.}~\bibnamefont {Kutepov}}, \ and\
  \bibinfo {author} {\bibfnamefont {G.}~\bibnamefont {Kotliar}},\ }\bibfield
  {title} {\enquote {\bibinfo {title} {{Correlation-Enhanced Electron-Phonon
  Coupling: Applications of $GW$ and Screened Hybrid Functional to Bismuthates,
  Chloronitrides, and Other High-${T}_{c}$ Superconductors}},}\ }\href
  {\doibase 10.1103/PhysRevX.3.021011} {\bibfield  {journal} {\bibinfo
  {journal} {Phys. Rev. X}\ }\textbf {\bibinfo {volume} {3}},\ \bibinfo {pages}
  {021011} (\bibinfo {year} {2013})}\BibitemShut {NoStop}%
\bibitem [{\citenamefont {Tomczyk}\ \emph {et~al.}(2015)\citenamefont
  {Tomczyk}, \citenamefont {Lu}, \citenamefont {Veazey}, \citenamefont {Huang},
  \citenamefont {Irvin}, \citenamefont {Ryu}, \citenamefont {Lee},
  \citenamefont {Eom}, \citenamefont {Hellberg},\ and\ \citenamefont
  {Levy}}]{nature14398}%
  \BibitemOpen
  \bibfield  {author} {\bibinfo {author} {\bibfnamefont {M.}~\bibnamefont
  {Tomczyk}}, \bibinfo {author} {\bibfnamefont {S.}~\bibnamefont {Lu}},
  \bibinfo {author} {\bibfnamefont {J.~P.}\ \bibnamefont {Veazey}}, \bibinfo
  {author} {\bibfnamefont {M.}~\bibnamefont {Huang}}, \bibinfo {author}
  {\bibfnamefont {P.}~\bibnamefont {Irvin}}, \bibinfo {author} {\bibfnamefont
  {S.}~\bibnamefont {Ryu}}, \bibinfo {author} {\bibfnamefont {H.}~\bibnamefont
  {Lee}}, \bibinfo {author} {\bibfnamefont {C.-B.}\ \bibnamefont {Eom}},
  \bibinfo {author} {\bibfnamefont {C.~S.}\ \bibnamefont {Hellberg}}, \ and\
  \bibinfo {author} {\bibfnamefont {J.}~\bibnamefont {Levy}},\ }\bibfield
  {title} {\enquote {\bibinfo {title} {{Electron pairing without
  superconductivity}},}\ }\href {\doibase 10.1038/nature14398} {\bibfield
  {journal} {\bibinfo  {journal} {Nature}\ }\textbf {\bibinfo {volume} {521}},\
  \bibinfo {pages} {196–199} (\bibinfo {year} {2015})}\BibitemShut {NoStop}%
\bibitem [{\citenamefont {Cheng}\ \emph {et~al.}(2016)\citenamefont {Cheng},
  \citenamefont {Tomczyk}, \citenamefont {Tacla}, \citenamefont {Lee},
  \citenamefont {Lu}, \citenamefont {Veazey}, \citenamefont {Huang},
  \citenamefont {Irvin}, \citenamefont {Ryu}, \citenamefont {Eom},
  \citenamefont {Daley}, \citenamefont {Pekker},\ and\ \citenamefont
  {Levy}}]{PhysRevX.6.041042}%
  \BibitemOpen
  \bibfield  {author} {\bibinfo {author} {\bibfnamefont {G.}~\bibnamefont
  {Cheng}}, \bibinfo {author} {\bibfnamefont {M.}~\bibnamefont {Tomczyk}},
  \bibinfo {author} {\bibfnamefont {A.~B.}\ \bibnamefont {Tacla}}, \bibinfo
  {author} {\bibfnamefont {H.}~\bibnamefont {Lee}}, \bibinfo {author}
  {\bibfnamefont {S.}~\bibnamefont {Lu}}, \bibinfo {author} {\bibfnamefont
  {J.~P.}\ \bibnamefont {Veazey}}, \bibinfo {author} {\bibfnamefont
  {M.}~\bibnamefont {Huang}}, \bibinfo {author} {\bibfnamefont
  {P.}~\bibnamefont {Irvin}}, \bibinfo {author} {\bibfnamefont
  {S.}~\bibnamefont {Ryu}}, \bibinfo {author} {\bibfnamefont {C.-B.}\
  \bibnamefont {Eom}}, \bibinfo {author} {\bibfnamefont {A.}~\bibnamefont
  {Daley}}, \bibinfo {author} {\bibfnamefont {D.}~\bibnamefont {Pekker}}, \
  and\ \bibinfo {author} {\bibfnamefont {J.}~\bibnamefont {Levy}},\ }\bibfield
  {title} {\enquote {\bibinfo {title} {{Tunable Electron-Electron Interactions
  in ${\mathrm{LaAlO}}_{3}/{\mathrm{SrTiO}}_{3}$ Nanostructures}},}\ }\href
  {\doibase 10.1103/PhysRevX.6.041042} {\bibfield  {journal} {\bibinfo
  {journal} {Phys. Rev. X}\ }\textbf {\bibinfo {volume} {6}},\ \bibinfo {pages}
  {041042} (\bibinfo {year} {2016})}\BibitemShut {NoStop}%
\bibitem [{\citenamefont {Tomczyk}\ \emph {et~al.}(2016)\citenamefont
  {Tomczyk}, \citenamefont {Cheng}, \citenamefont {Lee}, \citenamefont {Lu},
  \citenamefont {Annadi}, \citenamefont {Veazey}, \citenamefont {Huang},
  \citenamefont {Irvin}, \citenamefont {Ryu}, \citenamefont {Eom},\ and\
  \citenamefont {Levy}}]{PhysRevLett.117.096801}%
  \BibitemOpen
  \bibfield  {author} {\bibinfo {author} {\bibfnamefont {M.}~\bibnamefont
  {Tomczyk}}, \bibinfo {author} {\bibfnamefont {G.}~\bibnamefont {Cheng}},
  \bibinfo {author} {\bibfnamefont {H.}~\bibnamefont {Lee}}, \bibinfo {author}
  {\bibfnamefont {S.}~\bibnamefont {Lu}}, \bibinfo {author} {\bibfnamefont
  {A.}~\bibnamefont {Annadi}}, \bibinfo {author} {\bibfnamefont {J.~P.}\
  \bibnamefont {Veazey}}, \bibinfo {author} {\bibfnamefont {M.}~\bibnamefont
  {Huang}}, \bibinfo {author} {\bibfnamefont {P.}~\bibnamefont {Irvin}},
  \bibinfo {author} {\bibfnamefont {S.}~\bibnamefont {Ryu}}, \bibinfo {author}
  {\bibfnamefont {C.-B.}\ \bibnamefont {Eom}}, \ and\ \bibinfo {author}
  {\bibfnamefont {J.}~\bibnamefont {Levy}},\ }\bibfield  {title} {\enquote
  {\bibinfo {title} {{Micrometer-Scale Ballistic Transport of Electron Pairs in
  ${\mathrm{LaAlO}}_{3}/{\mathrm{SrTiO}}_{3}$ Nanowires}},}\ }\href {\doibase
  10.1103/PhysRevLett.117.096801} {\bibfield  {journal} {\bibinfo  {journal}
  {Phys. Rev. Lett.}\ }\textbf {\bibinfo {volume} {117}},\ \bibinfo {pages}
  {096801} (\bibinfo {year} {2016})}\BibitemShut {NoStop}%
\bibitem [{\citenamefont {Prawiroatmodjo}\ \emph {et~al.}(2017)\citenamefont
  {Prawiroatmodjo}, \citenamefont {Leijnse}, \citenamefont {Trier},
  \citenamefont {Chen}, \citenamefont {Christensen}, \citenamefont {von
  Soosten}, \citenamefont {Pryds},\ and\ \citenamefont
  {Jespersen}}]{s41467-017-00495-7}%
  \BibitemOpen
  \bibfield  {author} {\bibinfo {author} {\bibfnamefont {G.~E. D.~K.}\
  \bibnamefont {Prawiroatmodjo}}, \bibinfo {author} {\bibfnamefont
  {M.}~\bibnamefont {Leijnse}}, \bibinfo {author} {\bibfnamefont
  {F.}~\bibnamefont {Trier}}, \bibinfo {author} {\bibfnamefont
  {Y.}~\bibnamefont {Chen}}, \bibinfo {author} {\bibfnamefont {D.~V.}\
  \bibnamefont {Christensen}}, \bibinfo {author} {\bibfnamefont
  {M.}~\bibnamefont {von Soosten}}, \bibinfo {author} {\bibfnamefont
  {N.}~\bibnamefont {Pryds}}, \ and\ \bibinfo {author} {\bibfnamefont {T.~S.}\
  \bibnamefont {Jespersen}},\ }\bibfield  {title} {\enquote {\bibinfo {title}
  {{Transport and excitations in a negative-U quantum dot at the LaAlO3/SrTiO3
  interface}},}\ }\href {\doibase 10.1038/s41467-017-00495-7} {\bibfield
  {journal} {\bibinfo  {journal} {Nat. Commun.}\ }\textbf {\bibinfo {volume}
  {8}},\ \bibinfo {pages} {395} (\bibinfo {year} {2017})}\BibitemShut {NoStop}%
\bibitem [{\citenamefont {van~der Marel}\ and\ \citenamefont
  {Sawatzky}(1988)}]{PhysRevB.37.10674}%
  \BibitemOpen
  \bibfield  {author} {\bibinfo {author} {\bibfnamefont {D.}~\bibnamefont
  {van~der Marel}}\ and\ \bibinfo {author} {\bibfnamefont {G.~A.}\ \bibnamefont
  {Sawatzky}},\ }\bibfield  {title} {\enquote {\bibinfo {title}
  {{Electron-electron interaction and localization in d and f transition
  metals}},}\ }\href {\doibase 10.1103/PhysRevB.37.10674} {\bibfield  {journal}
  {\bibinfo  {journal} {Phys. Rev. B}\ }\textbf {\bibinfo {volume} {37}},\
  \bibinfo {pages} {10674--10684} (\bibinfo {year} {1988})}\BibitemShut
  {NoStop}%
\bibitem [{\citenamefont {Strand}(2014)}]{PhysRevB.90.155108}%
  \BibitemOpen
  \bibfield  {author} {\bibinfo {author} {\bibfnamefont {H.~U.~R.}\
  \bibnamefont {Strand}},\ }\bibfield  {title} {\enquote {\bibinfo {title}
  {{Valence-skipping and negative-$U$ in the $d$-band from repulsive local
  Coulomb interaction}},}\ }\href {\doibase 10.1103/PhysRevB.90.155108}
  {\bibfield  {journal} {\bibinfo  {journal} {Phys. Rev. B}\ }\textbf {\bibinfo
  {volume} {90}},\ \bibinfo {pages} {155108} (\bibinfo {year}
  {2014})}\BibitemShut {NoStop}%
\bibitem [{\citenamefont
  {Esslinger}(2010)}]{doi:10.1146/annurev-conmatphys-070909-104059}%
  \BibitemOpen
  \bibfield  {author} {\bibinfo {author} {\bibfnamefont {T.}~\bibnamefont
  {Esslinger}},\ }\bibfield  {title} {\enquote {\bibinfo {title}
  {{Fermi-Hubbard Physics with Atoms in an Optical Lattice}},}\ }\href
  {\doibase 10.1146/annurev-conmatphys-070909-104059} {\bibfield  {journal}
  {\bibinfo  {journal} {Annu. Rev. Condens. Matter Phys.}\ }\textbf {\bibinfo
  {volume} {1}},\ \bibinfo {pages} {129--152} (\bibinfo {year}
  {2010})}\BibitemShut {NoStop}%
\bibitem [{\citenamefont {Rubtsov}(2018)}]{PhysRevE.97.052120}%
  \BibitemOpen
  \bibfield  {author} {\bibinfo {author} {\bibfnamefont {A.~N.}\ \bibnamefont
  {Rubtsov}},\ }\bibfield  {title} {\enquote {\bibinfo {title} {{Fluctuating
  local field method probed for a description of small classical correlated
  lattices}},}\ }\href {\doibase 10.1103/PhysRevE.97.052120} {\bibfield
  {journal} {\bibinfo  {journal} {Phys. Rev. E}\ }\textbf {\bibinfo {volume}
  {97}},\ \bibinfo {pages} {052120} (\bibinfo {year} {2018})}\BibitemShut
  {NoStop}%
\bibitem [{\citenamefont {Rubtsov}\ \emph {et~al.}(2020)\citenamefont
  {Rubtsov}, \citenamefont {Stepanov},\ and\ \citenamefont
  {Lichtenstein}}]{PhysRevB.102.224423}%
  \BibitemOpen
  \bibfield  {author} {\bibinfo {author} {\bibfnamefont {A.~N.}\ \bibnamefont
  {Rubtsov}}, \bibinfo {author} {\bibfnamefont {E.~A.}\ \bibnamefont
  {Stepanov}}, \ and\ \bibinfo {author} {\bibfnamefont {A.~I.}\ \bibnamefont
  {Lichtenstein}},\ }\bibfield  {title} {\enquote {\bibinfo {title}
  {{Collective magnetic fluctuations in Hubbard plaquettes captured by
  fluctuating local field method}},}\ }\href {\doibase
  10.1103/PhysRevB.102.224423} {\bibfield  {journal} {\bibinfo  {journal}
  {Phys. Rev. B}\ }\textbf {\bibinfo {volume} {102}},\ \bibinfo {pages}
  {224423} (\bibinfo {year} {2020})}\BibitemShut {NoStop}%
\bibitem [{\citenamefont {Lyakhova}\ \emph {et~al.}(2022)\citenamefont
  {Lyakhova}, \citenamefont {Stepanov},\ and\ \citenamefont
  {Rubtsov}}]{PhysRevB.105.035118}%
  \BibitemOpen
  \bibfield  {author} {\bibinfo {author} {\bibfnamefont {Y.~S.}\ \bibnamefont
  {Lyakhova}}, \bibinfo {author} {\bibfnamefont {E.~A.}\ \bibnamefont
  {Stepanov}}, \ and\ \bibinfo {author} {\bibfnamefont {A.~N.}\ \bibnamefont
  {Rubtsov}},\ }\bibfield  {title} {\enquote {\bibinfo {title} {{Fluctuating
  local field approach to free energy of one-dimensional molecules with strong
  collective electronic fluctuations}},}\ }\href {\doibase
  10.1103/PhysRevB.105.035118} {\bibfield  {journal} {\bibinfo  {journal}
  {Phys. Rev. B}\ }\textbf {\bibinfo {volume} {105}},\ \bibinfo {pages}
  {035118} (\bibinfo {year} {2022})}\BibitemShut {NoStop}%
\bibitem [{\citenamefont {Lyakhova}\ and\ \citenamefont
  {Rubtsov}(2022)}]{s10948-022-06303-8}%
  \BibitemOpen
  \bibfield  {author} {\bibinfo {author} {\bibfnamefont {Y.~S.}\ \bibnamefont
  {Lyakhova}}\ and\ \bibinfo {author} {\bibfnamefont {A.~N.}\ \bibnamefont
  {Rubtsov}},\ }\bibfield  {title} {\enquote {\bibinfo {title} {{Fluctuating
  local field approach to the description of lattice models in the strong
  coupling regime}},}\ }\href {\doibase 10.1007/s10948-022-06303-8} {\bibfield
  {journal} {\bibinfo  {journal} {J. Supercond. Nov. Magn.}\ }\textbf {\bibinfo
  {volume} {35}},\ \bibinfo {pages} {2169–2173} (\bibinfo {year}
  {2022})}\BibitemShut {NoStop}%
\bibitem [{\citenamefont {Hohenberg}(1967)}]{PhysRev.158.383}%
  \BibitemOpen
  \bibfield  {author} {\bibinfo {author} {\bibfnamefont {P.~C.}\ \bibnamefont
  {Hohenberg}},\ }\bibfield  {title} {\enquote {\bibinfo {title} {{Existence of
  Long-Range Order in One and Two Dimensions}},}\ }\href {\doibase
  10.1103/PhysRev.158.383} {\bibfield  {journal} {\bibinfo  {journal} {Phys.
  Rev.}\ }\textbf {\bibinfo {volume} {158}},\ \bibinfo {pages} {383--386}
  (\bibinfo {year} {1967})}\BibitemShut {NoStop}%
\bibitem [{\citenamefont {Mermin}\ and\ \citenamefont
  {Wagner}(1966)}]{PhysRevLett.17.1133}%
  \BibitemOpen
  \bibfield  {author} {\bibinfo {author} {\bibfnamefont {N.~D.}\ \bibnamefont
  {Mermin}}\ and\ \bibinfo {author} {\bibfnamefont {H.}~\bibnamefont
  {Wagner}},\ }\bibfield  {title} {\enquote {\bibinfo {title} {{Absence of
  Ferromagnetism or Antiferromagnetism in One- or Two-Dimensional Isotropic
  Heisenberg Models}},}\ }\href {\doibase 10.1103/PhysRevLett.17.1133}
  {\bibfield  {journal} {\bibinfo  {journal} {Phys. Rev. Lett.}\ }\textbf
  {\bibinfo {volume} {17}},\ \bibinfo {pages} {1133--1136} (\bibinfo {year}
  {1966})}\BibitemShut {NoStop}%
\bibitem [{\citenamefont {Walker}\ and\ \citenamefont
  {Ruijgrok}(1968)}]{PhysRev.171.513}%
  \BibitemOpen
  \bibfield  {author} {\bibinfo {author} {\bibfnamefont {M.~B.}\ \bibnamefont
  {Walker}}\ and\ \bibinfo {author} {\bibfnamefont {Th.~W.}\ \bibnamefont
  {Ruijgrok}},\ }\bibfield  {title} {\enquote {\bibinfo {title} {{Absence of
  Magnetic Ordering in One and Two Dimensions in a Many-Band Model for
  Interacting Electrons in a Metal}},}\ }\href {\doibase
  10.1103/PhysRev.171.513} {\bibfield  {journal} {\bibinfo  {journal} {Phys.
  Rev.}\ }\textbf {\bibinfo {volume} {171}},\ \bibinfo {pages} {513--515}
  (\bibinfo {year} {1968})}\BibitemShut {NoStop}%
\bibitem [{SM()}]{SM}%
  \BibitemOpen
  \bibinfo {note} {See Supplemental Material for a detailed description of the
  MCFF approach applied to the extended Hubbard model, including a definition
  of the calculated quantities and a derivation of the variational construction
  of MCFF trial action.}\BibitemShut {Stop}%
\bibitem [{\citenamefont {Peierls}(1938)}]{PhysRev.54.918}%
  \BibitemOpen
  \bibfield  {author} {\bibinfo {author} {\bibfnamefont {R.}~\bibnamefont
  {Peierls}},\ }\bibfield  {title} {\enquote {\bibinfo {title} {{On a Minimum
  Property of the Free Energy}},}\ }\href {\doibase 10.1103/PhysRev.54.918}
  {\bibfield  {journal} {\bibinfo  {journal} {Phys. Rev.}\ }\textbf {\bibinfo
  {volume} {54}},\ \bibinfo {pages} {918--919} (\bibinfo {year}
  {1938})}\BibitemShut {NoStop}%
\bibitem [{\citenamefont {Bogolyubov}(1958)}]{Bogolyubov:1958zv}%
  \BibitemOpen
  \bibfield  {author} {\bibinfo {author} {\bibfnamefont {N.~N.}\ \bibnamefont
  {Bogolyubov}},\ }\bibfield  {title} {\enquote {\bibinfo {title} {{On a
  variational principle in the many-body problem}},}\ }\href@noop {} {\bibfield
   {journal} {\bibinfo  {journal} {Sov. Phys. Dokl.}\ }\textbf {\bibinfo
  {volume} {3}},\ \bibinfo {pages} {292--294} (\bibinfo {year}
  {1958})}\BibitemShut {NoStop}%
\bibitem [{\citenamefont {Feynman}(1972)}]{feynman1972}%
  \BibitemOpen
  \bibfield  {author} {\bibinfo {author} {\bibfnamefont {R.~P.}\ \bibnamefont
  {Feynman}},\ }\href@noop {} {\emph {\bibinfo {title} {{Statistical mechanics:
  A set of lectures}}}}\ (\bibinfo  {publisher} {Reading, Mass:
  Benjamin/Cummings},\ \bibinfo {year} {1972})\BibitemShut {NoStop}%
\bibitem [{\citenamefont {\ifmmode~\check{S}\else \v{S}\fi{}imkovic}\ \emph
  {et~al.}(2020)\citenamefont {\ifmmode~\check{S}\else \v{S}\fi{}imkovic},
  \citenamefont {LeBlanc}, \citenamefont {Kim}, \citenamefont {Deng},
  \citenamefont {Prokof'ev}, \citenamefont {Svistunov},\ and\ \citenamefont
  {Kozik}}]{PhysRevLett.124.017003}%
  \BibitemOpen
  \bibfield  {author} {\bibinfo {author} {\bibfnamefont {F.}~\bibnamefont
  {\ifmmode~\check{S}\else \v{S}\fi{}imkovic}}, \bibinfo {author}
  {\bibfnamefont {J.~P.~F.}\ \bibnamefont {LeBlanc}}, \bibinfo {author}
  {\bibfnamefont {A.~J.}\ \bibnamefont {Kim}}, \bibinfo {author} {\bibfnamefont
  {Y.}~\bibnamefont {Deng}}, \bibinfo {author} {\bibfnamefont {N.~V.}\
  \bibnamefont {Prokof'ev}}, \bibinfo {author} {\bibfnamefont {B.~V.}\
  \bibnamefont {Svistunov}}, \ and\ \bibinfo {author} {\bibfnamefont
  {E.}~\bibnamefont {Kozik}},\ }\bibfield  {title} {\enquote {\bibinfo {title}
  {{Extended Crossover from a Fermi Liquid to a Quasiantiferromagnet in the
  Half-Filled 2D Hubbard Model}},}\ }\href {\doibase
  10.1103/PhysRevLett.124.017003} {\bibfield  {journal} {\bibinfo  {journal}
  {Phys. Rev. Lett.}\ }\textbf {\bibinfo {volume} {124}},\ \bibinfo {pages}
  {017003} (\bibinfo {year} {2020})}\BibitemShut {NoStop}%
\bibitem [{\citenamefont {Itin}\ and\ \citenamefont
  {Neishtadt}(2014)}]{ITIN2014822}%
  \BibitemOpen
  \bibfield  {author} {\bibinfo {author} {\bibfnamefont {A.~P.}\ \bibnamefont
  {Itin}}\ and\ \bibinfo {author} {\bibfnamefont {A.~I.}\ \bibnamefont
  {Neishtadt}},\ }\bibfield  {title} {\enquote {\bibinfo {title} {{Effective
  Hamiltonians for fastly driven tight-binding chains}},}\ }\href {\doibase
  https://doi.org/10.1016/j.physleta.2014.01.007} {\bibfield  {journal}
  {\bibinfo  {journal} {Phys. Lett. A}\ }\textbf {\bibinfo {volume} {378}},\
  \bibinfo {pages} {822 -- 825} (\bibinfo {year} {2014})}\BibitemShut {NoStop}%
\bibitem [{\citenamefont {Itin}\ and\ \citenamefont
  {Katsnelson}(2015)}]{PhysRevLett.115.075301}%
  \BibitemOpen
  \bibfield  {author} {\bibinfo {author} {\bibfnamefont {A.~P.}\ \bibnamefont
  {Itin}}\ and\ \bibinfo {author} {\bibfnamefont {M.~I.}\ \bibnamefont
  {Katsnelson}},\ }\bibfield  {title} {\enquote {\bibinfo {title} {{Effective
  Hamiltonians for Rapidly Driven Many-Body Lattice Systems: Induced Exchange
  Interactions and Density-Dependent Hoppings}},}\ }\href {\doibase
  10.1103/PhysRevLett.115.075301} {\bibfield  {journal} {\bibinfo  {journal}
  {Phys. Rev. Lett.}\ }\textbf {\bibinfo {volume} {115}},\ \bibinfo {pages}
  {075301} (\bibinfo {year} {2015})}\BibitemShut {NoStop}%
\bibitem [{\citenamefont {Bukov}\ \emph {et~al.}(2016)\citenamefont {Bukov},
  \citenamefont {Kolodrubetz},\ and\ \citenamefont
  {Polkovnikov}}]{PhysRevLett.116.125301}%
  \BibitemOpen
  \bibfield  {author} {\bibinfo {author} {\bibfnamefont {M.}~\bibnamefont
  {Bukov}}, \bibinfo {author} {\bibfnamefont {M.}~\bibnamefont {Kolodrubetz}},
  \ and\ \bibinfo {author} {\bibfnamefont {A.}~\bibnamefont {Polkovnikov}},\
  }\bibfield  {title} {\enquote {\bibinfo {title} {{Schrieffer-Wolff
  Transformation for Periodically Driven Systems: Strongly Correlated Systems
  with Artificial Gauge Fields}},}\ }\href {\doibase
  10.1103/PhysRevLett.116.125301} {\bibfield  {journal} {\bibinfo  {journal}
  {Phys. Rev. Lett.}\ }\textbf {\bibinfo {volume} {116}},\ \bibinfo {pages}
  {125301} (\bibinfo {year} {2016})}\BibitemShut {NoStop}%
\bibitem [{\citenamefont {Dutreix}\ and\ \citenamefont
  {Katsnelson}(2017)}]{PhysRevB.95.024306}%
  \BibitemOpen
  \bibfield  {author} {\bibinfo {author} {\bibfnamefont {C.}~\bibnamefont
  {Dutreix}}\ and\ \bibinfo {author} {\bibfnamefont {M.~I.}\ \bibnamefont
  {Katsnelson}},\ }\bibfield  {title} {\enquote {\bibinfo {title} {{Dynamical
  control of electron-phonon interactions with high-frequency light}},}\ }\href
  {\doibase 10.1103/PhysRevB.95.024306} {\bibfield  {journal} {\bibinfo
  {journal} {Phys. Rev. B}\ }\textbf {\bibinfo {volume} {95}},\ \bibinfo
  {pages} {024306} (\bibinfo {year} {2017})}\BibitemShut {NoStop}%
\bibitem [{\citenamefont {Dutreix}\ \emph {et~al.}(2016)\citenamefont
  {Dutreix}, \citenamefont {Stepanov},\ and\ \citenamefont
  {Katsnelson}}]{PhysRevB.93.241404}%
  \BibitemOpen
  \bibfield  {author} {\bibinfo {author} {\bibfnamefont {C.}~\bibnamefont
  {Dutreix}}, \bibinfo {author} {\bibfnamefont {E.~A.}\ \bibnamefont
  {Stepanov}}, \ and\ \bibinfo {author} {\bibfnamefont {M.~I.}\ \bibnamefont
  {Katsnelson}},\ }\bibfield  {title} {\enquote {\bibinfo {title}
  {{Laser-induced topological transitions in phosphorene with inversion
  symmetry}},}\ }\href {\doibase 10.1103/PhysRevB.93.241404} {\bibfield
  {journal} {\bibinfo  {journal} {Phys. Rev. B}\ }\textbf {\bibinfo {volume}
  {93}},\ \bibinfo {pages} {241404(R)} (\bibinfo {year} {2016})}\BibitemShut
  {NoStop}%
\bibitem [{\citenamefont {Stepanov}\ \emph {et~al.}(2017)\citenamefont
  {Stepanov}, \citenamefont {Dutreix},\ and\ \citenamefont
  {Katsnelson}}]{PhysRevLett.118.157201}%
  \BibitemOpen
  \bibfield  {author} {\bibinfo {author} {\bibfnamefont {E.~A.}\ \bibnamefont
  {Stepanov}}, \bibinfo {author} {\bibfnamefont {C.}~\bibnamefont {Dutreix}}, \
  and\ \bibinfo {author} {\bibfnamefont {M.~I.}\ \bibnamefont {Katsnelson}},\
  }\bibfield  {title} {\enquote {\bibinfo {title} {{Dynamical and Reversible
  Control of Topological Spin Textures}},}\ }\href {\doibase
  10.1103/PhysRevLett.118.157201} {\bibfield  {journal} {\bibinfo  {journal}
  {Phys. Rev. Lett.}\ }\textbf {\bibinfo {volume} {118}},\ \bibinfo {pages}
  {157201} (\bibinfo {year} {2017})}\BibitemShut {NoStop}%
\bibitem [{\citenamefont {Valmispild}\ \emph {et~al.}(2020)\citenamefont
  {Valmispild}, \citenamefont {Dutreix}, \citenamefont {Eckstein},
  \citenamefont {Katsnelson}, \citenamefont {Lichtenstein},\ and\ \citenamefont
  {Stepanov}}]{PhysRevB.102.220301}%
  \BibitemOpen
  \bibfield  {author} {\bibinfo {author} {\bibfnamefont {V.~N.}\ \bibnamefont
  {Valmispild}}, \bibinfo {author} {\bibfnamefont {C.}~\bibnamefont {Dutreix}},
  \bibinfo {author} {\bibfnamefont {M.}~\bibnamefont {Eckstein}}, \bibinfo
  {author} {\bibfnamefont {M.~I.}\ \bibnamefont {Katsnelson}}, \bibinfo
  {author} {\bibfnamefont {A.~I.}\ \bibnamefont {Lichtenstein}}, \ and\
  \bibinfo {author} {\bibfnamefont {E.~A.}\ \bibnamefont {Stepanov}},\
  }\bibfield  {title} {\enquote {\bibinfo {title} {{Dynamically induced doublon
  repulsion in the Fermi-Hubbard model probed by a single-particle density of
  states}},}\ }\href {\doibase 10.1103/PhysRevB.102.220301} {\bibfield
  {journal} {\bibinfo  {journal} {Phys. Rev. B}\ }\textbf {\bibinfo {volume}
  {102}},\ \bibinfo {pages} {220301} (\bibinfo {year} {2020})}\BibitemShut
  {NoStop}%
\bibitem [{\citenamefont {Sentef}(2017)}]{PhysRevB.95.205111}%
  \BibitemOpen
  \bibfield  {author} {\bibinfo {author} {\bibfnamefont {M.~A.}\ \bibnamefont
  {Sentef}},\ }\bibfield  {title} {\enquote {\bibinfo {title} {{Light-enhanced
  electron-phonon coupling from nonlinear electron-phonon coupling}},}\ }\href
  {\doibase 10.1103/PhysRevB.95.205111} {\bibfield  {journal} {\bibinfo
  {journal} {Phys. Rev. B}\ }\textbf {\bibinfo {volume} {95}},\ \bibinfo
  {pages} {205111} (\bibinfo {year} {2017})}\BibitemShut {NoStop}%
\bibitem [{\citenamefont {Berger}\ \emph {et~al.}(1995)\citenamefont {Berger},
  \citenamefont {Val\'a\ifmmode~\check{s}\else \v{s}\fi{}ek},\ and\
  \citenamefont {von~der Linden}}]{PhysRevB.52.4806}%
  \BibitemOpen
  \bibfield  {author} {\bibinfo {author} {\bibfnamefont {E.}~\bibnamefont
  {Berger}}, \bibinfo {author} {\bibfnamefont {P.}~\bibnamefont
  {Val\'a\ifmmode~\check{s}\else \v{s}\fi{}ek}}, \ and\ \bibinfo {author}
  {\bibfnamefont {W.}~\bibnamefont {von~der Linden}},\ }\bibfield  {title}
  {\enquote {\bibinfo {title} {{Two-dimensional Hubbard-Holstein model}},}\
  }\href {\doibase 10.1103/PhysRevB.52.4806} {\bibfield  {journal} {\bibinfo
  {journal} {Phys. Rev. B}\ }\textbf {\bibinfo {volume} {52}},\ \bibinfo
  {pages} {4806--4814} (\bibinfo {year} {1995})}\BibitemShut {NoStop}%
\bibitem [{\citenamefont {Sangiovanni}\ \emph {et~al.}(2005)\citenamefont
  {Sangiovanni}, \citenamefont {Capone}, \citenamefont {Castellani},\ and\
  \citenamefont {Grilli}}]{PhysRevLett.94.026401}%
  \BibitemOpen
  \bibfield  {author} {\bibinfo {author} {\bibfnamefont {G.}~\bibnamefont
  {Sangiovanni}}, \bibinfo {author} {\bibfnamefont {M.}~\bibnamefont {Capone}},
  \bibinfo {author} {\bibfnamefont {C.}~\bibnamefont {Castellani}}, \ and\
  \bibinfo {author} {\bibfnamefont {M.}~\bibnamefont {Grilli}},\ }\bibfield
  {title} {\enquote {\bibinfo {title} {{Electron-Phonon Interaction Close to a
  Mott Transition}},}\ }\href {\doibase 10.1103/PhysRevLett.94.026401}
  {\bibfield  {journal} {\bibinfo  {journal} {Phys. Rev. Lett.}\ }\textbf
  {\bibinfo {volume} {94}},\ \bibinfo {pages} {026401} (\bibinfo {year}
  {2005})}\BibitemShut {NoStop}%
\bibitem [{\citenamefont {Werner}\ and\ \citenamefont
  {Millis}(2007)}]{PhysRevLett.99.146404}%
  \BibitemOpen
  \bibfield  {author} {\bibinfo {author} {\bibfnamefont {P.}~\bibnamefont
  {Werner}}\ and\ \bibinfo {author} {\bibfnamefont {A.~J.}\ \bibnamefont
  {Millis}},\ }\bibfield  {title} {\enquote {\bibinfo {title} {{Efficient
  Dynamical Mean Field Simulation of the Holstein-Hubbard Model}},}\ }\href
  {\doibase 10.1103/PhysRevLett.99.146404} {\bibfield  {journal} {\bibinfo
  {journal} {Phys. Rev. Lett.}\ }\textbf {\bibinfo {volume} {99}},\ \bibinfo
  {pages} {146404} (\bibinfo {year} {2007})}\BibitemShut {NoStop}%
\bibitem [{\citenamefont {Bednorz}\ and\ \citenamefont
  {Müller}(1986)}]{Bednorz1986}%
  \BibitemOpen
  \bibfield  {author} {\bibinfo {author} {\bibfnamefont {J.~G.}\ \bibnamefont
  {Bednorz}}\ and\ \bibinfo {author} {\bibfnamefont {K.~A.}\ \bibnamefont
  {Müller}},\ }\bibfield  {title} {\enquote {\bibinfo {title} {{Possible high
  ${T_c}$ superconductivity in the Ba-La-Cu-O system}},}\ }\href {\doibase
  10.1007/BF01303701} {\bibfield  {journal} {\bibinfo  {journal} {Z. Phys. B
  Condens. Matter}\ }\textbf {\bibinfo {volume} {64}},\ \bibinfo {pages}
  {189--193} (\bibinfo {year} {1986})}\BibitemShut {NoStop}%
\bibitem [{\citenamefont {Jorgensen}\ \emph {et~al.}(1988)\citenamefont
  {Jorgensen}, \citenamefont {Dabrowski}, \citenamefont {Pei}, \citenamefont
  {Hinks}, \citenamefont {Soderholm}, \citenamefont {Morosin}, \citenamefont
  {Schirber}, \citenamefont {Venturini},\ and\ \citenamefont
  {Ginley}}]{PhysRevB.38.11337}%
  \BibitemOpen
  \bibfield  {author} {\bibinfo {author} {\bibfnamefont {J.~D.}\ \bibnamefont
  {Jorgensen}}, \bibinfo {author} {\bibfnamefont {B.}~\bibnamefont
  {Dabrowski}}, \bibinfo {author} {\bibfnamefont {Shiyou}\ \bibnamefont {Pei}},
  \bibinfo {author} {\bibfnamefont {D.~G.}\ \bibnamefont {Hinks}}, \bibinfo
  {author} {\bibfnamefont {L.}~\bibnamefont {Soderholm}}, \bibinfo {author}
  {\bibfnamefont {B.}~\bibnamefont {Morosin}}, \bibinfo {author} {\bibfnamefont
  {J.~E.}\ \bibnamefont {Schirber}}, \bibinfo {author} {\bibfnamefont {E.~L.}\
  \bibnamefont {Venturini}}, \ and\ \bibinfo {author} {\bibfnamefont {D.~S.}\
  \bibnamefont {Ginley}},\ }\bibfield  {title} {\enquote {\bibinfo {title}
  {Superconducting phase of {La}$_{2}${CuO}$_{4+\ensuremath{\delta}}$: A
  superconducting composition resulting from phase separation},}\ }\href
  {\doibase 10.1103/PhysRevB.38.11337} {\bibfield  {journal} {\bibinfo
  {journal} {Phys. Rev. B}\ }\textbf {\bibinfo {volume} {38}},\ \bibinfo
  {pages} {11337--11345} (\bibinfo {year} {1988})}\BibitemShut {NoStop}%
\bibitem [{\citenamefont {Zaanen}\ and\ \citenamefont
  {Gunnarsson}(1989)}]{PhysRevB.40.7391}%
  \BibitemOpen
  \bibfield  {author} {\bibinfo {author} {\bibfnamefont {Jan}\ \bibnamefont
  {Zaanen}}\ and\ \bibinfo {author} {\bibfnamefont {Olle}\ \bibnamefont
  {Gunnarsson}},\ }\bibfield  {title} {\enquote {\bibinfo {title} {Charged
  magnetic domain lines and the magnetism of high-${T}_{c}$ oxides},}\ }\href
  {\doibase 10.1103/PhysRevB.40.7391} {\bibfield  {journal} {\bibinfo
  {journal} {Phys. Rev. B}\ }\textbf {\bibinfo {volume} {40}},\ \bibinfo
  {pages} {7391--7394} (\bibinfo {year} {1989})}\BibitemShut {NoStop}%
\bibitem [{\citenamefont {Emery}\ and\ \citenamefont
  {Kivelson}(1993)}]{EMERY1993597}%
  \BibitemOpen
  \bibfield  {author} {\bibinfo {author} {\bibfnamefont {V.J.}\ \bibnamefont
  {Emery}}\ and\ \bibinfo {author} {\bibfnamefont {S.A.}\ \bibnamefont
  {Kivelson}},\ }\bibfield  {title} {\enquote {\bibinfo {title} {{Frustrated
  electronic phase separation and high-temperature superconductors}},}\ }\href
  {\doibase https://doi.org/10.1016/0921-4534(93)90581-A} {\bibfield  {journal}
  {\bibinfo  {journal} {Physica C: Superconductivity}\ }\textbf {\bibinfo
  {volume} {209}},\ \bibinfo {pages} {597--621} (\bibinfo {year}
  {1993})}\BibitemShut {NoStop}%
\bibitem [{\citenamefont {Statt}\ \emph {et~al.}(1995)\citenamefont {Statt},
  \citenamefont {Hammel}, \citenamefont {Fisk}, \citenamefont {Cheong},
  \citenamefont {Chou}, \citenamefont {Johnston},\ and\ \citenamefont
  {Schirber}}]{PhysRevB.52.15575}%
  \BibitemOpen
  \bibfield  {author} {\bibinfo {author} {\bibfnamefont {B.~W.}\ \bibnamefont
  {Statt}}, \bibinfo {author} {\bibfnamefont {P.~C.}\ \bibnamefont {Hammel}},
  \bibinfo {author} {\bibfnamefont {Z.}~\bibnamefont {Fisk}}, \bibinfo {author}
  {\bibfnamefont {S-W.}\ \bibnamefont {Cheong}}, \bibinfo {author}
  {\bibfnamefont {F.~C.}\ \bibnamefont {Chou}}, \bibinfo {author}
  {\bibfnamefont {D.~C.}\ \bibnamefont {Johnston}}, \ and\ \bibinfo {author}
  {\bibfnamefont {J.~E.}\ \bibnamefont {Schirber}},\ }\bibfield  {title}
  {\enquote {\bibinfo {title} {Oxygen ordering and phase separation in
  {La}$_{2}${CuO}$_{4+\ensuremath{\delta}}$},}\ }\href {\doibase
  10.1103/PhysRevB.52.15575} {\bibfield  {journal} {\bibinfo  {journal} {Phys.
  Rev. B}\ }\textbf {\bibinfo {volume} {52}},\ \bibinfo {pages} {15575--15581}
  (\bibinfo {year} {1995})}\BibitemShut {NoStop}%
\bibitem [{\citenamefont {Fratini}\ \emph {et~al.}(2010)\citenamefont
  {Fratini}, \citenamefont {Poccia}, \citenamefont {Ricci}, \citenamefont
  {Campi}, \citenamefont {Burghammer}, \citenamefont {Aeppli},\ and\
  \citenamefont {Bianconi}}]{nature09260}%
  \BibitemOpen
  \bibfield  {author} {\bibinfo {author} {\bibfnamefont {M.}~\bibnamefont
  {Fratini}}, \bibinfo {author} {\bibfnamefont {N.}~\bibnamefont {Poccia}},
  \bibinfo {author} {\bibfnamefont {A.}~\bibnamefont {Ricci}}, \bibinfo
  {author} {\bibfnamefont {G.}~\bibnamefont {Campi}}, \bibinfo {author}
  {\bibfnamefont {M.}~\bibnamefont {Burghammer}}, \bibinfo {author}
  {\bibfnamefont {G.}~\bibnamefont {Aeppli}}, \ and\ \bibinfo {author}
  {\bibfnamefont {A.}~\bibnamefont {Bianconi}},\ }\bibfield  {title} {\enquote
  {\bibinfo {title} {Scale-free structural organization of oxygen interstitials
  in {La}$_{2}${CuO}$_{4+\mathrm{\ensuremath{y}}}$},}\ }\href {\doibase
  10.1038/nature09260} {\bibfield  {journal} {\bibinfo  {journal} {Nature}\
  }\textbf {\bibinfo {volume} {466}},\ \bibinfo {pages} {841--844} (\bibinfo
  {year} {2010})}\BibitemShut {NoStop}%
\bibitem [{\citenamefont {Poccia}\ \emph {et~al.}(2012)\citenamefont {Poccia},
  \citenamefont {Ricci}, \citenamefont {Campi}, \citenamefont {Fratini},
  \citenamefont {Puri}, \citenamefont {Gioacchino}, \citenamefont {Marcelli},
  \citenamefont {Reynolds}, \citenamefont {Burghammer}, \citenamefont {Saini},\
  and\ \citenamefont {Aeppli}}]{pnas.1208492109}%
  \BibitemOpen
  \bibfield  {author} {\bibinfo {author} {\bibfnamefont {N.}~\bibnamefont
  {Poccia}}, \bibinfo {author} {\bibfnamefont {A.}~\bibnamefont {Ricci}},
  \bibinfo {author} {\bibfnamefont {G.}~\bibnamefont {Campi}}, \bibinfo
  {author} {\bibfnamefont {M.}~\bibnamefont {Fratini}}, \bibinfo {author}
  {\bibfnamefont {A.}~\bibnamefont {Puri}}, \bibinfo {author} {\bibfnamefont
  {D.D.}\ \bibnamefont {Gioacchino}}, \bibinfo {author} {\bibfnamefont
  {A.}~\bibnamefont {Marcelli}}, \bibinfo {author} {\bibfnamefont
  {M.}~\bibnamefont {Reynolds}}, \bibinfo {author} {\bibfnamefont
  {M.}~\bibnamefont {Burghammer}}, \bibinfo {author} {\bibfnamefont {N.L.}\
  \bibnamefont {Saini}}, \ and\ \bibinfo {author} {\bibfnamefont
  {G.}~\bibnamefont {Aeppli}},\ }\bibfield  {title} {\enquote {\bibinfo {title}
  {{Optimum inhomogeneity of local lattice distortions in
  ${\mathrm{La}}_{2}$${\mathrm{CuO}}_{4+\mathrm{\ensuremath{y}}}$}},}\ }\href
  {\doibase 10.1073/pnas.1208492109} {\bibfield  {journal} {\bibinfo  {journal}
  {PNAS}\ }\textbf {\bibinfo {volume} {109}},\ \bibinfo {pages} {15685--15690}
  (\bibinfo {year} {2012})}\BibitemShut {NoStop}%
\bibitem [{\citenamefont {Campi}\ \emph {et~al.}(2015)\citenamefont {Campi},
  \citenamefont {Bianconi}, \citenamefont {Poccia}, \citenamefont {Bianconi},
  \citenamefont {Barba}, \citenamefont {Arrighetti}, \citenamefont {Innocenti},
  \citenamefont {Karpinski}, \citenamefont {Zhigadlo}, \citenamefont {Kazakov},
  \citenamefont {Burghammer}, \citenamefont {Zimmermann}, \citenamefont
  {Sprung},\ and\ \citenamefont {Ricci}}]{Campi2015}%
  \BibitemOpen
  \bibfield  {author} {\bibinfo {author} {\bibfnamefont {G.}~\bibnamefont
  {Campi}}, \bibinfo {author} {\bibfnamefont {A.}~\bibnamefont {Bianconi}},
  \bibinfo {author} {\bibfnamefont {N.}~\bibnamefont {Poccia}}, \bibinfo
  {author} {\bibfnamefont {G.}~\bibnamefont {Bianconi}}, \bibinfo {author}
  {\bibfnamefont {L.}~\bibnamefont {Barba}}, \bibinfo {author} {\bibfnamefont
  {G.}~\bibnamefont {Arrighetti}}, \bibinfo {author} {\bibfnamefont
  {D.}~\bibnamefont {Innocenti}}, \bibinfo {author} {\bibfnamefont
  {J.}~\bibnamefont {Karpinski}}, \bibinfo {author} {\bibfnamefont {N.~D.}\
  \bibnamefont {Zhigadlo}}, \bibinfo {author} {\bibfnamefont {S.~M.}\
  \bibnamefont {Kazakov}}, \bibinfo {author} {\bibfnamefont {M.}~\bibnamefont
  {Burghammer}}, \bibinfo {author} {\bibfnamefont {M.~v.}\ \bibnamefont
  {Zimmermann}}, \bibinfo {author} {\bibfnamefont {M.}~\bibnamefont {Sprung}},
  \ and\ \bibinfo {author} {\bibfnamefont {A.}~\bibnamefont {Ricci}},\
  }\bibfield  {title} {\enquote {\bibinfo {title} {{Inhomogeneity of
  charge-density-wave order and quenched disorder in a high-$T_c$
  superconductor}},}\ }\href {\doibase 10.1038/nature14987} {\bibfield
  {journal} {\bibinfo  {journal} {Nature}\ }\textbf {\bibinfo {volume} {525}},\
  \bibinfo {pages} {359--362} (\bibinfo {year} {2015})}\BibitemShut {NoStop}%
\bibitem [{\citenamefont {Wu}\ \emph {et~al.}(2013)\citenamefont {Wu},
  \citenamefont {Pelleg}, \citenamefont {Logvenov}, \citenamefont {Bollinger},
  \citenamefont {Sun}, \citenamefont {Boebinger}, \citenamefont {Vanević},
  \citenamefont {Radović},\ and\ \citenamefont
  {Božović}}]{10.1038/nmat3719}%
  \BibitemOpen
  \bibfield  {author} {\bibinfo {author} {\bibfnamefont {J.}~\bibnamefont
  {Wu}}, \bibinfo {author} {\bibfnamefont {O.}~\bibnamefont {Pelleg}}, \bibinfo
  {author} {\bibfnamefont {G.}~\bibnamefont {Logvenov}}, \bibinfo {author}
  {\bibfnamefont {A.~T.}\ \bibnamefont {Bollinger}}, \bibinfo {author}
  {\bibfnamefont {Y-J.}\ \bibnamefont {Sun}}, \bibinfo {author} {\bibfnamefont
  {G.~S.}\ \bibnamefont {Boebinger}}, \bibinfo {author} {\bibfnamefont
  {M.}~\bibnamefont {Vanević}}, \bibinfo {author} {\bibfnamefont
  {Z.}~\bibnamefont {Radović}}, \ and\ \bibinfo {author} {\bibfnamefont
  {I.}~\bibnamefont {Božović}},\ }\bibfield  {title} {\enquote {\bibinfo
  {title} {{Anomalous independence of interface superconductivity from carrier
  density}},}\ }\href {\doibase https://doi.org/10.1038/nmat3719} {\bibfield
  {journal} {\bibinfo  {journal} {Nat. Mat.}\ }\textbf {\bibinfo {volume}
  {12}},\ \bibinfo {pages} {877--881} (\bibinfo {year} {2013})}\BibitemShut
  {NoStop}%
\bibitem [{\citenamefont {Misawa}\ \emph {et~al.}(2016)\citenamefont {Misawa},
  \citenamefont {Nomura}, \citenamefont {Biermann},\ and\ \citenamefont
  {Imada}}]{sciadv.1600664}%
  \BibitemOpen
  \bibfield  {author} {\bibinfo {author} {\bibfnamefont {T.}~\bibnamefont
  {Misawa}}, \bibinfo {author} {\bibfnamefont {Y.}~\bibnamefont {Nomura}},
  \bibinfo {author} {\bibfnamefont {S.}~\bibnamefont {Biermann}}, \ and\
  \bibinfo {author} {\bibfnamefont {M.}~\bibnamefont {Imada}},\ }\bibfield
  {title} {\enquote {\bibinfo {title} {{Self-optimized superconductivity
  attainable by interlayer phase separation at cuprate interfaces}},}\ }\href
  {\doibase 10.1126/sciadv.1600664} {\bibfield  {journal} {\bibinfo  {journal}
  {Sci. Adv.}\ }\textbf {\bibinfo {volume} {2}},\ \bibinfo {pages} {e1600664}
  (\bibinfo {year} {2016})},\ \Eprint
  {http://arxiv.org/abs/https://www.science.org/doi/pdf/10.1126/sciadv.1600664}
  {https://www.science.org/doi/pdf/10.1126/sciadv.1600664} \BibitemShut
  {NoStop}%
\bibitem [{\citenamefont {Kamihara}\ \emph {et~al.}(2006)\citenamefont
  {Kamihara}, \citenamefont {Hiramatsu}, \citenamefont {Hirano}, \citenamefont
  {Kawamura}, \citenamefont {Yanagi}, \citenamefont {Kamiya},\ and\
  \citenamefont {Hosono}}]{ja063355c}%
  \BibitemOpen
  \bibfield  {author} {\bibinfo {author} {\bibfnamefont {Y.}~\bibnamefont
  {Kamihara}}, \bibinfo {author} {\bibfnamefont {H.}~\bibnamefont {Hiramatsu}},
  \bibinfo {author} {\bibfnamefont {M.}~\bibnamefont {Hirano}}, \bibinfo
  {author} {\bibfnamefont {R.}~\bibnamefont {Kawamura}}, \bibinfo {author}
  {\bibfnamefont {H.}~\bibnamefont {Yanagi}}, \bibinfo {author} {\bibfnamefont
  {T.}~\bibnamefont {Kamiya}}, \ and\ \bibinfo {author} {\bibfnamefont
  {H.}~\bibnamefont {Hosono}},\ }\bibfield  {title} {\enquote {\bibinfo {title}
  {{Iron-Based Layered Superconductor: LaOFeP}},}\ }\href {\doibase
  10.1021/ja063355c} {\bibfield  {journal} {\bibinfo  {journal} {J. Am. Chem.
  Soc.}\ }\textbf {\bibinfo {volume} {128}},\ \bibinfo {pages} {10012--10013}
  (\bibinfo {year} {2006})},\ \Eprint
  {http://arxiv.org/abs/https://doi.org/10.1021/ja063355c}
  {https://doi.org/10.1021/ja063355c} \BibitemShut {NoStop}%
\bibitem [{\citenamefont {Takahashi}\ \emph {et~al.}(2008)\citenamefont
  {Takahashi}, \citenamefont {Igawa}, \citenamefont {Arii}, \citenamefont
  {Kamihara}, \citenamefont {Hirano},\ and\ \citenamefont
  {Hosono}}]{nature06972}%
  \BibitemOpen
  \bibfield  {author} {\bibinfo {author} {\bibfnamefont {H.}~\bibnamefont
  {Takahashi}}, \bibinfo {author} {\bibfnamefont {K.}~\bibnamefont {Igawa}},
  \bibinfo {author} {\bibfnamefont {K.}~\bibnamefont {Arii}}, \bibinfo {author}
  {\bibfnamefont {Y.}~\bibnamefont {Kamihara}}, \bibinfo {author}
  {\bibfnamefont {M.}~\bibnamefont {Hirano}}, \ and\ \bibinfo {author}
  {\bibfnamefont {H.}~\bibnamefont {Hosono}},\ }\bibfield  {title} {\enquote
  {\bibinfo {title} {{Superconductivity at 43 K in an iron-based layered
  compound ${\mathrm{LaO}}_{1-x}$${\mathrm{F}}_{x}$${\mathrm{FeAs}}$}},}\
  }\href {\doibase 10.1038/nature06972} {\bibfield  {journal} {\bibinfo
  {journal} {Nature}\ }\textbf {\bibinfo {volume} {453}},\ \bibinfo {pages}
  {376--378} (\bibinfo {year} {2008})},\ \Eprint
  {http://arxiv.org/abs/https://doi.org/10.1038/nature06972}
  {https://doi.org/10.1038/nature06972} \BibitemShut {NoStop}%
\bibitem [{\citenamefont {Ricci}\ \emph {et~al.}(2011)\citenamefont {Ricci},
  \citenamefont {Poccia}, \citenamefont {Campi}, \citenamefont {Joseph},
  \citenamefont {Arrighetti}, \citenamefont {Barba}, \citenamefont {Reynolds},
  \citenamefont {Burghammer}, \citenamefont {Takeya}, \citenamefont
  {Mizuguchi}, \citenamefont {Takano}, \citenamefont {Colapietro},
  \citenamefont {Saini},\ and\ \citenamefont {Bianconi}}]{PhysRevB.84.060511}%
  \BibitemOpen
  \bibfield  {author} {\bibinfo {author} {\bibfnamefont {A.}~\bibnamefont
  {Ricci}}, \bibinfo {author} {\bibfnamefont {N.}~\bibnamefont {Poccia}},
  \bibinfo {author} {\bibfnamefont {G.}~\bibnamefont {Campi}}, \bibinfo
  {author} {\bibfnamefont {B.}~\bibnamefont {Joseph}}, \bibinfo {author}
  {\bibfnamefont {G.}~\bibnamefont {Arrighetti}}, \bibinfo {author}
  {\bibfnamefont {L.}~\bibnamefont {Barba}}, \bibinfo {author} {\bibfnamefont
  {M.}~\bibnamefont {Reynolds}}, \bibinfo {author} {\bibfnamefont
  {M.}~\bibnamefont {Burghammer}}, \bibinfo {author} {\bibfnamefont
  {H.}~\bibnamefont {Takeya}}, \bibinfo {author} {\bibfnamefont
  {Y.}~\bibnamefont {Mizuguchi}}, \bibinfo {author} {\bibfnamefont
  {Y.}~\bibnamefont {Takano}}, \bibinfo {author} {\bibfnamefont
  {M.}~\bibnamefont {Colapietro}}, \bibinfo {author} {\bibfnamefont {N.~L.}\
  \bibnamefont {Saini}}, \ and\ \bibinfo {author} {\bibfnamefont
  {A.}~\bibnamefont {Bianconi}},\ }\bibfield  {title} {\enquote {\bibinfo
  {title} {{Nanoscale phase separation in the iron chalcogenide superconductor
  K${}_{0.8}$Fe${}_{1.6}$Se${}_{2}$ as seen via scanning nanofocused x-ray
  diffraction}},}\ }\href {\doibase 10.1103/PhysRevB.84.060511} {\bibfield
  {journal} {\bibinfo  {journal} {Phys. Rev. B}\ }\textbf {\bibinfo {volume}
  {84}},\ \bibinfo {pages} {060511} (\bibinfo {year} {2011})}\BibitemShut
  {NoStop}%
\bibitem [{\citenamefont {Bendele}\ \emph {et~al.}(2014)\citenamefont
  {Bendele}, \citenamefont {Barinov}, \citenamefont {Joseph}, \citenamefont
  {Innocenti}, \citenamefont {Iadecola}, \citenamefont {Bianconi},
  \citenamefont {Takeya}, \citenamefont {Mizuguchi}, \citenamefont {Takano},
  \citenamefont {Noji}, \citenamefont {Hatakeda}, \citenamefont {Koike},
  \citenamefont {Horio}, \citenamefont {Fujimori}, \citenamefont {Ootsuki},
  \citenamefont {Mizokawa},\ and\ \citenamefont {Saini}}]{Bendele2014}%
  \BibitemOpen
  \bibfield  {author} {\bibinfo {author} {\bibfnamefont {M.}~\bibnamefont
  {Bendele}}, \bibinfo {author} {\bibfnamefont {A.}~\bibnamefont {Barinov}},
  \bibinfo {author} {\bibfnamefont {B.}~\bibnamefont {Joseph}}, \bibinfo
  {author} {\bibfnamefont {D.}~\bibnamefont {Innocenti}}, \bibinfo {author}
  {\bibfnamefont {A.}~\bibnamefont {Iadecola}}, \bibinfo {author}
  {\bibfnamefont {A.}~\bibnamefont {Bianconi}}, \bibinfo {author}
  {\bibfnamefont {H.}~\bibnamefont {Takeya}}, \bibinfo {author} {\bibfnamefont
  {Y.}~\bibnamefont {Mizuguchi}}, \bibinfo {author} {\bibfnamefont
  {Y.}~\bibnamefont {Takano}}, \bibinfo {author} {\bibfnamefont
  {T.}~\bibnamefont {Noji}}, \bibinfo {author} {\bibfnamefont {T.}~\bibnamefont
  {Hatakeda}}, \bibinfo {author} {\bibfnamefont {Y.}~\bibnamefont {Koike}},
  \bibinfo {author} {\bibfnamefont {M.}~\bibnamefont {Horio}}, \bibinfo
  {author} {\bibfnamefont {A.}~\bibnamefont {Fujimori}}, \bibinfo {author}
  {\bibfnamefont {D.}~\bibnamefont {Ootsuki}}, \bibinfo {author} {\bibfnamefont
  {T.}~\bibnamefont {Mizokawa}}, \ and\ \bibinfo {author} {\bibfnamefont
  {N.~L.}\ \bibnamefont {Saini}},\ }\bibfield  {title} {\enquote {\bibinfo
  {title} {{Spectromicroscopy of electronic phase separation in
  K${}_{x}$Fe${}_{2-y}$Se${}_{2}$ superconductor}},}\ }\href {\doibase
  10.1038/srep05592} {\bibfield  {journal} {\bibinfo  {journal} {Sci. Rep.}\
  }\textbf {\bibinfo {volume} {4}},\ \bibinfo {pages} {5592} (\bibinfo {year}
  {2014})}\BibitemShut {NoStop}%
\bibitem [{\citenamefont {Simonelli}\ \emph {et~al.}(2014)\citenamefont
  {Simonelli}, \citenamefont {Mizokawa}, \citenamefont {Sala}, \citenamefont
  {Takeya}, \citenamefont {Mizuguchi}, \citenamefont {Takano}, \citenamefont
  {Garbarino}, \citenamefont {Monaco},\ and\ \citenamefont
  {Saini}}]{PhysRevB.90.214516}%
  \BibitemOpen
  \bibfield  {author} {\bibinfo {author} {\bibfnamefont {L.}~\bibnamefont
  {Simonelli}}, \bibinfo {author} {\bibfnamefont {T.}~\bibnamefont {Mizokawa}},
  \bibinfo {author} {\bibfnamefont {M.~Moretti}\ \bibnamefont {Sala}}, \bibinfo
  {author} {\bibfnamefont {H.}~\bibnamefont {Takeya}}, \bibinfo {author}
  {\bibfnamefont {Y.}~\bibnamefont {Mizuguchi}}, \bibinfo {author}
  {\bibfnamefont {Y.}~\bibnamefont {Takano}}, \bibinfo {author} {\bibfnamefont
  {G.}~\bibnamefont {Garbarino}}, \bibinfo {author} {\bibfnamefont
  {G.}~\bibnamefont {Monaco}}, \ and\ \bibinfo {author} {\bibfnamefont {N.~L.}\
  \bibnamefont {Saini}},\ }\bibfield  {title} {\enquote {\bibinfo {title}
  {Temperature dependence of iron local magnetic moment in phase-separated
  superconducting chalcogenide},}\ }\href {\doibase 10.1103/PhysRevB.90.214516}
  {\bibfield  {journal} {\bibinfo  {journal} {Phys. Rev. B}\ }\textbf {\bibinfo
  {volume} {90}},\ \bibinfo {pages} {214516} (\bibinfo {year}
  {2014})}\BibitemShut {NoStop}%
\bibitem [{\citenamefont {Ricci}\ \emph {et~al.}(2015)\citenamefont {Ricci},
  \citenamefont {Poccia}, \citenamefont {Joseph}, \citenamefont {Innocenti},
  \citenamefont {Campi}, \citenamefont {Zozulya}, \citenamefont {Westermeier},
  \citenamefont {Schavkan}, \citenamefont {Coneri}, \citenamefont {Bianconi},
  \citenamefont {Takeya}, \citenamefont {Mizuguchi}, \citenamefont {Takano},
  \citenamefont {Mizokawa}, \citenamefont {Sprung},\ and\ \citenamefont
  {Saini}}]{PhysRevB.91.020503}%
  \BibitemOpen
  \bibfield  {author} {\bibinfo {author} {\bibfnamefont {A.}~\bibnamefont
  {Ricci}}, \bibinfo {author} {\bibfnamefont {N.}~\bibnamefont {Poccia}},
  \bibinfo {author} {\bibfnamefont {B.}~\bibnamefont {Joseph}}, \bibinfo
  {author} {\bibfnamefont {D.}~\bibnamefont {Innocenti}}, \bibinfo {author}
  {\bibfnamefont {G.}~\bibnamefont {Campi}}, \bibinfo {author} {\bibfnamefont
  {A.}~\bibnamefont {Zozulya}}, \bibinfo {author} {\bibfnamefont
  {F.}~\bibnamefont {Westermeier}}, \bibinfo {author} {\bibfnamefont
  {A.}~\bibnamefont {Schavkan}}, \bibinfo {author} {\bibfnamefont
  {F.}~\bibnamefont {Coneri}}, \bibinfo {author} {\bibfnamefont
  {A.}~\bibnamefont {Bianconi}}, \bibinfo {author} {\bibfnamefont
  {H.}~\bibnamefont {Takeya}}, \bibinfo {author} {\bibfnamefont
  {Y.}~\bibnamefont {Mizuguchi}}, \bibinfo {author} {\bibfnamefont
  {Y.}~\bibnamefont {Takano}}, \bibinfo {author} {\bibfnamefont
  {T.}~\bibnamefont {Mizokawa}}, \bibinfo {author} {\bibfnamefont
  {M.}~\bibnamefont {Sprung}}, \ and\ \bibinfo {author} {\bibfnamefont {N.~L.}\
  \bibnamefont {Saini}},\ }\bibfield  {title} {\enquote {\bibinfo {title}
  {Direct observation of nanoscale interface phase in the superconducting
  chalcogenide
  ${\mathrm{k}}_{x}{\mathrm{fe}}_{2\ensuremath{-}y}{\mathrm{se}}_{2}$ with
  intrinsic phase separation},}\ }\href {\doibase 10.1103/PhysRevB.91.020503}
  {\bibfield  {journal} {\bibinfo  {journal} {Phys. Rev. B}\ }\textbf {\bibinfo
  {volume} {91}},\ \bibinfo {pages} {020503} (\bibinfo {year}
  {2015})}\BibitemShut {NoStop}%
\end{thebibliography}%


\begin{thebibliography}{7}%
\makeatletter
\providecommand \@ifxundefined [1]{%
 \@ifx{#1\undefined}
}%
\providecommand \@ifnum [1]{%
 \ifnum #1\expandafter \@firstoftwo
 \else \expandafter \@secondoftwo
 \fi
}%
\providecommand \@ifx [1]{%
 \ifx #1\expandafter \@firstoftwo
 \else \expandafter \@secondoftwo
 \fi
}%
\providecommand \natexlab [1]{#1}%
\providecommand \enquote  [1]{``#1''}%
\providecommand \bibnamefont  [1]{#1}%
\providecommand \bibfnamefont [1]{#1}%
\providecommand \citenamefont [1]{#1}%
\providecommand \href@noop [0]{\@secondoftwo}%
\providecommand \href [0]{\begingroup \@sanitize@url \@href}%
\providecommand \@href[1]{\@@startlink{#1}\@@href}%
\providecommand \@@href[1]{\endgroup#1\@@endlink}%
\providecommand \@sanitize@url [0]{\catcode `\\12\catcode `\$12\catcode
  `\&12\catcode `\#12\catcode `\^12\catcode `\_12\catcode `\%12\relax}%
\providecommand \@@startlink[1]{}%
\providecommand \@@endlink[0]{}%
\providecommand \url  [0]{\begingroup\@sanitize@url \@url }%
\providecommand \@url [1]{\endgroup\@href {#1}{\urlprefix }}%
\providecommand \urlprefix  [0]{URL }%
\providecommand \Eprint [0]{\href }%
\providecommand \doibase [0]{http://dx.doi.org/}%
\providecommand \selectlanguage [0]{\@gobble}%
\providecommand \bibinfo  [0]{\@secondoftwo}%
\providecommand \bibfield  [0]{\@secondoftwo}%
\providecommand \translation [1]{[#1]}%
\providecommand \BibitemOpen [0]{}%
\providecommand \bibitemStop [0]{}%
\providecommand \bibitemNoStop [0]{.\EOS\space}%
\providecommand \EOS [0]{\spacefactor3000\relax}%
\providecommand \BibitemShut  [1]{\csname bibitem#1\endcsname}%
\let\auto@bib@innerbib\@empty
%</preamble>
\bibitem [{\citenamefont {{Linn{\'e}r}}\ \emph {et~al.}(2022)\citenamefont
  {{Linn{\'e}r}}, \citenamefont {{Lichtenstein}}, \citenamefont {{Biermann}},\
  and\ \citenamefont {{Stepanov}}}]{2022arXiv221005540L}%
  \BibitemOpen
  \bibfield  {author} {\bibinfo {author} {\bibfnamefont {E.}~\bibnamefont
  {{Linn{\'e}r}}}, \bibinfo {author} {\bibfnamefont {A.~I.}\ \bibnamefont
  {{Lichtenstein}}}, \bibinfo {author} {\bibfnamefont {S.}~\bibnamefont
  {{Biermann}}}, \ and\ \bibinfo {author} {\bibfnamefont {E.~A.}\ \bibnamefont
  {{Stepanov}}},\ }\bibfield  {title} {\enquote {\bibinfo {title}
  {{Multi-channel fluctuating field approach to competing instabilities in
  interacting electronic systems}},}\ }\href@noop {} {\bibfield  {journal}
  {\bibinfo  {journal} {arXiv e-prints}\ } (\bibinfo {year} {2022})},\ \Eprint
  {http://arxiv.org/abs/2210.05540} {arXiv:2210.05540} \BibitemShut {NoStop}%
\bibitem [{\citenamefont {Yang}(1989)}]{PhysRevLett.63.2144}%
  \BibitemOpen
  \bibfield  {author} {\bibinfo {author} {\bibfnamefont {C.~N.}\ \bibnamefont
  {Yang}},\ }\bibfield  {title} {\enquote {\bibinfo {title} {{\ensuremath{\eta}
  pairing and off-diagonal long-range order in a Hubbard model}},}\ }\href
  {\doibase 10.1103/PhysRevLett.63.2144} {\bibfield  {journal} {\bibinfo
  {journal} {Phys. Rev. Lett.}\ }\textbf {\bibinfo {volume} {63}},\ \bibinfo
  {pages} {2144--2147} (\bibinfo {year} {1989})}\BibitemShut {NoStop}%
\bibitem [{\citenamefont {Zhang}(1990)}]{PhysRevLett.65.120}%
  \BibitemOpen
  \bibfield  {author} {\bibinfo {author} {\bibfnamefont {S.}~\bibnamefont
  {Zhang}},\ }\bibfield  {title} {\enquote {\bibinfo {title} {{Pseudospin
  symmetry and new collective modes of the Hubbard model}},}\ }\href {\doibase
  10.1103/PhysRevLett.65.120} {\bibfield  {journal} {\bibinfo  {journal} {Phys.
  Rev. Lett.}\ }\textbf {\bibinfo {volume} {65}},\ \bibinfo {pages} {120--122}
  (\bibinfo {year} {1990})}\BibitemShut {NoStop}%
\bibitem [{\citenamefont {Peierls}(1938)}]{PhysRev.54.918}%
  \BibitemOpen
  \bibfield  {author} {\bibinfo {author} {\bibfnamefont {R.}~\bibnamefont
  {Peierls}},\ }\bibfield  {title} {\enquote {\bibinfo {title} {{On a Minimum
  Property of the Free Energy}},}\ }\href {\doibase 10.1103/PhysRev.54.918}
  {\bibfield  {journal} {\bibinfo  {journal} {Phys. Rev.}\ }\textbf {\bibinfo
  {volume} {54}},\ \bibinfo {pages} {918--919} (\bibinfo {year}
  {1938})}\BibitemShut {NoStop}%
\bibitem [{\citenamefont {Bogolyubov}(1958)}]{Bogolyubov:1958zv}%
  \BibitemOpen
  \bibfield  {author} {\bibinfo {author} {\bibfnamefont {N.~N.}\ \bibnamefont
  {Bogolyubov}},\ }\bibfield  {title} {\enquote {\bibinfo {title} {{On a
  variational principle in the many-body problem}},}\ }\href@noop {} {\bibfield
   {journal} {\bibinfo  {journal} {Sov. Phys. Dokl.}\ }\textbf {\bibinfo
  {volume} {3}},\ \bibinfo {pages} {292--294} (\bibinfo {year}
  {1958})}\BibitemShut {NoStop}%
\bibitem [{\citenamefont {Feynman}(1972)}]{feynman1972}%
  \BibitemOpen
  \bibfield  {author} {\bibinfo {author} {\bibfnamefont {R.~P.}\ \bibnamefont
  {Feynman}},\ }\href@noop {} {\emph {\bibinfo {title} {{Statistical mechanics:
  A set of lectures}}}}\ (\bibinfo  {publisher} {Reading, Mass:
  Benjamin/Cummings},\ \bibinfo {year} {1972})\BibitemShut {NoStop}%
\bibitem [{\citenamefont {Rubtsov}\ \emph {et~al.}(2020)\citenamefont
  {Rubtsov}, \citenamefont {Stepanov},\ and\ \citenamefont
  {Lichtenstein}}]{PhysRevB.102.224423}%
  \BibitemOpen
  \bibfield  {author} {\bibinfo {author} {\bibfnamefont {A.~N.}\ \bibnamefont
  {Rubtsov}}, \bibinfo {author} {\bibfnamefont {E.~A.}\ \bibnamefont
  {Stepanov}}, \ and\ \bibinfo {author} {\bibfnamefont {A.~I.}\ \bibnamefont
  {Lichtenstein}},\ }\bibfield  {title} {\enquote {\bibinfo {title}
  {{Collective magnetic fluctuations in Hubbard plaquettes captured by
  fluctuating local field method}},}\ }\href {\doibase
  10.1103/PhysRevB.102.224423} {\bibfield  {journal} {\bibinfo  {journal}
  {Phys. Rev. B}\ }\textbf {\bibinfo {volume} {102}},\ \bibinfo {pages}
  {224423} (\bibinfo {year} {2020})}\BibitemShut {NoStop}%
\end{thebibliography}%

\end{document}

\title{
Supplemental Material\\[0.5cm]
Coexistence of $s$-wave superconductivity and phase separation in the \\half-filled extended Hubbard model with attractive interactions
}

\author{E. Linn\'er}
\affiliation{CPHT, CNRS, {\'E}cole polytechnique, Institut Polytechnique de Paris, 91120 Palaiseau, France}

\author{C. Dutreix}
\affiliation{Univ. Bordeaux, CNRS, LOMA, UMR 5798, F-33405 Talence, France}

\author{S. Biermann}
\affiliation{CPHT, CNRS, {\'E}cole polytechnique, Institut Polytechnique de Paris, 91120 Palaiseau, France}
\affiliation{Coll\`ege de France, 11 place Marcelin Berthelot, 75005 Paris, France}
\affiliation{Department of Physics, Division of Mathematical Physics, Lund University, Professorsgatan 1, 22363 Lund, Sweden}
\affiliation{European Theoretical Spectroscopy Facility, 91128 Palaiseau, France}

\author{E. A. Stepanov}
\affiliation{CPHT, CNRS, {\'E}cole polytechnique, Institut Polytechnique de Paris, 91120 Palaiseau, France}

\maketitle

\onecolumngrid

\vspace{-0.8cm}

\section{Model}
In order to apply the multi-channel fluctuating field (MCFF) method~\cite{2022arXiv221005540L} in the following sections, it is convenient to work within the action formalism.
The action for a single-band extended Hubbard model is the following:  
\begin{align}
\mathcal{S} = - \frac{1}{\beta N}\sum_{\bf{k},\nu ,\sigma} c_{\bf{k}\nu\sigma}^{*} \mathcal{G}_{\bf{k}\nu}^{-1} c^{\phantom{*}}_{\bf{k}\nu\sigma} + \frac{U}{\beta N} \sum_{\bf{q},\omega} \rho_{\bf{q}\omega\uparrow} \rho_{-\bf{q},-\omega\downarrow}
+ \frac{1}{2 \beta N}\sum_{\bf{q},\omega,\sigma\sigma'}V_{\bf{q}} \rho_{\bf{q}\omega\sigma}\rho_{-\bf{q},-\omega\sigma'},
\label{Eq:EHubbard_Action}
\end{align}
with the inverse temperature ${\beta=1/T}$ and number of lattice sites $N$. 
Grassmann variables $c^{(*)}$ correspond to the annihilation (creation) of electrons, where the subscripts denote the momentum $\bf{k}$ and fermionic Matsubara frequency $\nu$.
The inverse of the bare (non-interacting) Green's function is defined as ${\mathcal{G}^{-1}_{\bf{k}\nu} = i\nu + \mu - \epsilon_{\bf{k}}}$, where $\mu$ is the chemical potential and ${\epsilon_{\bf{k}}=-2t(\cos{k_x} + \cos{k_y})}$ is the bare dispersion due to nearest-neighbor hopping amplitudes $t$ on a two-dimensional square lattice.
The interaction is modeled by the on-site $U$ and the nearest-neighbor ${V_{\bf{q}} = 2V(\cos{q_x} + \cos{q_y})}$ interactions.
For convenience, the interaction terms in Eq.~\eqref{Eq:EHubbard_Action} are written for the shifted electronic densities ${\rho_{\bf{q}\omega\sigma} = n_{\bf{q}\omega\sigma} - \langle n_{\bf{q}\omega\sigma}\rangle\delta_{\bf{q},\bf{0}}\delta_{\omega,0}}$ with ${n_{{\bf q}\omega\sigma} = \sum_{{\bf k},\nu}c^{*}_{{\bf k+q},\nu+\omega\sigma} c^{\phantom{*}}_{{\bf k}\nu\sigma}}$, where ${\bf q}$ and $\omega$ are the momentum and bosonic Matsubara frequency indices, respectively.
The reason for this shift is discussed in detail in Ref.~\cite{2022arXiv221005540L}.
Our considerations are limited to attractive $U$, while $V$ may be both, repulsive and attractive.
However, we also note, that the extended Hubbard model with attractive local interaction is analogues, by the staggered particle-hole transformation associate with $\eta$-pairing~\cite{PhysRevLett.63.2144, PhysRevLett.65.120}, to the repulsive Hubbard model with an additional $V_{\bf q}m^{z}_{{\bf{q}}\omega}m^{z}_{{-\bf{q}},-\omega}$ term describing a nearest-neighbor ferromagnetic (FM) or AFM exchange coupling in the spin $z$-direction, with $m^{z}_{{\bf{q}}\omega} = \rho_{{\bf{q}}\omega \uparrow}-\rho_{{\bf{q}}\omega \downarrow}$.

\section{Multi-channel fluctuating field approach}
Our aim of the current section is to give a detailed account of the MCFF approach applied to the half-filled two-dimensional extended Hubbard model in the attractive $U$ regime.

\begin{center}
{\bf Trial action}
\end{center}
We now employ the MCFF approach by constructing a trial action:
\begin{align}
\mathcal{S}^{*} =  &- \frac{1}{\beta N}\sum_{\bf{k},\nu,\sigma} c_{\bf{k}\nu\sigma}^* \mathcal{G}_{\bf{k}\nu}^{-1} c^{\phantom{*}}_{\bf{k}\nu\sigma} + \sum_{\bf{Q},\varsigma} \left[ \phi_{\bf{Q}}^\varsigma \rho^\varsigma_{-\bf{Q}} - \frac{1}{2}\frac{\beta N}{J^\varsigma_{\bf{Q}}}\phi_{\bf{Q}}^\varsigma \phi_{-\bf{Q}}^\varsigma \right]
\label{Eq:EHubbard_FF_Action}
\end{align}
that explicitly takes into account fluctuations in the charge and superconducting channels by a set of classical vector pseudo-spin (${\varsigma \in \{x,y,z\}}$) fields $\phi^\varsigma_{\bf{Q}}$ coupled to the composite variables ${\rho^{\varsigma}_{\bf{Q}} = n^{\varsigma}_{\bf{Q}} - \langle n^{\varsigma}_{\bf{Q}}\rangle\delta^{\phantom{*}}_{\bf{Q},\bf{0}}}$.
These variables are associated with the classical (${\omega=0}$) order parameters and defined as:
\begin{align}
n^{\varsigma}_{{\bf{Q}}} & \equiv \frac{1}{\beta{}N}\sum_{{\bf k}, \nu, \sigma\sigma'} \psi^{*}_{\bf{k+Q},\nu\sigma}\sigma^{\varsigma}_{\sigma\sigma'}\psi^{\phantom{*}}_{{\bf k}\nu\sigma'}.
\end{align}
In this expression, ${\bf Q}$ is the ordering vector, $\sigma^{\varsigma}$ is the Pauli spin matrix, and the Grassman variables $\psi^{(*)}$ are the Nambu spinors: {${\psi_{\bf{k},\omega,\uparrow} =  c_{\bf{k}\omega\uparrow}}$}, {${\psi_{\bf{k},\omega,\downarrow} =  c^*_{-\bf{k}+\bf{Q},-\omega\downarrow}}$}, {${\psi^*_{\bf{k},\omega,\uparrow} =  c^*_{\bf{k}\omega\uparrow}}$}, and {${\psi^*_{\bf{k},\omega,\downarrow} =  c_{-\bf{k}+\bf{Q},-\omega\downarrow}}$}.
Nambu spinors are introduced for a clear exhibition of the pseudo-spin structure arising at half-filling in the effective action~\eqref{Eq:EHubbard_FF_Action} in the absence of the non-local interaction $V$~\cite{PhysRevLett.63.2144, PhysRevLett.65.120}.
Hence, $n^{\varsigma}_{{\bf{Q}}}$ combines $s$-wave pairing ($\varsigma \in \{x,y\}$), and charge ($\varsigma \in \{z\}$) fluctuations.
Note, that the notations for the electronic $\rho_{\bf{q}\omega\sigma}$ and the pseudo-spin $\rho^{\varsigma}_{\bf{Q}}$ densities differ by the superscript.
The amplitude of the fluctuations of the classical fields $\phi^\varsigma_{\bf{Q}}$ is governed by a set of stiffness constants $J^{\varsigma}_{\bf{Q}}$ that are determined by the Peierls-Feynman-Bogoliubov variational principle described below~\cite{PhysRev.54.918, Bogolyubov:1958zv, feynman1972}.

\begin{center}
{\bf Free energy and order parameters}
\end{center}
Due to the Gaussian form with respect to the Grassmann variables $c^{(*)}$ of the action~\eqref{Eq:EHubbard_FF_Action}, one may construct an effective action for the classical degrees of freedom by analytically integrating out the fermionic degrees of freedom. 
The effective action for the classical fields is of the following form:
\begin{align}
\mathcal{S}_{\phi} = &- \Tr \ln \left[ \mathcal{G}^{-1}_{{\bf k}\nu}\delta^{\phantom{*}}_{{\bf Q},0}\delta^{\phantom{*}}_{\sigma,\sigma'} - \sum_{\varsigma}\phi_{\bf{Q}}^\varsigma \sigma^{\varsigma}_{\sigma\sigma'} \right] - \frac{1}{2}\sum_{\bf{Q},\varsigma}\frac{\beta N}{J^\varsigma_{\bf{Q}}}\phi_{\bf{Q}}^\varsigma \phi_{-\bf{Q}}^\varsigma.
\label{Eq:EHubbard_FF_Action_FluctuatingFields}
\end{align}
Here, the trace is taken over the momenta ${\bf k, Q}$, frequency $\nu$, and spin ${\sigma, \sigma'}$ indices.
The effective action $\mathcal{S}_{\phi}$ allows for a simplified description of the interplay between the collective electronic fluctuations.
In order to study the interplay between different fluctuations, one may construct a single-channel free energy
\begin{align}
{\cal F}(\phi^{a}) \equiv - \frac{1}{\beta N}\ln \left[\int d\phi^{b}\,\exp\big\{-\mathcal{S}_{\phi} \big\} \right],
\label{Eq_FreeEnergy}
\end{align}
where all classical fields $\phi^{b}$, except for a single field of interest $\phi^{a}$, are integrated out numerically exactly.
For a few collective modes, as holds in the current work, the numerical integration over the fields $\phi^{b}$ may be performed by the trapezoidal rule over a sufficiently dense grid.
Phase transitions are then identified by the development of a global minima of the free energy~\eqref{Eq_FreeEnergy} at ${\phi^{a} \neq 0}$.
In contrast, metastable fluctuations can be recognized by a local minima of the free energy~\eqref{Eq_FreeEnergy} at ${\phi^{a} \neq 0}$.
To retrieve further insight into the interplay between collective ordering, we find it useful to calculate the corresponding order parameters $\langle n^\varsigma_{\bf{Q}}\rangle$.
This can be done by substituting all the ${{\cal F}(\phi^{a})}$ saddle-point values of the classical fields ${\phi^{a}}$ in the effective fermionic action appearing within the trace of the logarithm in the action $\mathcal{S}_{\phi}$ in Eq.~\eqref{Eq:EHubbard_FF_Action_FluctuatingFields}.
The quantity $\langle n^\varsigma_{\bf{Q}}\rangle$ is thus associated with the global minima which the system fluctuates around.

\begin{center}
{\bf Determination of the stiffness parameters via a variational principle}
\end{center}
In order to calculate the quantities introduced in the previous section, one is required to apply the Peierls-Feynman-Bogoliubov variational principle~\cite{PhysRev.54.918,Bogolyubov:1958zv,feynman1972} to map the initial problem~\eqref{Eq:EHubbard_Action} on the trial action~\eqref{Eq:EHubbard_FF_Action_Fermion}.
This mapping is associated with the determination of the stiffness parameters $J^\varsigma_{\bf{Q}}$ by the variational principle following the work in Ref.~\onlinecite{PhysRevB.102.224423,2022arXiv221005540L}. 
This variational principle corresponds to the minimization of the functional
\begin{align}
\mathcal{F}(J^\varsigma_{\bf{Q}}) & = \mathcal{F}_{c}(J^\varsigma_{\bf{Q}}) + \frac{1}{\beta N}\left\langle \mathcal{S} -\mathcal{S}_{c} \right\rangle_{{\cal S}_c}
\label{Eq:F_variational}
\end{align}
by varying $J^\varsigma_{\bf{Q}}$, allowing to construct a unique and unambiguous set of $J^\varsigma_{\bf{Q}}$.
Here, $\langle \ldots \rangle_{{\cal S}_c}$ denotes the expectation value with respect to the effective fermionic action $\mathcal{S}_{c}$:
\begin{align}
\mathcal{S}_{c} & = - \frac{1}{\beta N}\sum_{\bf{k},\nu,\sigma} c_{\bf{k}\nu\sigma}^* \mathcal{G}_{\bf{k}\nu}^{-1} c^{\phantom{*}}_{\bf{k}\nu\sigma} + \frac{1}{2} \sum_{\bf{Q},\varsigma}\frac{J^\varsigma_{\bf{Q}}}{\beta N}\rho_{\bf{Q}}^\varsigma \rho_{-\bf{Q}}^\varsigma,\label{Eq:EHubbard_FF_Action_Fermion}
\end{align}
in which the classical field degrees of freedom are integrated out in the action $\mathcal{S}^{*}$ in Eq.~\eqref{Eq:EHubbard_FF_Action}.
In addition, the free energy ${\mathcal{F}_{c}(J^\varsigma_{\bf{Q}}) = - \ln {({\cal Z}_{c}})}/{\beta N}$ is introduced, where ${\cal Z}_{c}$ is the partition function of the action $\mathcal{S}_{c}$.
A useful explicit relation, allowing to evaluate $\langle \ldots \rangle_{{\cal S}_c}$, is the following~\cite{PhysRevB.102.224423,2022arXiv221005540L}:
\begin{equation}
    \langle ... \rangle_{{\cal S}_c} = \langle \langle ... \rangle_{{\cal S}_e}\rangle_{{\cal S}_\phi},\label{Eq:ExpectionValue}
\end{equation}
where the inner expectation value is taken with respect to the fermionic part of the trial action~\eqref{Eq:EHubbard_FF_Action}:
\begin{align}
\mathcal{S}_{e} & =  - \frac{1}{\beta N}\sum_{\bf{k},\nu,\sigma} c_{\bf{k}\nu\sigma}^* \mathcal{G}_{\bf{k}\nu}^{-1} c^{\phantom{*}}_{\bf{k}\nu\sigma} + \sum_{\bf{Q},\varsigma} \phi_{\bf{Q}}^\varsigma \rho^\varsigma_{-\bf{Q}},
\label{Eq:EHubbard_FF_Action_Fermion_Inner}
\end{align}
which depends on the classical fields $\phi_{\bf{Q}}^\varsigma$.
Due to inner expectation value being determined by the $\mathcal{S}_{e}$, which is Gaussian with respect to the fermions, Wick's theorem applies.

Our aim is now to identify the stiffness parameters $J^\varsigma_{\bf{Q}}$ in the charge and s-wave pairing channels.
Following Ref.~\onlinecite{2022arXiv221005540L}, the free energy can be explicitly rewritten as:
\begin{align}
\mathcal{F}(J^\varsigma_{\bf{Q}}) & = \mathcal{F}_{c}(J^\varsigma_{\bf{Q}}) + \frac{1}{\beta N} \left\langle \frac{U}{\beta N}\sum_{{\bf{q}},\omega} \rho_{\bf{q}\omega\uparrow} \rho_{-\bf{q},-\omega\downarrow} + \frac{1}{2}\sum_{{\bf{q}},\omega}\frac{V_{\bf{q}}}{\beta N} \rho_{\bf{q}\omega}\rho_{-\bf{q},-\omega} - \sum_{\varsigma}\frac{1}{2}\frac{J^\varsigma_{\bf{Q}}}{\beta N}\rho_{\bf{Q}}^\varsigma \rho_{\bf{Q}}^\varsigma \right\rangle_{{\cal S}_c}
\label{eq:F_app}
\end{align}
Now exploiting Eq.~\eqref{Eq:ExpectionValue}, we may rewrite the local interaction term explicitly using Wick's theorem as:
\begin{align}
    U \left\langle n_{j\tau\uparrow} n_{j\tau\downarrow}\right\rangle_{{\cal S}_c} & = U \left\langle\left\langle c^{\dagger}_{j\tau\uparrow}c_{j\tau\uparrow} c^{\dagger}_{j\tau\downarrow}c_{j\tau\downarrow}\right\rangle_{{\cal S}_e}\right\rangle_{{\cal S}_\phi}
     = U \left\langle\left\langle c^{\dagger}_{j\tau\uparrow}c_{j\tau\uparrow}\right\rangle_{{\cal S}_e} \left\langle c^{\dagger}_{j\tau\downarrow}c_{j\tau\downarrow}\right\rangle_{{\cal S}_e}+\left\langle c^{\dagger}_{j\tau\uparrow}c^{\dagger}_{j\tau\downarrow}\right\rangle_{{\cal S}_e} \left\langle c_{j\tau\downarrow}c_{j\tau\uparrow}\right\rangle_{{\cal S}_e} \right\rangle_{{\cal S}_\phi}
     = \frac{U}{4}\left\langle\left\langle \vec{n}^{\varsigma}_{j\tau}\right\rangle_{{\cal S}_e}^2 \right\rangle_{{\cal S}_\phi},
\end{align}
where the construction of the pseudo-spin density $n^{\varsigma}_{j\tau}$ employs the Nambu spinor formalism, i.e. {${\psi_{j\tau\uparrow} =  c_{j\tau\uparrow}}$}, {${\psi_{j\tau\downarrow} =  (-1)^{j}c^{\dagger}_{j\tau\downarrow}}$}, {${\psi^{\dagger}_{j\tau\uparrow} =  c^{\dagger}_{j\tau\uparrow}}$}, and {${\psi^{\dagger}_{j\tau\downarrow} =  (-1)^{j}c_{j\tau\downarrow}}$}, and with the real-space representation being employed.
Now employing the Fourier transform, we obtain the following:
\begin{align}
    \sum_{j,\tau} U \left\langle \rho_{j\tau\uparrow} \rho_{j\tau\downarrow}\right\rangle_{{\cal S}_c} = \frac{1}{\beta N} \sum_{\bf{q},\omega} \frac{U}{4}\left\langle \left\langle\vec{\rho}^{\varsigma}_{\bf{q}\omega}\right\rangle_{{\cal S}_e} \cdot \left\langle \vec{\rho}^{\varsigma}_{-\bf{q},-\omega}\right\rangle_{{\cal S}_e} \right\rangle_{{\cal S}_\phi}.
\end{align}
Equivalently to Ref.~\onlinecite{2022arXiv221005540L}, the non-local interaction term may be rewritten approximately using Wick's theorem as:
\begin{align}
    \frac{1}{2} V_{ij}\left\langle n_{i\tau}n_{j\tau} \right\rangle_{{\cal S}_c} & = \frac{1}{2} V_{ij} \sum_{\sigma\sigma'}\left\langle\left\langle c^{\dagger}_{i\tau\sigma}c^{\phantom{\dagger}}_{i\tau\sigma}c^{\dagger}_{j\tau\sigma^\prime}c^{\phantom{\dagger}}_{j\tau\sigma^\prime} \right\rangle_{{\cal S}_e}\right\rangle_{{\cal S}_\phi}
     = \frac{1}{2} V_{ij} \sum_{\sigma\sigma'}\left\langle\left\langle c^{\dagger}_{i\tau\sigma}c_{i\tau\sigma}\right\rangle_{{\cal S}_e}\left\langle c^{\dagger}_{j\tau\sigma^\prime}c_{j\tau\sigma^\prime} \right\rangle_{{\cal S}_e} + \left\langle c^{\dagger}_{i\tau\sigma} c^{\dagger}_{j\tau\sigma^\prime} \right\rangle_{{\cal S}_e}\left\langle c_{j\tau\sigma^\prime} c_{i\tau\sigma} \right\rangle_{{\cal S}_e} \right\rangle_{{\cal S}_\phi}\nonumber \\
    & \approx \frac{1}{2} V_{ij}\left\langle\left\langle n^{z}_{i\tau}\right\rangle_{{\cal S}_e} \left\langle n^{z}_{j\tau}\right\rangle_{{\cal S}_e} \right\rangle_{{\cal S}_\phi},
\end{align}
where ${i\neq j}$ and with sub-leading non-local expectation values scaling as $1/N$ being dropped, see Ref.~\onlinecite{PhysRevB.102.224423}.
Now employing the Fourier transform, we obtain the following:
\begin{align}
    \frac{1}{2}\sum_{ij,\tau} V_{ij}\left\langle \rho_{i\tau} \rho_{j\tau} \right\rangle_{{\cal S}_c} & \approx \frac{1}{2\beta N}\sum_{\bf{q},\omega} V_{\bf{q}} \left\langle\left\langle \rho^{z}_{\bf{q}\omega}\right\rangle_{{\cal S}_e} \left\langle \rho^{z}_{-\bf{q},-\omega} \right\rangle_{{\cal S}_e}\right\rangle_{{\cal S}_\phi}.
\end{align}
Finally, the expectation value of the interaction term in the MCFF action is approximately:
\begin{align}
    \frac{1}{2}\frac{J_{\bf{Q}}^{\varsigma}}{\beta N}\left\langle {\rho_{\bf{Q}}^\varsigma}^2 \right\rangle_{{\cal S}_c} & \approx \frac{1}{2}\frac{J_{\bf{Q}}^{\varsigma}}{\beta N}\left\langle\left\langle \rho_{\bf{Q}}^\varsigma \right\rangle^2_{{\cal S}_e}\right\rangle_{\cal{S}_\phi},
\end{align}
where again the sub-leading non-local expectation values scaling as $1/N$ have been dropped~\cite{PhysRevB.102.224423}.
The form of the MCFF action ${\cal S}_{e}$~\eqref{Eq:EHubbard_FF_Action_Fermion_Inner} only allows for certain quasi-momentum modes of the local and non-local interaction terms to contribute to the free energy, i.e. the components with ${\omega=0}$ and ${{\bf q = Q}}$ contribute to the average of the shifted density: ${\langle \rho^{\varsigma}_{{\bf q}\omega} \rangle_{{\cal S}_{c}} = \langle \rho^{\varsigma}_{{\bf Q}} \rangle_{{\cal S}_{c}}}$.
This allows to rewrite the free energy~\eqref{eq:F_app} in the following form:
\begin{align}
    \mathcal{F}(J^\varsigma_{\bf{Q}}) & \approx \mathcal{F}_{c}(J^\varsigma_{\bf{Q}}) + \frac{1}{(\beta N)^{2}}\left(\frac{U}{4} + \frac{V_{\bf{Q}}}{2} -\frac{J_{\bf{Q}}^{z}}{2} \right) \left\langle\left\langle \rho^{z}_{{\bf{Q}}}\right\rangle_{{\cal S}_e}^2\right\rangle_{{\cal S}_\phi} + \frac{1}{(\beta N)^2} \sum_{i=x,y}\left(\frac{U}{4} - \frac{J_{\bf{Q}}^{i}}{2} \right) \left\langle\left\langle \rho^{i}_{{\bf{Q}}}\right\rangle_{
    {\cal S}_e}^2 \right\rangle_{{\cal S}_\phi}.
\end{align}
For the leading instabilities that dominate the collective electronic behavior in the attractive $U$ case we thus retrieve: ${J^{z}_{(\pi,\pi)} = U/2 - 4V}$ for CDW fluctuations in agreement with Ref.~\onlinecite{2022arXiv221005540L}, ${J^{x/y}_{(\pi,\pi)} = U/2}$ for \mbox{s-SC} fluctuations, and ${J^{z}_{(0,0)^+} = U/2 + 4V}$ for PS fluctuations.
Note the equivalence ${J^{x/y}_{(\pi,\pi)} = J^{z}_{(\pi,\pi)}}$ for ${V=0}$ associated with the emergence of the pseudo-spin symmetry in the MCFF action~\eqref{Eq:EHubbard_FF_Action}.
The choice to keep only the main ${\bf Q}$ classical ${\omega=0}$ mode for each fluctuation is argued to be sufficient for predicting the phase boundaries~\cite{2022arXiv221005540L}.

\bibliography{Bib_MCFF}